\documentclass{aa}  
\usepackage{graphicx}
\usepackage[varg]{txfonts}
\usepackage{appendix}
\def\cm#1{\ifmmode {\,{\rm cm^{-#1}}}                  
        \else \hbox{$\,${\rm cm$^{\rm -#1}$}}\fi}
\def\raw {\ifmmode\rightarrow\else$\rightarrow$\fi}
\def\ex#1{\ifmmode {\times 10^{#1}}         
        \else \hbox{{$\times 10^{\rm #1}$}}\fi}
\def\xmol#1{\ifmmode {\mbox{X({\rm #1})}}
            \else
            \hbox{\mbox{$X$(#1)}}
           \fi}
\def\snu#1{\ifmmode {S_\nu\,\propto\,\nu^{#1}}
          \else \hbox{$S_\nu$\,$\propto$\,$\nu^{#1}$}\fi}

\newcommand{\iram}{IRAM\,30\,m}

\newcommand{\kms}{\mbox{km~s$^{-1}$}}

\newcommand{\mloss}{\mbox{$\dot{M}$}}
\newcommand{\mpagb}{\mbox{$\dot{M}_{\rm pAGB}$}}
\newcommand{\my}{\mbox{$M_{\odot}$~yr$^{-1}$}}
\newcommand{\ls}{\mbox{$L_{\odot}$}}
\newcommand{\msun}{\mbox{$M_{\odot}$}}

\newcommand{\rs}{\mbox{$R_{\star}$}}
\newcommand{\rin}{\mbox{$R_{\rm in}$}}
\newcommand{\rout}{\mbox{$R_{\rm out}$}}

\newcommand{\vexp}{\mbox{$V_{\rm exp}$}} 
\newcommand{\vrot}{\mbox{$V_{\rm rot}$}}

\newcommand{\dv}{\mbox{$\Delta \upsilonup$}}

\newcommand{\vsys}{\mbox{$V_{\rm sys}$}} 
\newcommand{\vlsr}{\mbox{$V_{\rm LSR}$}} 

\newcommand{\vorbu}{\mbox{$V^1_{\rm orb}$}} 
 
\newcommand{\h}{$^{\rm h}$}
\newcommand{\m}{$^{\rm m}$}
\newcommand{\stime}{$^{\rm s}$}

\newcommand{\ta}{\mbox{$T^*_{\rm A}$}}
\newcommand{\tl}{\mbox{$T_{\rm L}$}}

\newcommand{\tmb}{\mbox{$T_{\rm MB}$}}

\newcommand{\teff}{\mbox{$T_{\rm eff}$}}

\newcommand{\te}{\mbox{$T_{\rm e}$}}

\newcommand{\dense}{\mbox{$n_{\rm e}$}}

\newcommand{\hal}{\mbox{H$\alpha$}}
\newcommand{\docetto}{\mbox{$^{12}$CO\,($J$=2$-$1)}}

\newcommand{\trece}{$^{13}$CO\,($J$=1$-$0)}

\newcommand{\inttl}{\mbox{$\int{T_{\rm L}d\upsilonup}$}}

\begin{document} 

   \title{A pilot search for mm-wavelength recombination lines from emerging
     ionized winds in pre-planetary nebulae candidates}


   \author{ C.~S\'anchez Contreras\inst{1}
          \and
          A.~B\'aez-Rubio\inst{2,5}
          \and 
          J.~Alcolea\inst{3} 
          \and 
          V.~Bujarrabal\inst{4}
          \and
          J. Mart\'in-Pintado\inst{5}
          }

  \institute{Centro de Astrobiolog{\'i}a (CSIC-INTA), Postal address:
  ESAC, Camino Bajo del Castillo s/n, Urb. Villafranca del Castillo,
  E-28691 Villanueva de la Ca\~nada, Madrid, Spain\\ \email{csanchez@cab.inta-csic.es}
  \and
  Instituto de Astronom{\'i}a, Universidad Nacional Aut{\'o}noma de M{\'e}xico,
  Postal address: Apartado Postal 70-264, 04510, M\'exico, CDMX, Mexico
  \and Observatorio Astron\'omico Nacional (IGN), Alfonso XII
  No 3, 28014 Madrid, Spain
  \and Observatorio Astron\'omico Nacional
  (IGN), Ap 112, 28803 Alcal\'a de Henares, Madrid, Spain
  \and Centro de Astrobiolog{\'i}a (CSIC-INTA), Ctra de Torrejón a Ajalvir, km 4, 28850 Torrej\'on de Ardoz, Madrid, Spain}
   \date{Received; accepted}

 
  \abstract{We report the results from a pilot search for radio
    recombination line (RRL) emission at millimeter wavelengths in a
    small sample of pre-planetary nebulae (pPNe) and young PNe (yPNe)
    with emerging central ionized regions. Observations of the
    H30$\alpha$, H31$\alpha$, H39$\alpha$, H41$\alpha$, H48$\beta$,
    H49$\beta$, H51$\beta$, and H55$\gamma$ lines at $\sim$1 and
    $\sim$3\,mm have been performed with the
    \iram\ radio telescope. These lines are excellent probes of the
    dense inner (\protect\la\,150\,au) and heavily obscured regions of these
    objects, where the yet unknown agents for PN-shaping originate.  We
    detected mm-RRLs in three objects: CRL\,618, MWC\,922, and
    M\,2-9. For CRL\,618, the only pPN with previous published
    detections of H41$\alpha$, H35$\alpha$, and H30$\alpha$ emission,
    we find significant changes in the line profiles indicating that
    current observations are probing regions of the ionized wind with
    larger expansion velocities and mass-loss rate than $\sim$29 years
    ago. In the case of MWC\,922, we observe a drastic transition from
    single-peaked profiles at 3\,mm (H39$\alpha$ and H41$\alpha$) to
    double-peaked profiles at 1\,mm (H31$\alpha$ and H30$\alpha$), which is
    consistent with maser amplification of the highest frequency
    lines; the observed line profiles are compatible with rotation and
    expansion of the ionized gas, probably arranged in a disk+wind
    system around a $\sim$5-10\,\msun\ central mass. In M\,2-9, the
    mm-RRL emission appears to be tracing a recent mass outburst by
    one of the stars of the central binary system.
    We present the
    results from non-LTE line and continuum radiative transfer models,
    which enables us to constrain the structure, kinematics, and physical
    conditions (electron temperature and density) of the ionized cores
    of our sample. We find temperatures \te$\sim$6000-17,000\,K, mean
    densities \dense$\sim$\,10$^{5}$-10$^{8}$\cm3, radial density
    gradients \dense$\propto r^{-\alpha_n}$ with
    $\alpha_n$$\sim$2-3.5, and motions with velocities of
    $\sim$10-30\,\kms\ in the ionized wind regions traced by these
    mm-wavelength observations. We deduce mass-loss rates of
    \mpagb\,$\approx$10$^{-6}$-$10^{-7}$\,\my, which are significantly
    higher than the values adopted by stellar evolution models
    currently in use and would result in a transition from the
    asymptotic giant branch to the PN phase faster than hitherto
    assumed. }


\titlerunning{Mm-wave RRLs in pre-PNe}
\authorrunning{S\'anchez Contreras et al.}
   \maketitle
%

\section{Introduction}
\label{intro}
The yet unknown physical mechanisms responsible for the onset of
asphericity and polar acceleration in planetary nebulae (PNe) are
already active in early stages of the evolution beyond the asymptotic
giant branch (AGB). Therefore, pre-PNe (pPNe) and young PNe (yPNe)
hold the key to understanding the complex and fast ($\sim$1000\,yr)
nebular evolution from the AGB toward the PN phase. Some studies of
pPNe support the idea that the multiple lobes and high-velocities
observed are produced by the impact of collimated fast winds (CFWs or
$jets$) on the spherical and slowly expanding circumstellar envelopes
(CSEs) formed in the previous AGB phase \citep[see, e.g.,][for a
  review]{bal02}. However, this jet+`AGB CSE' two-wind interaction
scenario remains unconfirmed by direct characterization of the
post-AGB jets themselves and of the central nebular regions from which
these jets would be launched (within $\sim$few$\times$100\,au).
Studying these central regions is difficult because of their small
(sub-arcsec) angular sizes and, most importantly, because they are
usually heavily obscured by optically thick circumstellar dust shells
or disks.  Moreover, post-AGB ejections, believed to be the main
PN-shaping agents, are short-duration phenomena (with lifetimes of
\la\,100\,yr) and they are already inactive (or, at least, much less
energetic) in evolved PNe. Therefore, progress requires sensitive
observations at long wavelengths of the crucial, short-lived pPN/yPN
phases.

Central stars of pPNe start ionizing their surroundings at the mid-stages of their post-AGB evolution toward the central star of PN
(CSPN) phase, typically when they reach a B-type spectral
classification.\,Optical spectroscopic observations of pPNe and yPNe
\citep[e.g.,][and references therein]{san02,arr03,san08} have revealed
the widespread presence of broad (FWZI$\approx$\,100-1000\,\kms)
H$\alpha$ emission, often with blue-shifted absorption features
(P-cygni profiles) that are produced at their nuclei. In some pPNe, the
H$\alpha$ emission
suggests the presence of active fast shocks in the
stellar vicinity that could be linked, for example, to ongoing jets
sculpting the inner regions of slower and older mass ejecta.
Owing to the large amount of dust in most pPNe/yPNe, which is mainly concentrated in the
equatorial regions, at optical wavelengths the light from the ionized
nucleus can only be observed {\sl indirectly} after it is reflected by
the dust in the lobes. This situation not only impedes line
detections but also complicates the interpretation of the
observed scattered-H$\alpha$ profiles  enormously  \citep[][]{san01,arr05}.

The just emerging central ionized cores of pPNe/yPNe, however, can be
traced by radio recombination line (RRL) emission with the important
advantage that dust extinction effects are minimal. The theory of RRL
emission is well understood \citep[e.g.,][]{bro78}, however, RRL
observations are challenging, especially in the cm-wavelength range
where these lines are extremely weak in comparison to the continuum
level (typically $\approx$1\%). However, in the mm-wavelength range
larger line-to-continuum flux ratios are expected, i.e.,  above $\sim$80\%\ and $\sim$10\%\ 
at 1 and 3\,mm, respectively.
To date, some
mm-wavelength, but mainly cm-wavelength RRL studies have been carried
out for some of the best-known and most luminous evolved PNe
\cite[e.g.,][]{roe91,bach92,vaz99}. However, this field remains largely
unexplored for pPNe/yPNe, where central ionized regions are in an
early stage of development. 

To our knowledge, RRLs have only been detected in the mm-wavelength
range (H41$\alpha$, H35$\alpha$, and H30$\alpha$) toward one pPN,
i.e., the C-rich CRL 618 \citep{mar88}.  As shown by these authors,
mm-wavelength RRLs (mm-RRLs)
are excellent tools to study the structure, physical
conditions (electron temperature, \te, and density, \dense), and
kinematics of the dense, central ionized regions of pPNe/yPNe. Indeed,
even though single-dish observations do not enable us to spatially
resolve the structure of the compact ionized cores, to some extent it
is possible to constrain the properties of the current (post-AGB)
mass-loss process from detailed radiative transfer modeling of RRL
observations. Then, RRLs allow us to study post-AGB mass-loss, which
remains very poorly known even when it is key to understanding the
physics mediating the AGB-to-PN evolution.

In this paper, we report the results from our pilot study of the
emerging central ionized regions of a sample of pPNe/yPNe candidates
by means of single-dish observations of RRLs at mm-wavelengths. In
Section \ref{sample}, we describe our sample and provide a brief, yet
somewhat detailed, introduction to the three sources with mm-RRL
detections.  The observations and observational results are
described in Sections \ref{obs} and \ref{res}, respectively. The
analysis of the data, which includes radiative transfer model of the
line and continuum emission, and the derived results are reported in
Section \ref{moreli}. Results are interpreted and discussed in
Sections \ref{dis1} and \ref{dis2} and our main conclusions are summarized
in Section \ref{summ}.

\section{Sample}
\label{sample}
A small sample of eight targets (Table\,\ref{t-buj}) was selected as
follows. First, we chose pPNe/yPNe candidates with compact
H$\alpha$ emission signalling active ionized winds at their
cores \citep{san08,arr03,vdS00} observable with the \iram\ dish.
In contrast to evolved PNe, radio continuum flux measurements are
lacking for the vast majority of pPNe/yPNe, however, we carefully
dug into the literature and publicly available data archives to
build the spectral energy distributions (SEDs) of our targets. We
preferentially selected objects with indications of positive spectral
indexes in the mm-to-cm range (\snu\alpha, with $\alpha>0$),
indicative of partially optically thick continuum emission. A positive
spectral index is expected in the case of an isothermal spherical wind
with a power-law radial density gradient
\citep[e.g.,][]{pan75,rey86}. In this case, the peak-line intensities
of RRLs steeply increase with frequency, optimizing the chances of
detection at mm-wavelengths.  We also made a final sieve by choosing
objects with the strongest H$\alpha$ line and continuum fluxes.  All
targets in our sample have relatively hot central stars with
effective temperatures in the range \teff$\sim$20,000-40,000\,K, which
correspond to early-B to late-O spectral types (references are given
in Table\,\ref{t-buj}).

In Figure\,\ref{f-seds}, we show the SEDs of the targets in our
sample.  In the mm-wavelength region, the continuum emission is a
mixture of thermal dust and free-free emission.
In the case of
the well studied pPN CRL\,618, with a strong variability of the radio
continuum flux reported over the past three decades \citep[e.g.,][and
  \S\,\ref{intro-crl618}]{kwo81,mar88,san04b,taf13}, we have used
coeval and almost coeval mm-to-cm data as compiled by \cite{taf13} and
\cite{planck}.  For the rest of the sources, ancillary data are not
necessarily coeval and a certain degree of variability of the radio
continuum 
cannot be ruled out; this could partially explain the
scatter of the cm-continuum data points in some targets. Indeed, this
is most likely the case of He\,3-1475 as noticed by
\cite{cerri11}, who found a periodic pattern in the
3.6\,cm continuum light curve of this object.

\begin{table*}
  \caption{Parameters of the sources observed in our pilot survey.}
   \small
\label{t-buj}
\centering 
\begin{tabular}{lccccccc}
\hline\hline  
Source & RA(J2000) & Dec(J2000) & \multicolumn{2}{c}{Continuum (mJy)\tablefootmark{a}}  & Distance & Luminosity\tablefootmark{b} & References  \\
 name & (\h\ \m\ \stime) & (\degr\ \arcmin\ \arcsec) & 2.7\,mm & 1.3\,mm & (kpc) & (10$^3$ \ls) & \\  
\hline    
\multicolumn{6}{l}{\it ~~RRL detections} \\
\object{CRL 618} & 04:42:53.64 & +36:06:53.4   & 2395$\pm$17 & 2670$\pm$50  & 0.9 & 8.5 & 10, 19, 20, 21, 25, 29 \\ 
\object{M 2-9}    & 17:05:37.96 & $-$10:08:32.5 & 121$\pm$3 & 180$\pm$10 & 0.65,1.3\tablefootmark{\dag} & 0.7,3.0\tablefootmark{\dag} & 5, 6, 8, 17 \\ 
\object{MWC 922}  & 18:21:15.91 & $-$13:01:27.1 & 109$\pm$2 & 203$\pm$5  & 1.7,3.0\tablefootmark{\dag} & 18,59\tablefootmark{\dag} & 2, 15, 16, 28 \\ 
\multicolumn{6}{l}{\it ~~RRL non-detections} \\
\object{Hen 3-1475}        & 17:45:14.19 & $-$17:56:46.9 & 6$\pm$2 & 31$\pm$10  & 5 & 9 & 11, 14, 18 \\ 
\object{PN M 1-91}           & 19:32:57.69 & +26:52:43.1   & 12$\pm$1 & 23$\pm$3  & 2.1  & 0.3 & 6, 12, 17, 27 \\ 
\object{PN M 1-92}           & 19:36:18.90 & +29:32:49.9   & 18$\pm$2 & 86$\pm$2 & 2.5 & 10 & 1, 3, 4, 9, 17, 22 \\ 
\object{IRAS 20462+3416}  & 20:48:16.63 & +34:27:24.3   & 6$\pm$2 & 16$\pm$3  & 3.5 & 0.8 &  13, 22, 24, 26 \\ 
\object{PN M 2-56}          & 23:56:36.38 & +70:48:17.9   & 6$\pm$1 & 39$\pm$4  & 2.1 & 5.5 &  7, 17, 23 \\ 
\hline
\end{tabular}
\tablefoot{
  All targets have stellar effective temperatures in the range
  \teff$\sim$20,000-40,000\,K, which correspond to early-B to late-O
  spectral types.  \\
    \tablefoottext{a}{Continuum flux measured by us near the
   H39$\alpha$ (2.7\,mm) and H41$\alpha$ (1.3\,mm) RRLs. The formal errors of the continuum fluxes, averaging all line-free channels within the observed sub-band, do not include 
the absolute flux calibration uncertainties, which can be up to 20 and 30\% at 3 and 1\,mm, respectively. } 
\tablefoottext{b}{Obtained integrating the SED and adopting the distance given in this table.} 
\tablefoottext{\dag}{Uncertain distance: two possible values and the corresponding luminosities are given (see \S\,\ref{intro-m29} and \ref{intro-mwc922}).}
}
\tablebib{~(1) \citet{alc07}; (2) \citet{all76}; (3) \citet{arr05}; (4) \citet{buj98}; (5) \citet{buj01}; (6) \citet{cal78}; (7) \citet{cc02}; (8) \citet{cc12}; (9) \citet{coh77}; (10) \citet{goo91}; (11) \citet{hug04}; (12) \citet{lee07}; (13) \citet{par93}; (14) \citet{rie95}; (15) \citet{rod12};  (16) \citet{rud92}; (17) \citet{san98}; (18) \citet{san01}; (19) \citet{san02}; (20) \citet{san04a}; (21) \citet{san04b}; (22) \citet{san08}; (23) \citet{san10}; (24) \citet{san12}; (25) \citet{sch81}; (26) \citet{sua06}; (27) \citet{tor10}; (28) \citet{tut07}; (29) \citet{wes75}.}

\end{table*}

In the following subsections, we provide a broad introduction
to the three sources with mm-RRL detections.
   
\subsection{CRL\,618}
\label{intro-crl618}

CRL\,618 (aka\,RAFGL 618 = IRAS\,04395+3601 = Westbrook Nebula) is
a well-studied C-rich pPN 
currently going through the very early stages of PN development
\citep[e.g$.,$][]{wes75,wyn77,kwo81}. The optical nebula consists of
multiple elongated ({\it jet/finger-like}) lobes, roughly oriented in
the E-W direction, composed of shock-excited gas rapidly outflowing
from the central star with velocities of up to
\vexp$\sim$180\,\kms\,\citep{san02,tra02,rie11,bal13}. The lobes seem
to be `cavities' excavated in the AGB CSE via an interaction with a spray
of clumps/bullets that resulted from brief, episodic, and asymmetric
ejection events along multiple axes about a century ago
\citep{bal13,vel14,hua16}. In the optical, the remnant of the AGB CSE
is seen as a large-scale, roughly round reflection halo around the
lobes \citep{tra93,san02}.

Most of the nebular material in CRL\,618 is in the form of molecular
gas ($\sim$0.3\,\msun), which has been very well studied based on
high-angular resolution submm/mm-wavelength emission maps of CO and
other molecules \citep{san04a,san04b,lee13a,lee13b,nak07}. The
molecular envelope includes fast compact outflows that are aligned with the
optical lobes with expansion velocities increasing linearly with the
distance to the nebula center, up to \vexp$\sim$340\,\kms\ (at the
lobe tips).  In addition, there are, at least, three distinct
structures expanding at low velocity: a large equatorial torus with a
dense compact core (\vexp\,\la\,12\,\kms), a thin-walled, bipolar
shell encompassing the shock-excited optical lobes
(\vexp$\sim$22\,\kms), and an extended tenuous halo
(\vexp$\sim$16\,\kms) surrounding all other components.

The central star, which is hidden from direct view at optical
wavelengths by a dusty equatorial torus, has been classified as
$\sim$B0 with an effective temperature of
30,000\,\la\teff\la\,40,000\,K \citep{wes75,sch81,tra93,bal14}.

Around the central star, there is a compact \ion{H}{ii} region that
produces free-free continuum emission \citep[e.g.,][]{wyn77,kwo81,kwo84,mar88}.
At cm-wavelengths, the ionized region shows an overall elliptical
brightness distribution elongated in the E-W direction
\citep{mar93}.
The continuum flux and size of the ionized core at 
cm-wavelengths have been increasing monotonically in the past three
decades. 
In their discovery paper, \cite{kwo81} interpreted this result as the advance of the
ionization front within the molecular envelope as a consequence of the
increasing stellar \teff. The rapid angular expansion of the nuclear \ion{H}{ii} region, which
had reached an angular size of $\sim$0\farcs5$\times$0\farcs2
($\sim$\,450$\times$180\,au at $d$=900\,pc) at 22\,GHz in 2008, has
recently been studied in detail by \cite{taf13}. These authors use 
high-angular resolution maps of the $\sim$1.5-43\,GHz continuum emission at multiple epochs spanning over 25 years. 
\cite{taf13} demonstrate that the advance of the ionization front
cannot explain the current expansion of the optically thick
continuum-emitting region.
These authors conclude that it is the optical depth of the free-free continuum at a given distance that
increases with time, which is attributed to the rise of the
electron density with time.
This supports the idea that CRL\,618 is
expelling gas in the form of an ionized wind whose mass-loss rate has
been increasing, from \mloss$\sim$4\ex{-6} to 
6\ex{-6}\,\my, during the last century.
\cite{taf13} establish the onset of the ionization around 1971. 

The free-free continuum at mm-wavelengths, which is optically thin
and, thus, probes denser and deeper (\la\,150\,au) regions than the
continuum at cm-wavelengths (optically thick), displays much abrupt
and unpredictable flux changes that cannot be simply attributed to the
smooth growth of the optically thick layers of the \ion{H}{ii} region
but most likely denote alterations in the activity and/or physical
conditions of the post-AGB wind at its core \citep[][see also
  Fig.\,\ref{f-crl6181mm}]{san04a,san04b}.

CRL\,618 is the only pPN in which mm-RRL emission has 
been reported prior to this work: the H41$\alpha$, H35$\alpha$, and
H30$\alpha$ transitions (at 3.5, 2, and 1.3\,mm, respectively) have
been detected by \cite{mar88}. From the analysis of their single-dish
spectra, which included LTE radiative transfer modeling of the line
and continuum emission, these authors conclude that mm-RRLs trace a
\te$\sim$13,000\,K stellar wind with a terminal expansion velocity of
\vexp$\approx$20\,\kms\ and a mass-loss rate of
\mloss$\sim$7.6\ex{-6}$(\frac{d}{900 {\rm pc}})^2$\,\my.

The distance to CRL\,618 adopted in this work is $d$=900\,pc, as
determined using independent techniques by \cite{goo91}, from a
total luminosity criteria, and by \cite{san04a}, from the analysis of
proper motions of the fast expansion of the molecular gas in the
lobes. The inclination of the nebular axis with respect to the plane
of the sky is $i$$\sim$30\degr\ \citep{san02,san04a}.

\subsection{M\,2-9} 
\label{intro-m29}
M\,2-9, also known as the ``Butterfly'' or the ``Twin Jet'' nebula, is
a bright bipolar nebula discovered by \cite{min47} and considered by
many authors to be in the earliest stages of becoming a PN.  In the
optical, it shows a bright compact core from which two bilobed nested
structures emerge oriented nearly north-south \citep{sch97,cly15}. The
brightness and morphology of the optical nebula have been changing
notably on a timescale of a few years or less since its discovery
\citep[e.g.,][]{all72,vdB74,koh80,doy00}.  In particular, the main
bright condensations ({\it knots}) in the lobes progressively shift in
the west-east direction, which has been most recently interpreted as
produced by a rotating collimated spray of high velocity particles
(jet) from the central binary system, which excites the walls of the
inner cavity of M\,2-9 \citep{corr11}. From the rotation period of the
beam particles, these authors infer an orbital period of
$P$$\sim$90\,yr.

Optical spectroscopic observations point to an expansive kinematics in
the lobes, as for most pPNe and young PNe with expansion velocities
ranging from $\sim$23\,\kms\ in the inner nebula up to $\sim$164\,\kms\ in the
outer lobe tips \citep{sch97,tor10,cly15}. The central core is dominated by
\hal\ emission with $\sim$1600\,\kms-wide wings. The broad
\hal\ profile could indicate ongoing ultra-fast winds, although \hal\ wing
emission could be partially (or totally) broadened by other
mechanisms, such as Raman scattering \citep{tor10,arr03}.

The central exciting star has been estimated to be of spectral type
B1 and possibly late O, \teff\,$\sim$35,000\,K \citep{cal78,swi79}.
As in the case of CRL\,618, the large extinction and high-infrared
excess observed are attributed to a dusty torus in our line of sight
that prevents us from seeing the central star directly but allows it
to illuminate the lobes. The binarity of the nucleus of M\,2-9 has
been observationally established by \cite{cc12}, hereafter
CC12. Interferometric \docetto\ emission maps with
subarcsecond-resolution by these authors show two coaxial, slightly
off-centered rings lying in the equatorial plane of the nebula. These
rings
have been interpreted as the result of two short mass ejections
produced at different positions in the binary orbit.  The rings lie at
a plane that is inclined with respect to the line of sight by
$\sim$17$\pm$1\degr\ \citep{cc17}.

M\,2-9 is known to be a source of thermal radio continuum emission.
According to \cite{kwok85}, two components are needed to properly fit
the cm-wavelength data: a compact ionized wind at the center that
produces free-free emission following a S$_{\nu}$$\propto$$\nu$$^{\sim
  [0.6-0.7]}$ law, and high-density condensations located in the
extended ionized lobes that produce a nearly flat continuum emission
distribution. At mm-wavelengths, the contribution from the extended
lobes to the observed continuum is small and the free-free emission
from the compact ionized core together with thermal emission by dust
dominate the observed continuum flux \citep[][and
  Fig\,\ref{f-seds}]{san98}. At cm-wavelengths, the continuum
brightness distribution is elongated along the N-S (lobe's) direction
\citep{kwok85,lim00,lim03}. At mm-wavelengths, the ionized core is
expected to be very compact, \la\,0\farcs1, but its geometry is
unknown. Based on VLA cm-continuum emission maps obtained in 1982,
\cite{kwok85} deduce a major-to-minor axis ratio that decreases
monotonically as the frequency increases, implying a nearly spherical
brightness distribution at $\sim$80\,GHz and beyond. However, based on
a different set of VLA cm-continuum maps obtained in 1999,
\cite{lim03} report a frequency-independent major-to-minor axis ratio
of $\sim$2.5 (in the 1.3-to-6\,cm wavelength range) that, if
extrapolated to mm-wavelengths, would imply an elongated geometry at
these high frequencies.

The distance to M\,2-9 is very uncertain. Values spanning almost two
orders of magnitude, from 50\,pc to 3\,kpc, are found in the
literature.  The two most accepted values derived from recent works
are $d$$\sim$\,650\,pc, from the analysis of the proper motions of the
expanding molecular rings at the nebula equator (CC12), and
$d$=1.3$\pm$0.2\,kpc, from the analysis of the proper motions of the
dusty blobs in the lobes \citep{corr11}.

At $d$$\sim$1.3\,kpc, the total luminosity of M\,2-9 is
$\sim$\,3000\,\ls\ \citep{lyk11}. At $d$$\sim$\,650\,pc, the luminosity is  a
factor four smaller and, therefore, lower than the low
limit of $\sim$1000\,\ls\ expected for the least massive AGB and
post-AGB stars \citep[][see also Fig.\,\ref{f-tracks}]{blo95}.

\subsection{MWC\,922}
\label{intro-mwc922}

The evolutionary status of this, yet barely studied,
dust/gas-enshrouded B[e] star is unclear. An evolved nature has been
suggested based on similarities of its SED and nebular morphology with
another well-known pPN, the Red Rectangle
\citep[e.g.,][]{coh75,wat98,buj16}.  Based on the appearance of its
surrounding nebulosity in the near-infrared, MWC 922 has been
nicknamed the Red Square Nebula.  \cite{tut07} report $H$-band
adaptive optics imaging showing an X-shaped nebular structure, which
extends for about 5\arcsec, with an equatorial dark band (running
along PA$\sim$46\degr) at the center.  A similar square-like structure
is observed in the mid-infrared \citep{lag11}. This nebular appearance
may result from the projection of two twin opposing biconical lobes
excavated in a dense dust-scattered halo, although illumination
effects could be significant; for example,  if
the light from the central source is
blocked by an inner dusty torus or disk at the nucleus.

The optical and infrared spectra of MWC\,922 show a great amount of
emission lines, including many recombination and forbidden
\ion{Fe}{II} lines but also transitions from neutral elements,
e.g.,\,\ion{O}{I}, H$_2$, etc \citep{all76,rud92,per03,weh14}.
A B\,3-B\,6 spectral class for the exciting star has been suggested by
\cite{rud92} based on the nebular emission line spectrum. These
authors acknowledge that this classification is very uncertain (in
fact, it would imply \teff$\sim$\,12,000-17,000\,K, which is not
sufficient to produce significant photoionization) and propose an
upper limit for the stellar temperature of \teff$<$31,000\,K based on
the strength of the \ion{He}{I}\,$\lambda$10830\AA\ line. A symbiotic
nature is not favored by these authors given the lack of molecular
absorption bands (very prominent in late-type giants) in the spectrum
of MWC\,922.

\begin{figure*}[htbp]
   \centering
\includegraphics[width=0.44\textwidth]{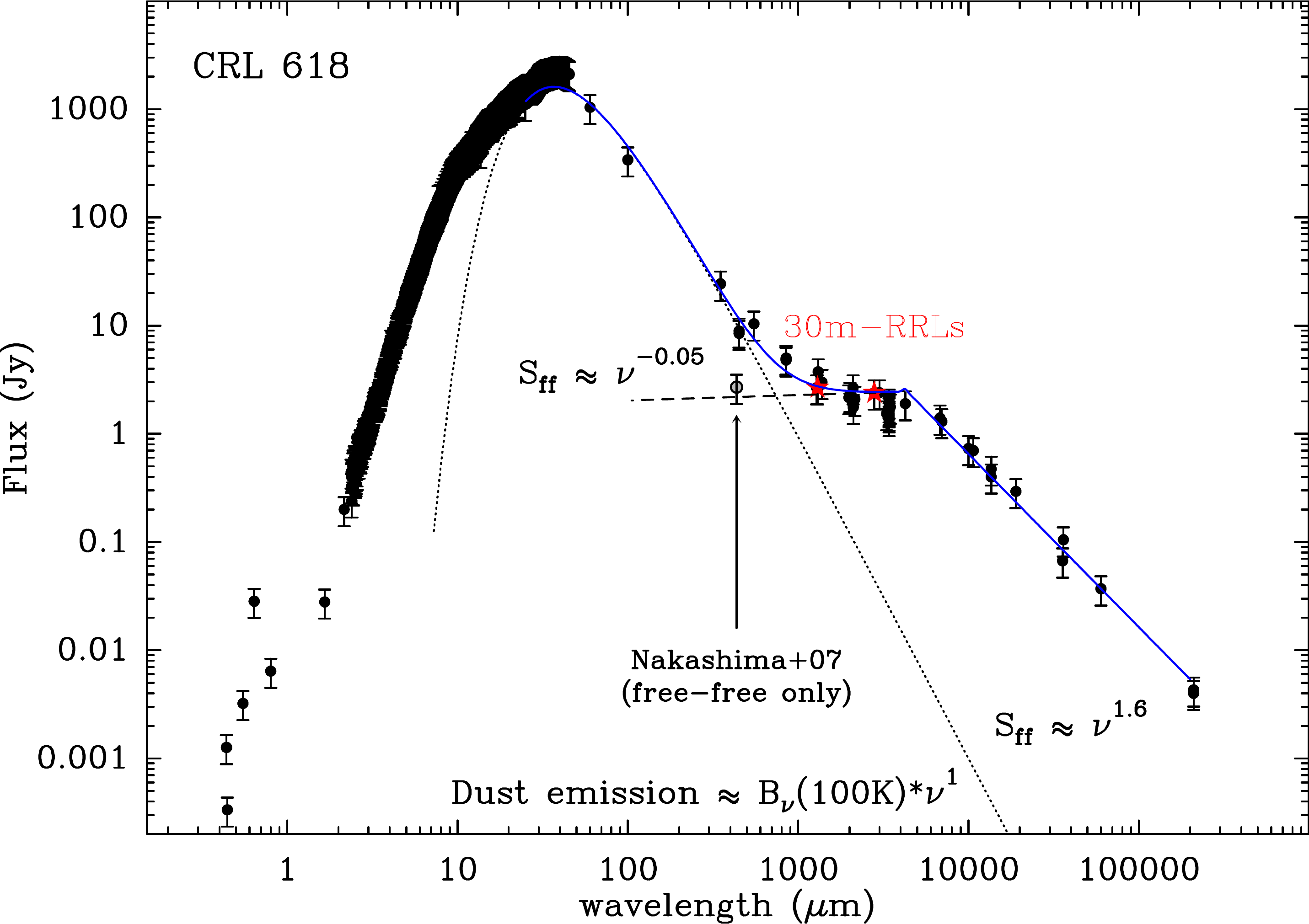}
\includegraphics[width=0.44\textwidth]{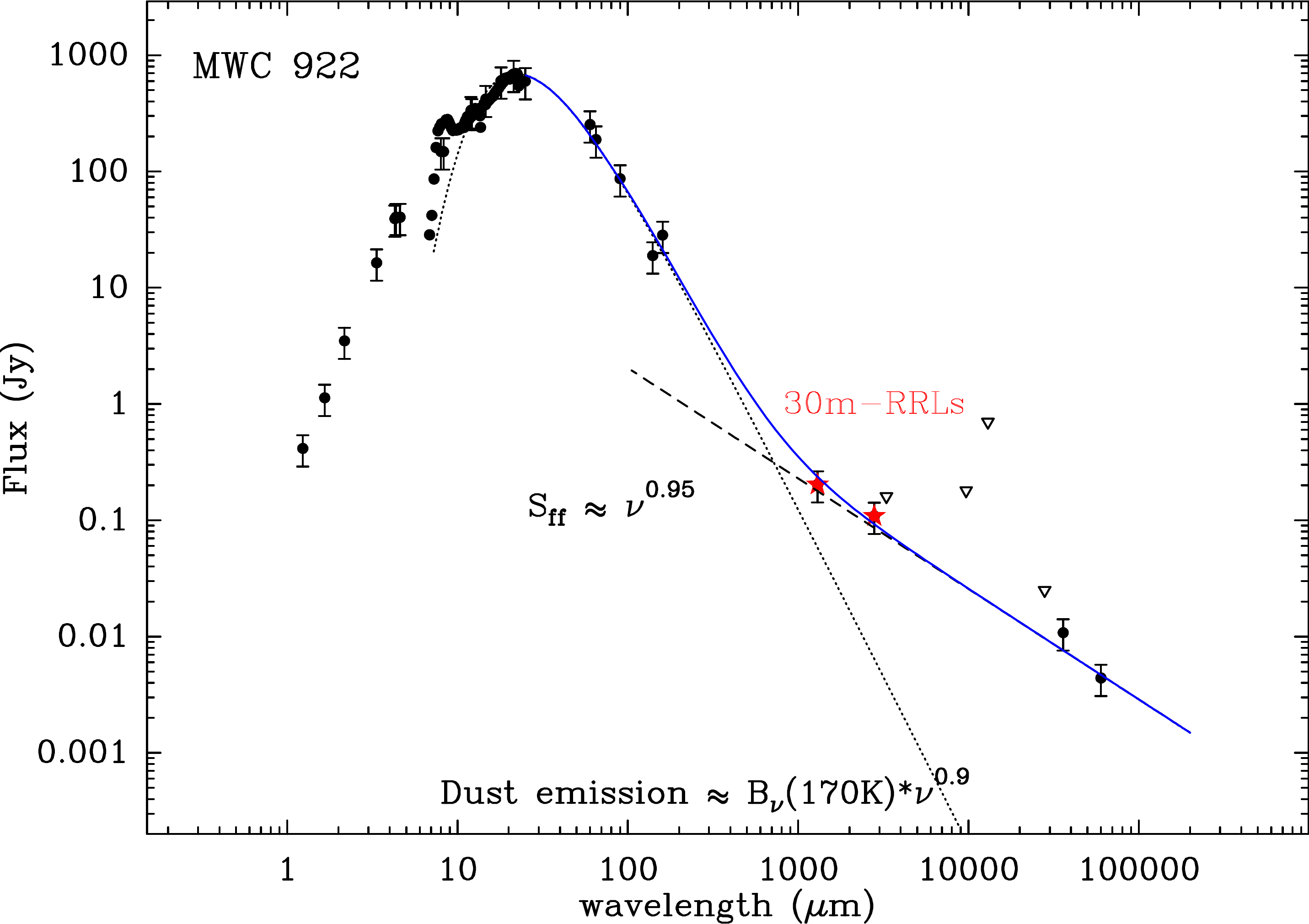}
\includegraphics[width=0.44\textwidth]{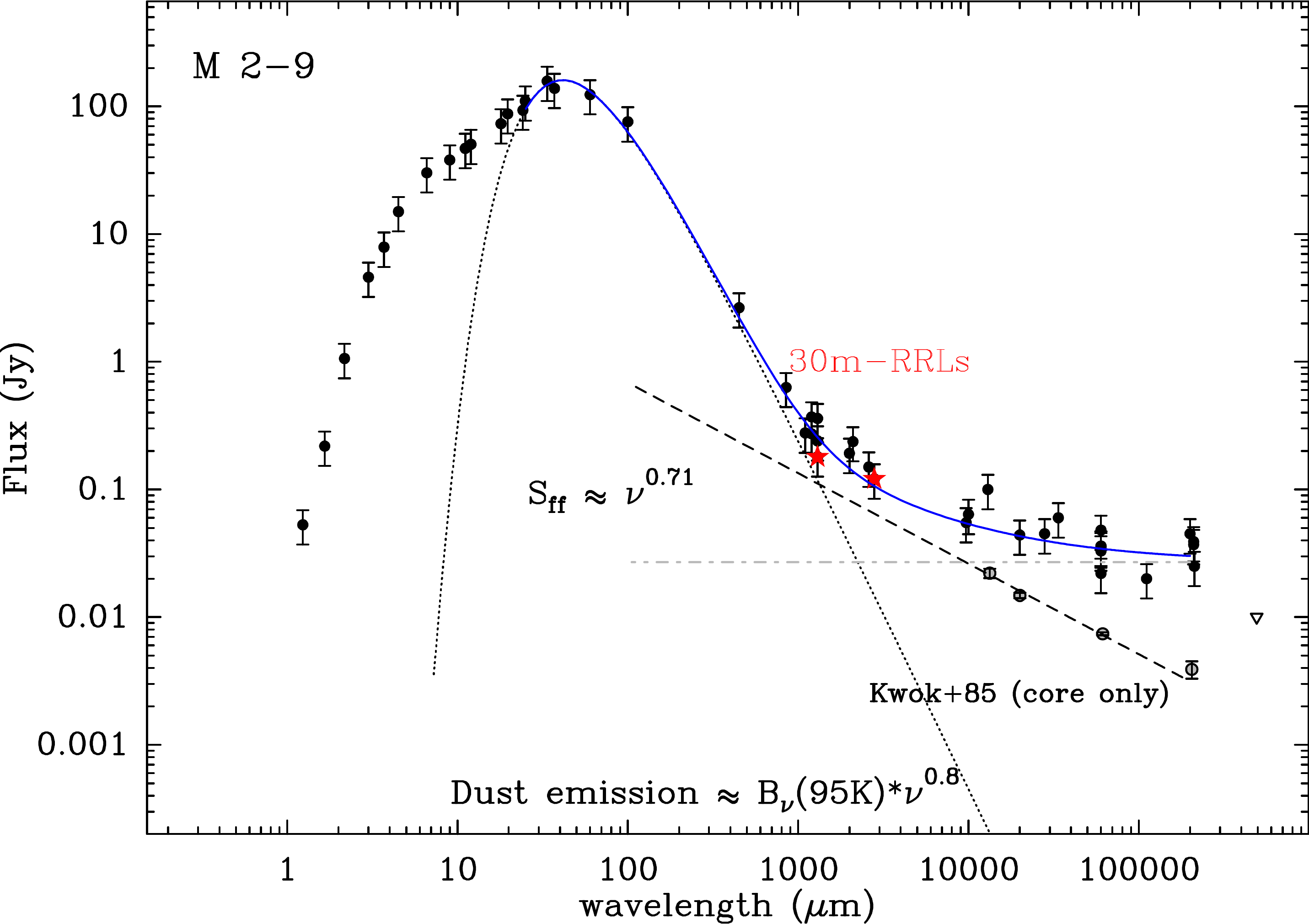}
\includegraphics[width=0.44\textwidth]{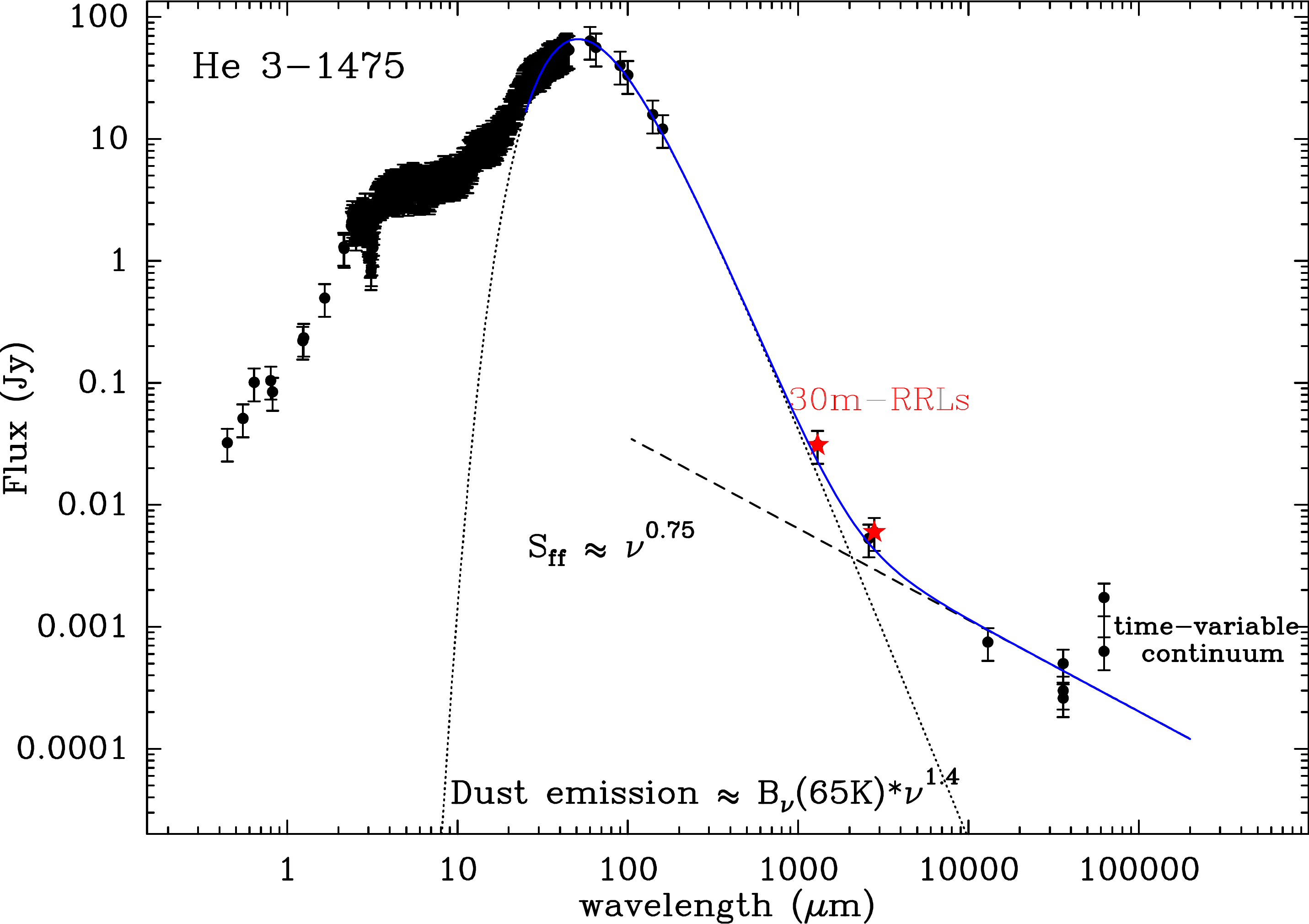}
\includegraphics[width=0.44\textwidth]{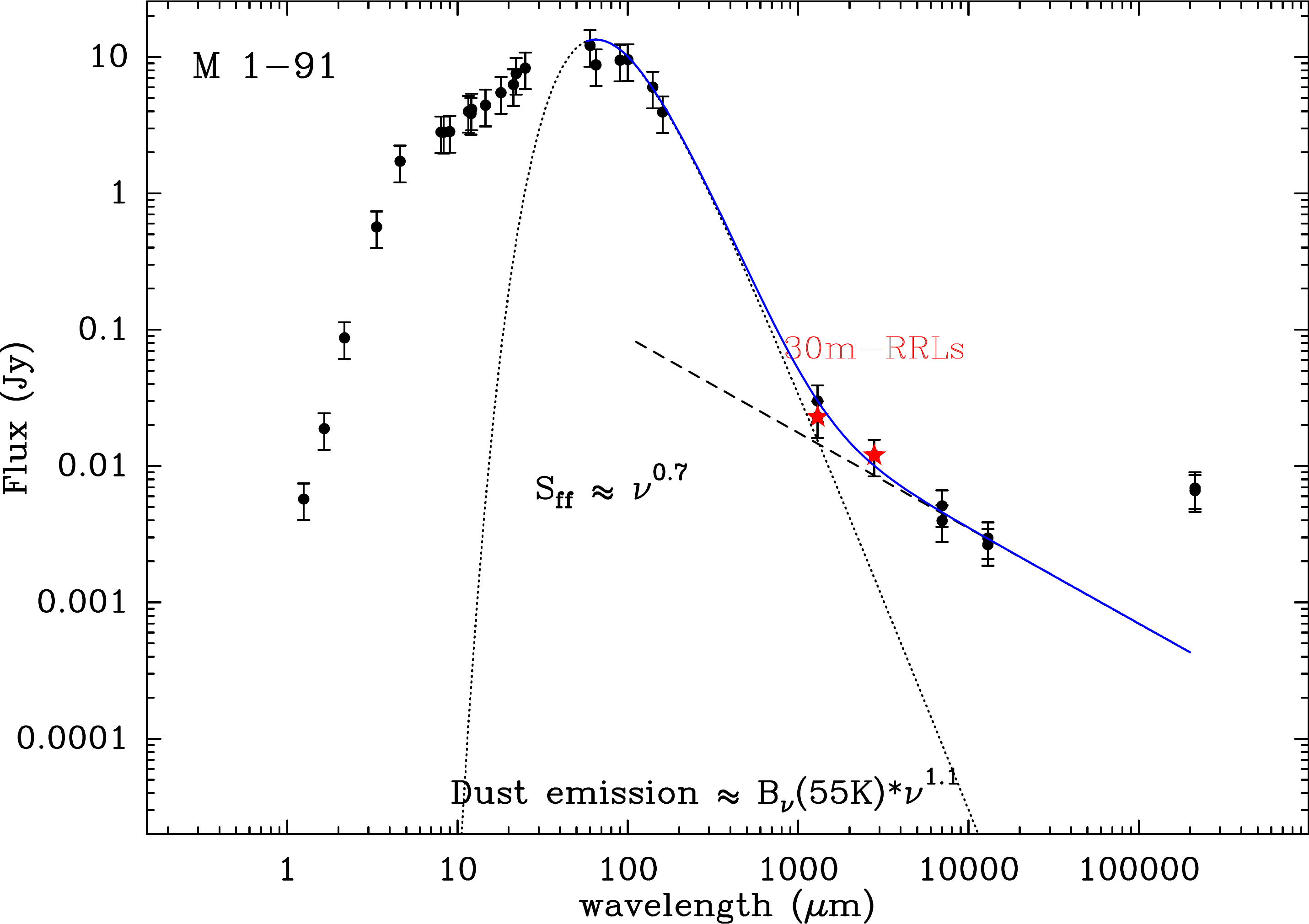}
\includegraphics[width=0.44\textwidth]{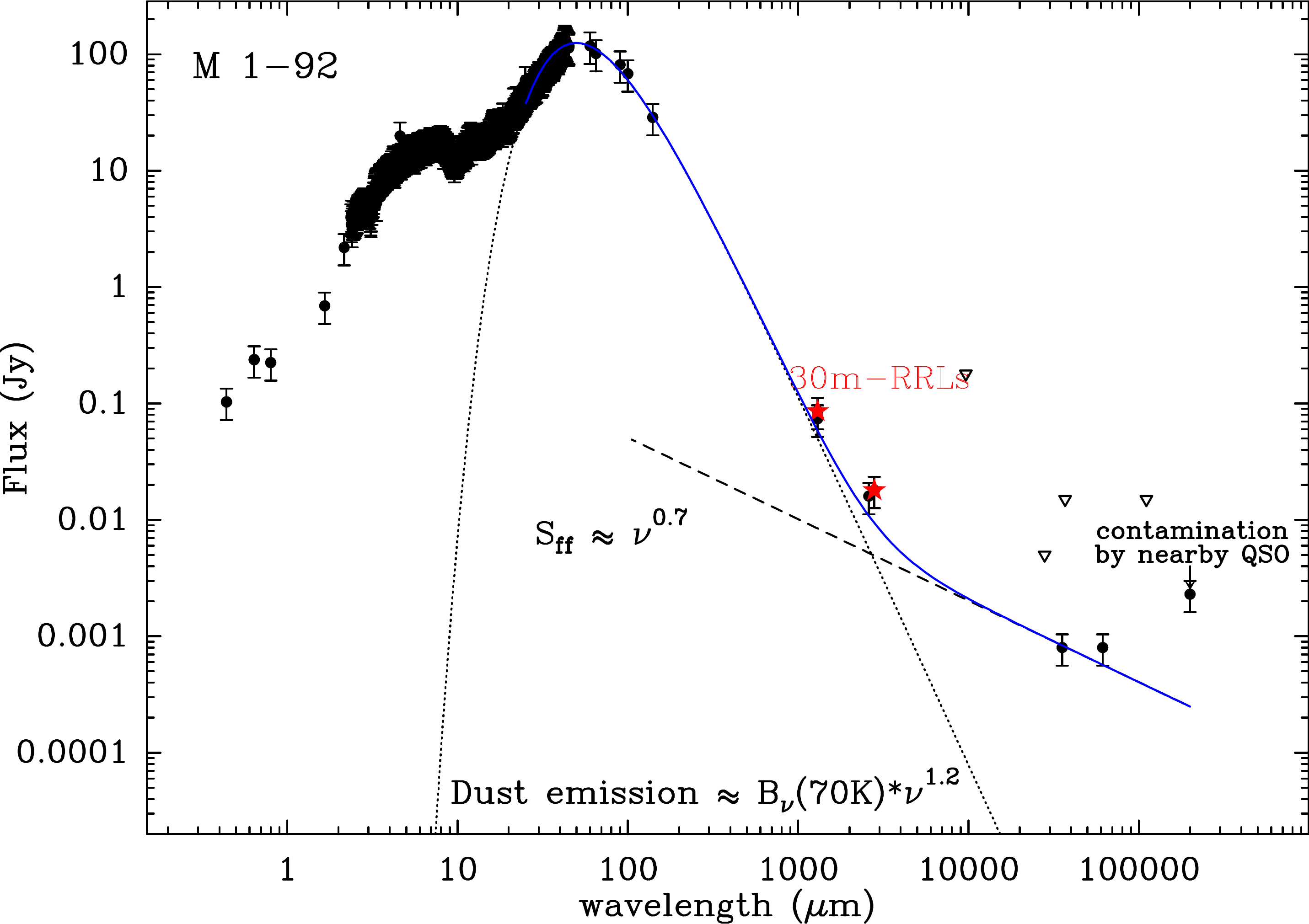}
\includegraphics[width=0.44\textwidth]{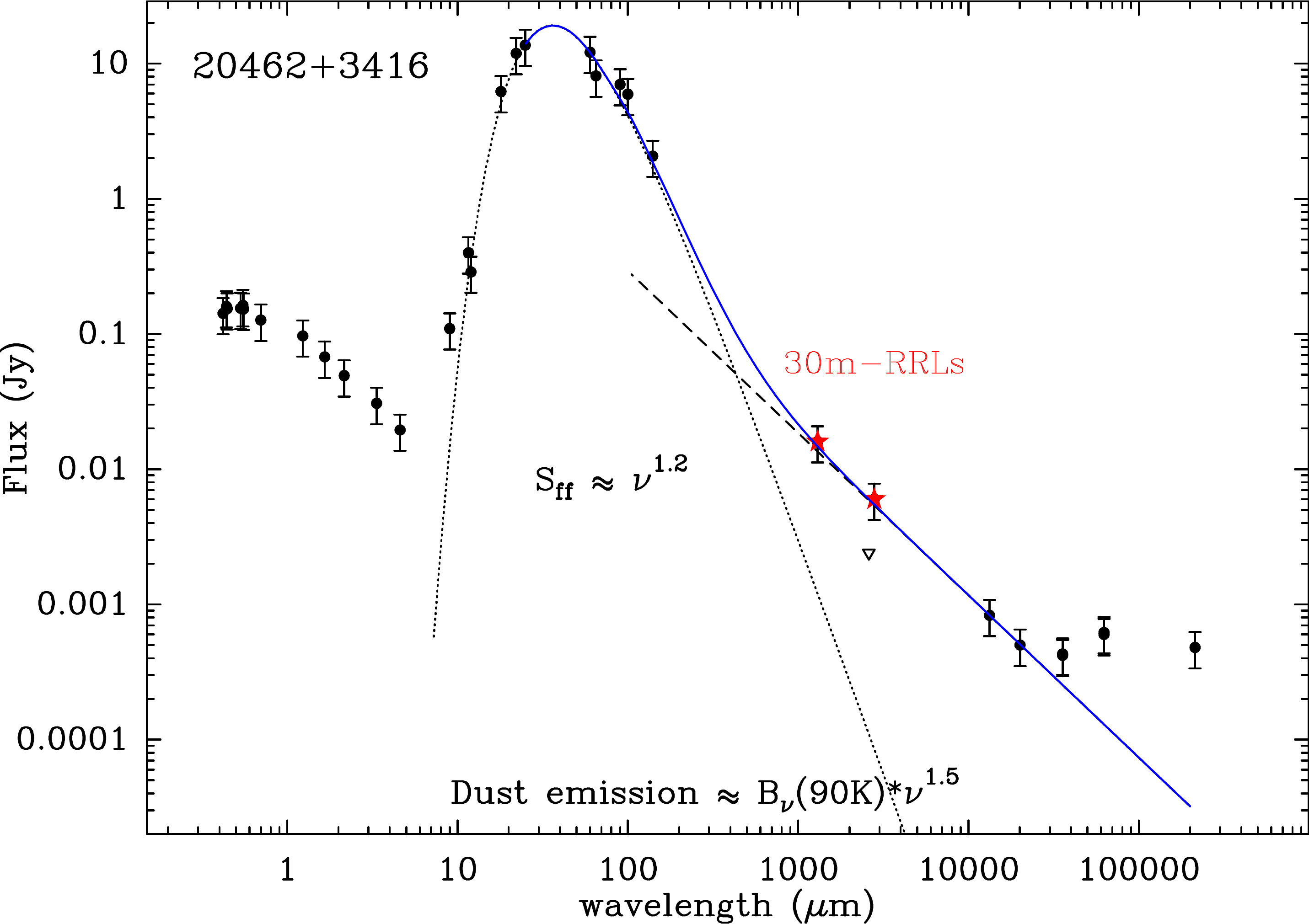}
\includegraphics[width=0.44\textwidth]{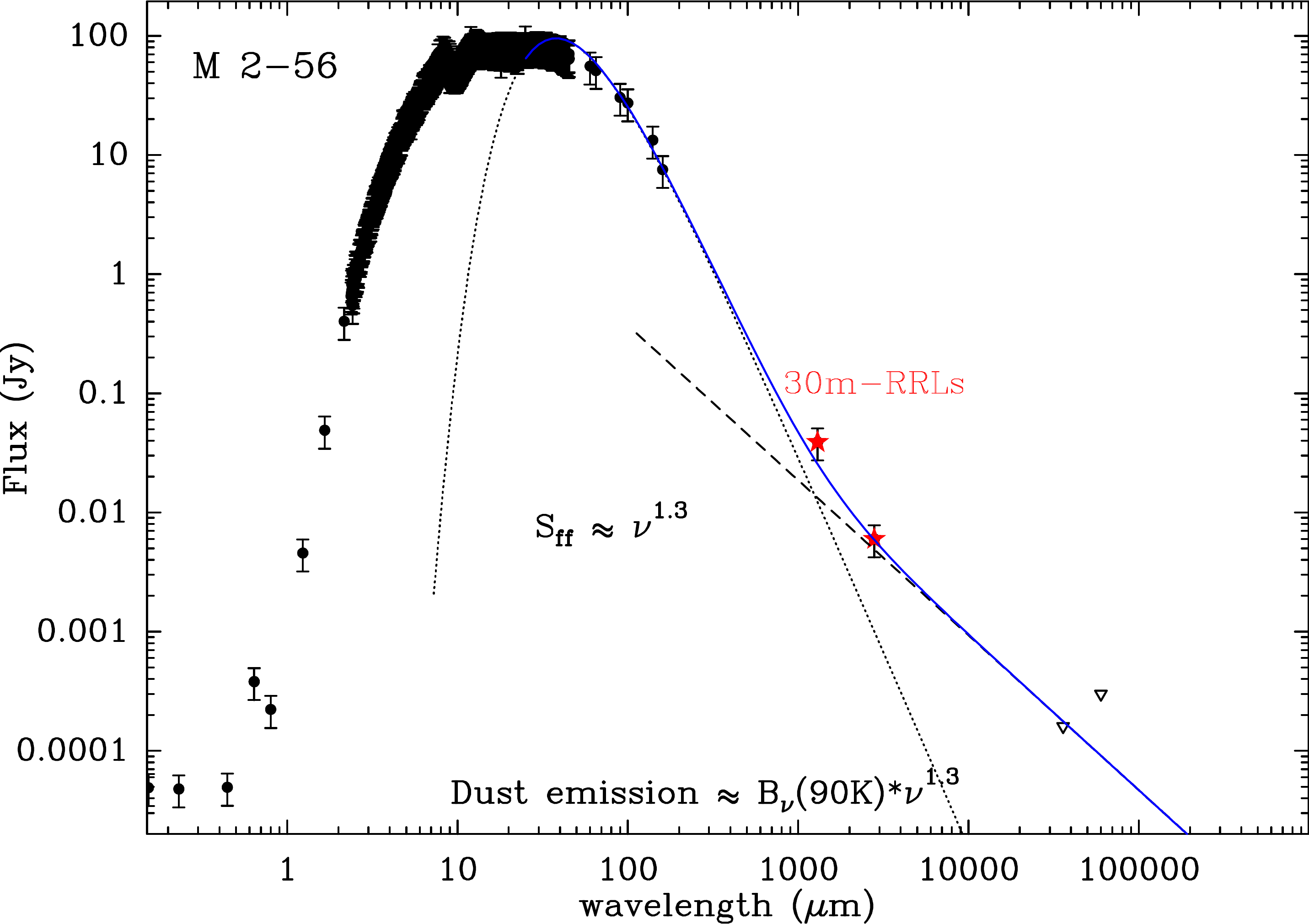}
   \caption{Spectral energy distributions of our targets
     showing literature photometry data (black circles and triangles indicate detections and upper limits, respectively) and the 1.3
     and 2.7\,mm continuum fluxes measured by us (star-like
     symbols; Table\,\ref{t-buj}). The error bars represent a conservative
     30\%\ flux uncertainty assigned to all continuum data points. The
     dotted line indicates a modified blackbody fit to the far-IR thermal dust
     emission, which is subtracted from the total (dust+free-free) emission observed at mm-wavelengths.
     No attempt has been made to fit the warm dust emission blueward of the SED peak. 
     The dashed line represents a fit of the type S$_{\rm ff}$\,$\propto\nu^\alpha$ to the mm-to-cm data points. The
     blue solid line represents the addition of these two fits.}
   \label{f-seds}
   \end{figure*}

There is a bright compact radio continuum source associated with
MWC\,922, which has been studied by \cite{rod12} using a
0\farcs95$\times$0\farcs66 angular resolution 
3.6\,cm radio continuum VLA map. In these maps, the ionized core is
angularly resolved with deconvolved dimensions
$\sim$0\farcs18$\times$0\farcs20.
Continuum observations at 6\,cm from the CORNISH project exist
\citep{pur13}; the corresponding flux is included in the SED shown in
Fig.\,\ref{f-seds}. The 6\,cm radio continuum source is unresolved with
$\sim$1\farcs5-angular resolution.

The distance to MWC\,922 is unknown. As mentioned by \cite{tut07}, if
it lies within the Ser OB1 association then the distance is
$d$=1.7\,kpc. If this (speculative) association is real then it may be
consistent with a pre-main sequence nature for this object, which
would still be linked to other young massive OB stars (but see
discussion in \S\,\ref{h-mwc922}).
We estimate the kinematic distance to MWC\,922 from its radial
velocity, \vlsr=$+$32.5\,\kms\ (deduced from the centroids of the RRLs; see \S\,\ref{res}), and its galactic coordinates ($l$=17.9275\degr,
$b$=+00.6336\degr) by assuming a simple galactic rotation law and
adopting a value for the A Oort constant of 14.4\,\kms\,kpc$^{-1}$ and
a galactocentric radius of 8.5\,kpc \citep{kerr86}. We deduce values
of $\sim$3 and $\sim$13\,kpc, for the near and far kinematic distance,
respectively.

The bolometric flux of MWC\,922 computed by integrating its SED from
$\sim$1\,$\mu$m to cm-wavelengths  is
$\sim$5800\,\ls\,kpc$^{-2}$. At the far kinematic distance, this would
imply a total luminosity on the order of a million \ls, approaching
that of the most luminous and massive ($\approx$100\,\msun) stars
known in the Galaxy. At $d$=1.7-3\,kpc, the total luminosity is
$\sim$[1.7-5.4]\ex{4}\,\ls, which is at the high end of the range for
post-AGB objects \citep{blo95}.  The foreground-ISM extinction toward
the 7\arcmin$\times$7\arcmin\ region around MWC\,922 is
$A_V$$\sim$1.7-2.7\,mag at a distance of $d$$\sim$2-3\,kpc
\citep{zas15}. After applying the corresponding ISM extinction
correction, the total luminosity deduced at $d$=1.7-3.0\,kpc is $\sim$[1.8-5.9]\ex{4}\,\ls.

%
%

\section{Observations and data reduction}
\label{obs}
Observations have been carried out with the \iram\ radiotelescope
(Pico Veleta, Granada, Spain) using the new generation heterodyne
Eight MIxer Receiver \citep[EMIR;][]{car12}. Spectra were taken in two
observational campaigns in June/July and September 2015 (project 050-15).

We simultaneously used the E090 (at 3\,mm) and E230 (at 1\,mm)
bands operated in dual sideband (2SB) mode.  For all targets, we
observed the $\sim$8\,GHz-wide upper side band in dual (H+V)
polarization. This enabled us to cover the $\sim$105-113 and
$\sim$226-234\,GHz frequency ranges simultaneously and, thus, to
observe H39$\alpha$, H48$\beta$, H49$\beta$, H55$\gamma$, and
\trece\ at 3\,mm together with H30$\alpha$ and \docetto\ at 1\,mm. For
CRL\,618, MWC\,922, and M\,2-9, we also used a different receiver
setup to survey the lower and upper side bands of each receiver at the
same time in single (either V or H) polarization. This enabled us to
expand the frequency coverage to the $\sim$89-97 and
$\sim$210-218\,GHz ranges, which include H41$\alpha$ and H51$\beta$ at
3\,mm and H31$\alpha$ at 1\,mm.

The image sideband was rejected with an average sideband rejection of
$\sim$14\,dB. Each receiver was connected to the Fast Fourier
Transform Spectrometer (FTS) in its 195\,kHz spectral resolution mode
(\dv$\sim$0.25 and 0.5\,\kms\ at 1 and 3\,mm, respectively).
Observations were performed in wobbler switching mode with a wobbler throw
of 120\arcsec\ and frequencies of 0.5\,Hz.  Calibration scans on the
standard two load system were taken every $\sim$15\,min. Pointing and
focus were checked regularly on nearby continuum sources. 
After
pointing corrections, the typical pointing accuracy was
$\sim$2\arcsec-3\arcsec.  Average system temperatures were 
130-200\,K at 3\,mm and 400-1000\,K at 1.3 mm. On-source integration
times are in the range $\sim$\,2-6.5\,h per target and receiver setup.

The beam parameters of the \iram\ telescope used in this work,
that is, the half-power beam width (HPBW), the main beam efficiency
($\eta_{\rm eff}$), and the point source sensitivity ($S$/\ta, i.e.,\,
the K-to-Jy conversion factor), are described to a good accuracy as a
function of the frequency ($\nu$, in GHz) by

\begin{equation} \label{telparms}
\left \{ 
\begin{array}{l l}
\mathrm{HPBW}(\arcsec)=2460/\nu  & \\
~
\eta_{\rm eff}=0.9\exp{-(\frac{\nu}{399.7})^2} & \\ 
~
S/\ta=5.44+(\frac{\nu}{147.148})^2 & \\ 
\end{array} 
\right . 
,\end{equation}

\noindent 
according to measurement updates performed in August 2013\footnote{\tt
  http://www.iram.es/IRAMES/mainWiki/EmirforAstronomers}.

We reduced the data
     using CLASS\footnote{CLASS is a
  world-wide software used to process, reduce and analyze heterodyne
  line observations maintained by the Institut de Radioastronomie
  Millimetrique (IRAM) and distributed with the GILDAS software, see
  {\tt http://www.iram.fr/IRAMFR/GILDAS}.} following the standard
procedure, which includes killing bad channels, subtracting baseline,
and averaging individual good-quality scans to produce the final
spectra.  We obtained and presented the spectra in antenna-temperature
(\ta) scale, which can be converted to main-beam temperature (\tmb)
applying \tmb=\ta/$\eta_{\rm eff}$ or to a flux scale using the
K-to-Jy conversion factor given above.

The uncertainty in the relative calibration of our observations was estimated by comparing the spectra in the same spectral ranges,
including the \docetto\ and \trece\, transitions, every day. We estimate that  
the total line calibration uncertainty is $\la$20\% and $\la$15\%\ at 1\,mm
and 3\,mm, respectively. The profiles of the bonus \docetto\ and
\trece\ transitions observed in this project are presented in Figures\,\ref{f-co1mm} and \ref{f-13co3mm} of
the Appendix.

\subsection{Continuum measurements}

We measured the continuum emission level (in a \ta\ scale) within
the different frequency ranges covered in these observations by
fitting a low-order polynomial function to the spectral baseline in
our data, which were obtained using the wobbler-switching method
described above.  For the sources that were observed in two epochs
(June/July and September 2015), namely, CRL\,618, MWC\,922, M\,2-9,
and M\,1-92, we obtained continuum measurements separately for both
epochs and we find that the continuum levels in the two epochs are in
agreement within 30\%\ at 1\,mm. The continuum flux uncertainty at
3\,mm is probably lower than 20\%.

For CRL\,618, which is a strong ($>$2\,Jy) millimeter continuum
emitter, we also measured the continuum from cross-scans in azimuth
and elevation direction performed for pointing.  After discarding a
few obvious bad/noisy scans, we obtained final, weighted average
scans\footnote{We used 1/$\sigma^2$ weights, where $\sigma$ is the rms
  of the baseline of the cross-scan.} for the four frequency ranges
observed, centered at 93, 109, 214, and 230\,GHz. 
Gaussian fits were applied to the average scans to derive the antenna
temperature at the peak, which was converted to flux density using
the frequency-dependent conversion in Eq.\,(\ref{telparms}).

In CRL\,618, the line emission flux contribution to the total flux in the continuum
backend (8\,GHz-wide) used for the pointing cross-scans is
estimated $\sim$10\%\ at $\sim$230\,GHz, where the strong
\docetto\ line lies, and less than 4\%\ at the lower frequency ranges
observed, and has been considered to derive the (line-free) continuum
flux level. The value of the continuum flux obtained from the two
methods (baseline fitting and cross-scans) agree to a 20\% level.

%


\section{Observational results}
\label{res}

\subsection{Continuum}

   \begin{figure}[tbp]
   \centering 
\includegraphics[width=0.44\textwidth]{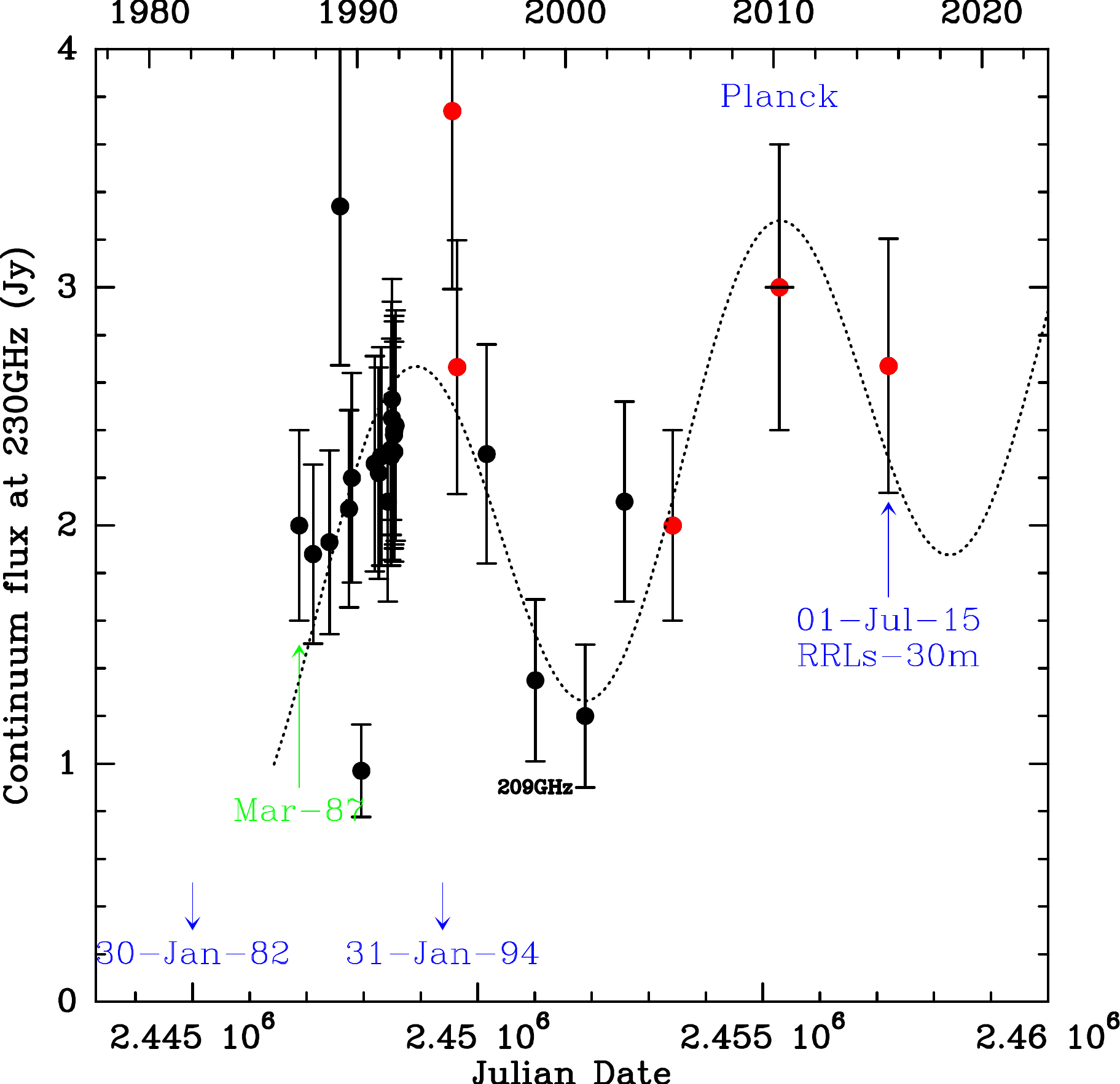}
   \caption{Light curve of the 1\,mm-continuum of CRL\,618 using
     ancillary data compiled by S\'anchez Contreras et al.\ (2004; {\it black circles});
     new measurements are added from \cite{reu97}, \cite{nak07},
     \cite{planck}, and this work ({\it red circles}).
     Some
     reference dates are indicated. Together with significant abrupt
     changes in short timescales of less than 0.1 years, a periodic
     pattern with a period of $\sim$17 years is hinted (dotted line)
     on top of an overall, long-term smooth increase.}
   \label{f-crl6181mm}
   \end{figure}


Our continuum flux measurements at frequencies adjacent to the
H39$\alpha$ and H30$\alpha$ lines, that is, at 2.7 and 1.3\,mm,
respectively, are given in Table\,\ref{t-buj} and are shown in
Fig.\,\ref{f-seds} (red symbols). In all cases, we find continuum
fluxes that are in good agreement with previous measurements at
similar wavelengths (whenever available) within the absolute flux
uncertainties. A fit to the mm-to-cm free-free continuum fluxes, after
removal of the dust thermal emission contribution, was performed and
is shown in Fig.\,\ref{f-seds} (dashed line). In the case of M\,2-9,
two free-free continuum components, one nearly flat and another one as
\snu{0.7}, were used for the fit based on previous results at
cm-wavelengths \cite[][and \S\,\ref{intro-m29}]{kwok85}.  To our
knowledge, our \iram\ data represent the first detection of the
continuum flux at mm-wavelengths for MWC\,922, M\,1-91
\citep[tentatively detected at $\sim$1\,mm by][]{san98},
IRAS\,20462+3416, and M\,2-56. For these sources, we find that the
mm-to-cm free-free continuum is consistent with \snu\alpha, with
values of $\alpha$ in the range $\sim$0.7 and 1.3.

The mm- and cm-wavelength continuum of CRL\,618 has been observed many
times during the last decades and is known to be variable, alternating
periods of brightening and weakening at mm-wavelengths
(\S\,\ref{intro-crl618}).  In Fig.\,\ref{f-crl6181mm} we show an
updated light curve of the continuum near 1mm including our recent
measurement from this work.

The free-free continuum of CRL\,618 is nearly flat
at mm-wavelengths, which is consistent with optically
thin emission and follows \snu{1.6} at longer wavelengths (Fig.\,\ref{f-seds}). Our
data (and model presented in \S\,\ref{moreli}) are consistent with a turnover
frequency $\nu_{\rm t}$\la\,80-90\,GHz in 2015, in agreement with
the value of $\nu_{\rm t}$ obtained in 1998 \citep{wyr03}. At earlier
epochs, the turnover frequency of the free-free continuum was larger,
in particular, $\nu_{\rm t}$$\sim$100-200\,GHz in 1987
\citep{mar88}. The decrease of $\nu_{\rm t}$ with time could imply a
decrease of the emission measure from EM$\approx$(1.6$\pm$0.9)\ex{11}
to $\approx$5.5\ex{10}\cm6 pc (adopting \te=13,000\,K). However, this
simplistic interpretation is only correct if the
physical conditions in, and geometry of, the ionized core have remained unaltered, which
seems unlikely given the rapid unforeseen changes of the mm-continuum
emission (Fig.\,\ref{f-crl6181mm} and \S\,\ref{intro-crl618}); for example,
  a variation of the electron temperature from \te$\sim$10,000 to
  18,000K would result in a shift of $\nu_{\rm t}$ from $\sim$150 to
  $\sim$100\,GHz for the same value of EM$\approx$1.1\ex{11}\,\cm6
  pc.

In the case of the pPN M\,2-9, our continuum data (both at 1 and
3\,mm) appear to be systematically lower than all previous published
measurements at these wavelengths. 
This, together with the scatter of the
continuum (ancillary) data points in the mm- and cm-wavelength domain 
could reflect time variability of
the free-free emission in M\,2-9. 

The pPN Hen\,3-1475 is another object in our sample with 
flux-density variations. A possibly periodic pattern and an
inversion of its spectral index on a timescale of a few years have
been suggested from previous multifrequency high-angular resolution
radio observations \citep{cerri11}. 

In the case of M\,1-92, M\,1-91, and IRAS\,20462$+$3416, the continuum flux at
\ga\,3-4\,cm clearly exceeds that expected by extrapolating the trends
observed at lower wavelengths, including the mm-range. 
This could be due to contaminating foreground/background emission
sources, strong continuum variability, or additional superimposed
free-free emission components with different contributions at cm- and
mm-wavelengths (e.g.,\,analog to the faint, extended emission from
the lobes of M\,2-9, \S\,\ref{intro-m29}).

\subsection{Recombination lines}
\label{resRRLs}
\begin{figure*}[htbp]
   \centering
\includegraphics*[bb=-10   0 820  520,width=0.475\textwidth]{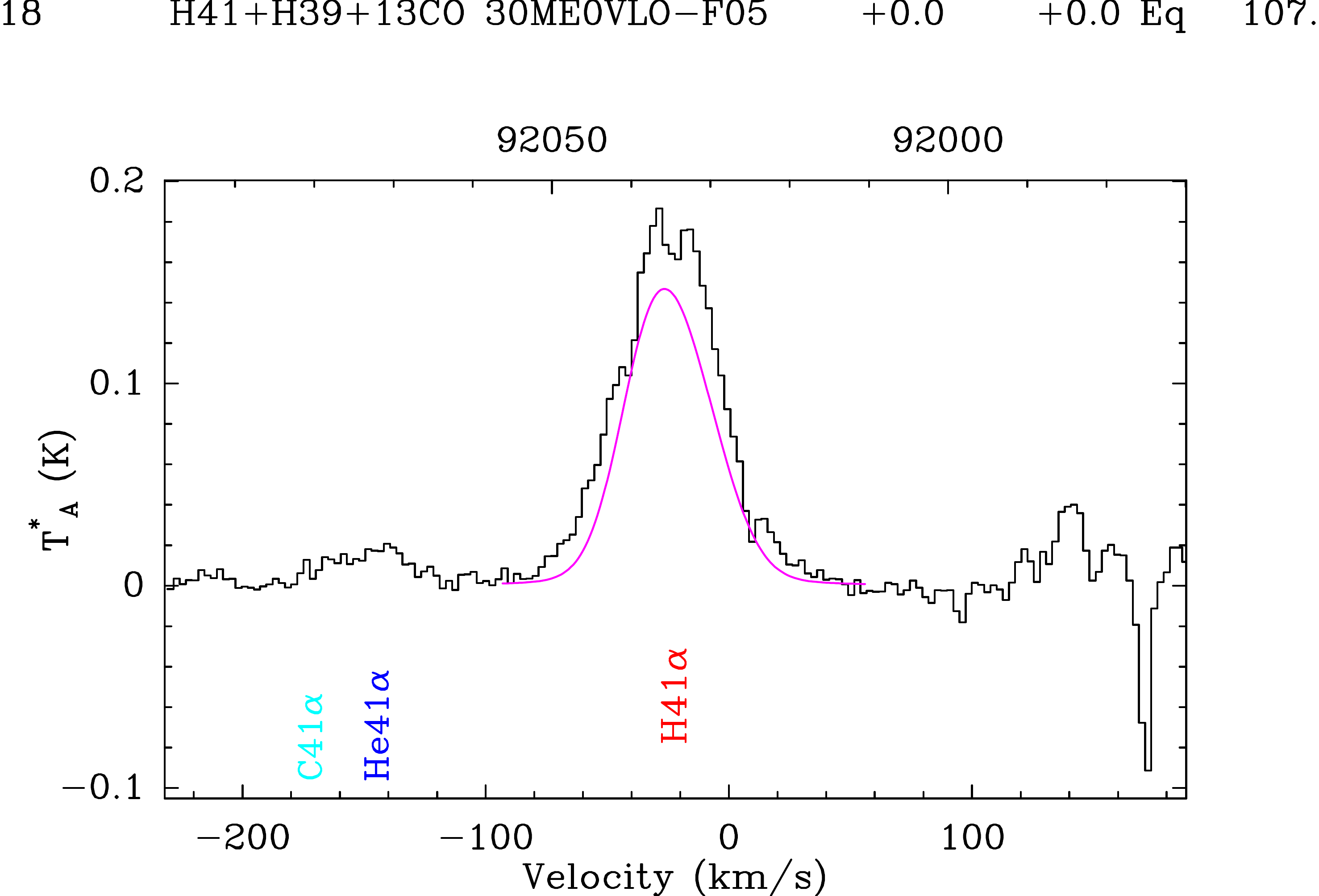}
\includegraphics*[bb=-10   0 820  520,width=0.475\textwidth]{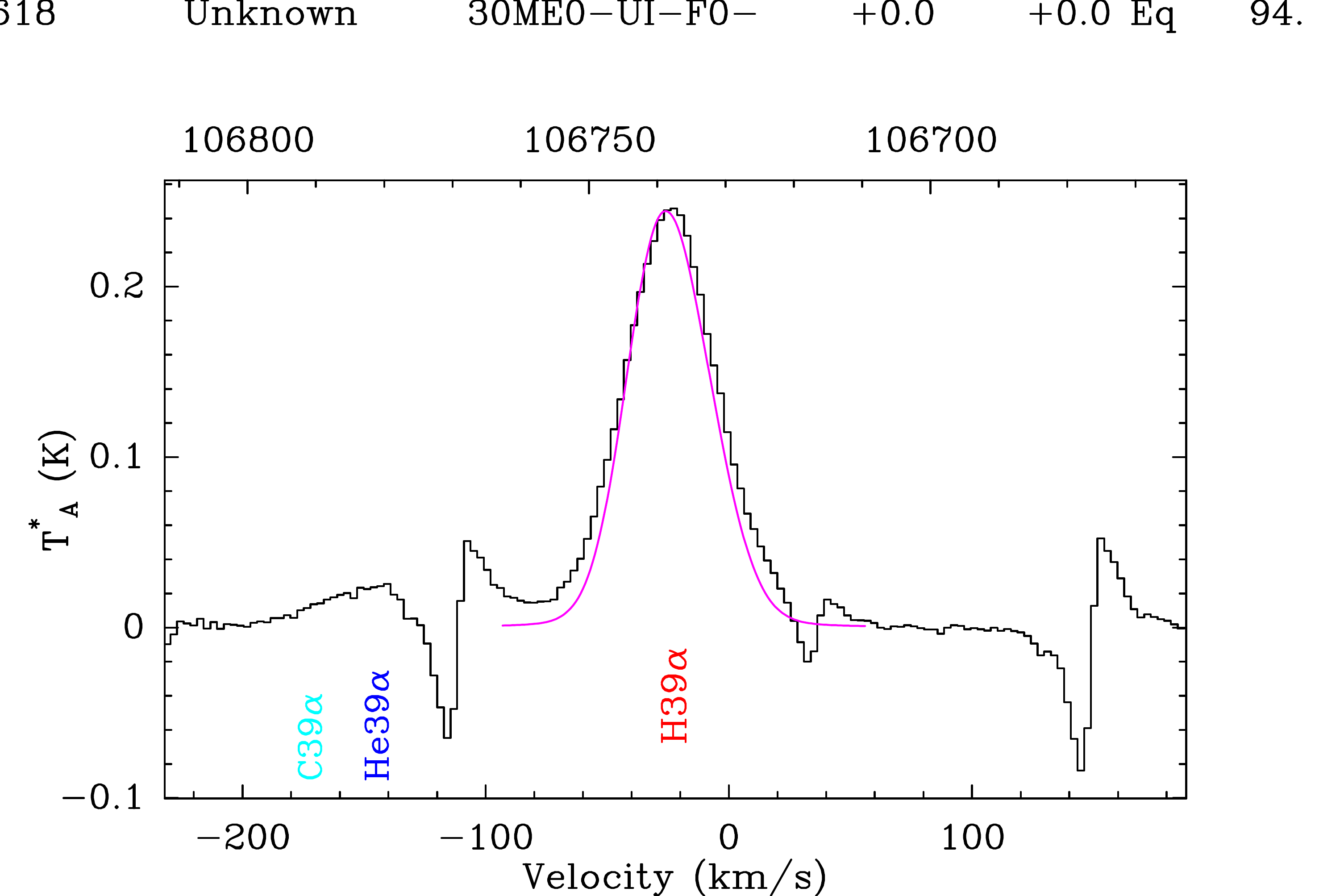}
\includegraphics*[bb=-10   0 820  520,width=0.475\textwidth]{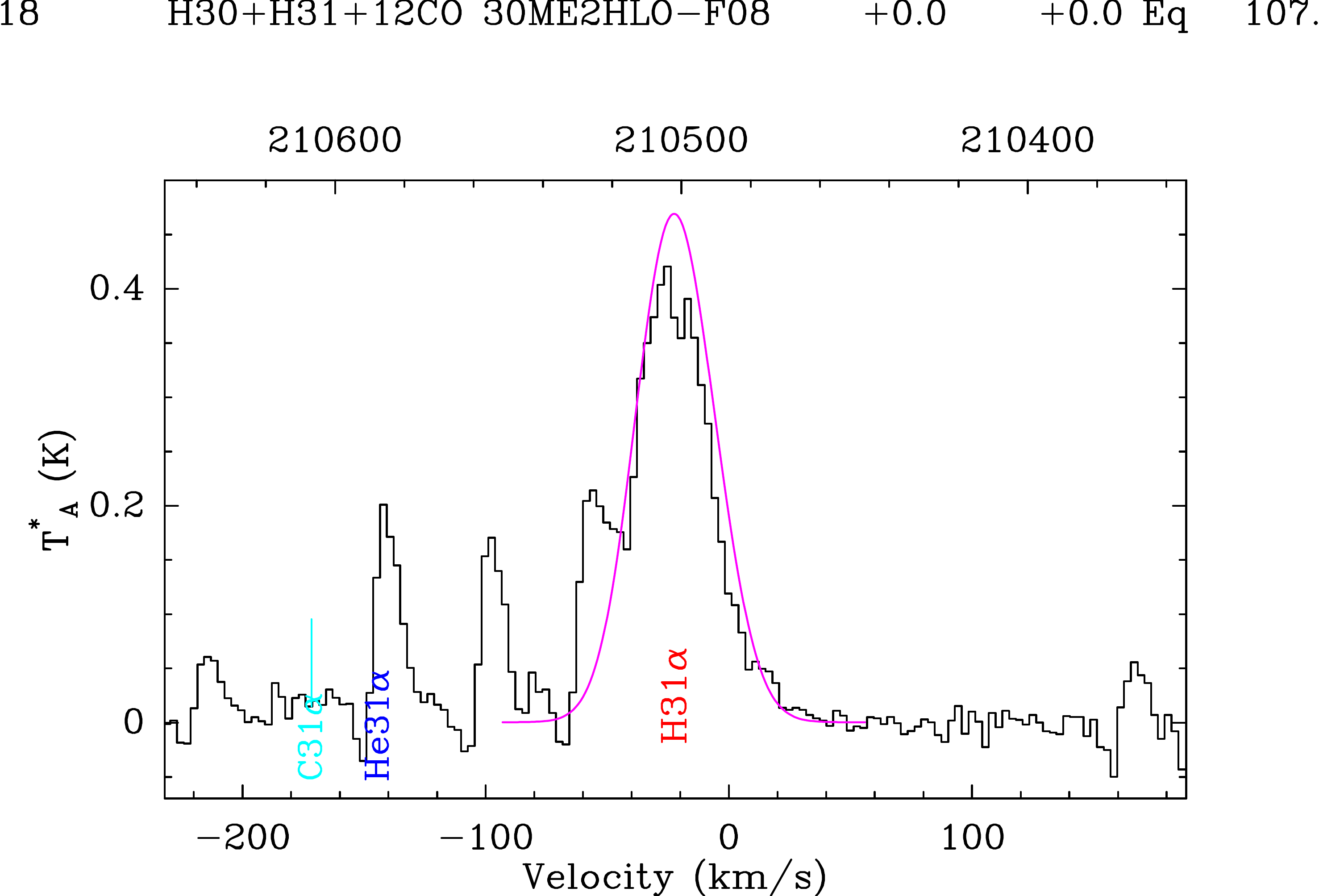}
\includegraphics*[bb=-10   0 820  520,width=0.475\textwidth]{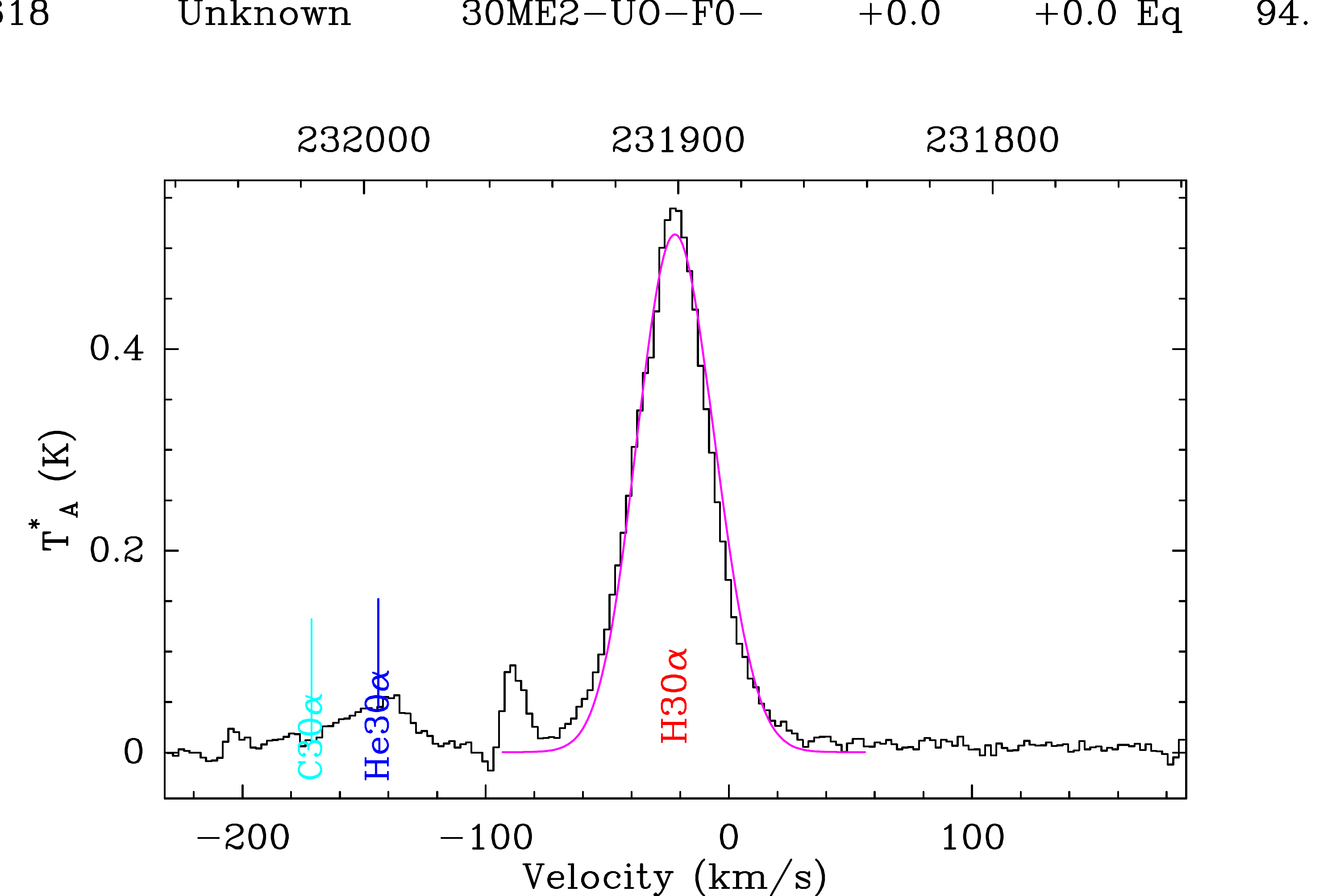} 

\vspace{1cm}
\includegraphics*[bb=-10 0 820 520,width=0.475\textwidth]{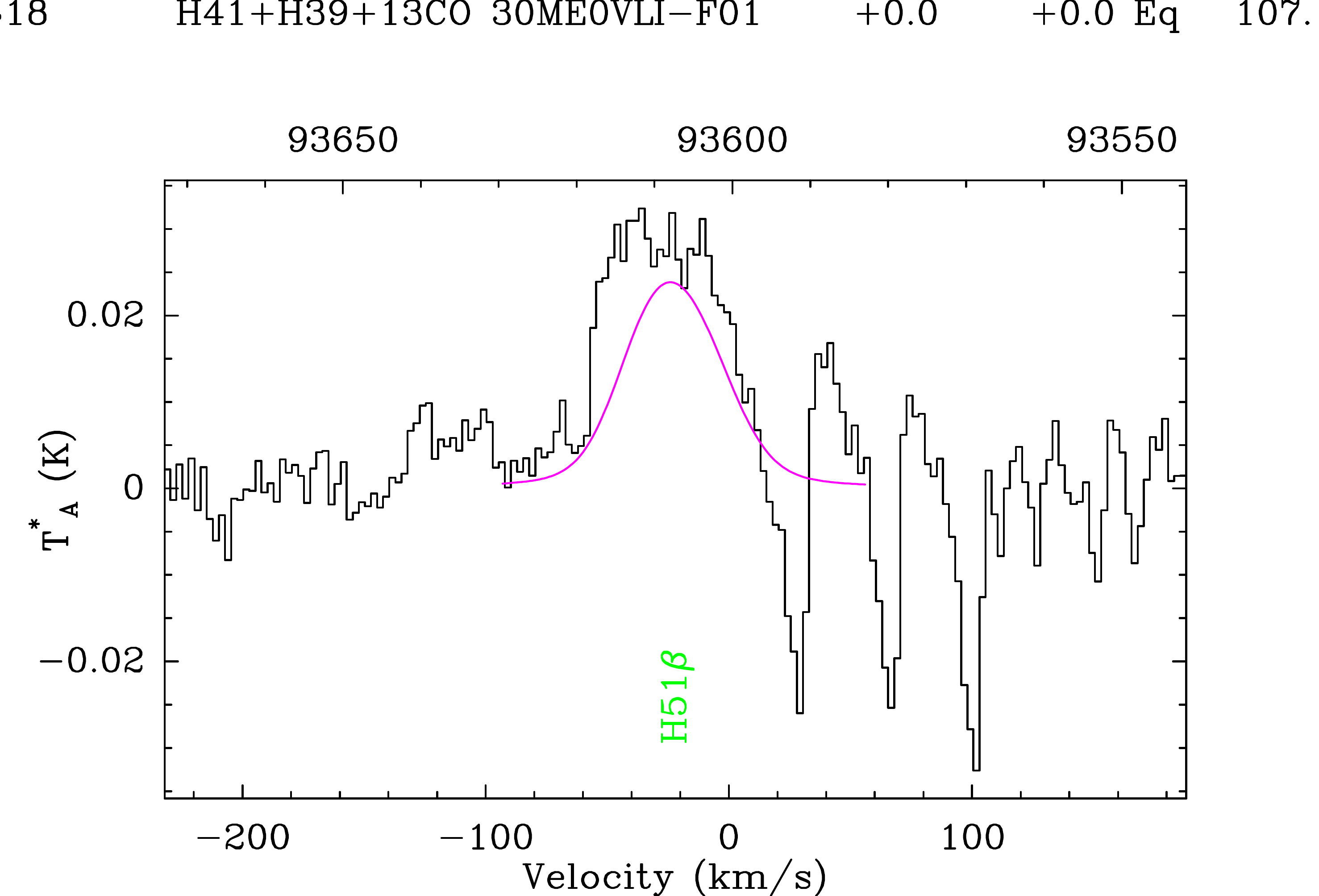}
\includegraphics*[bb=-10 0 820 520,width=0.475\textwidth]{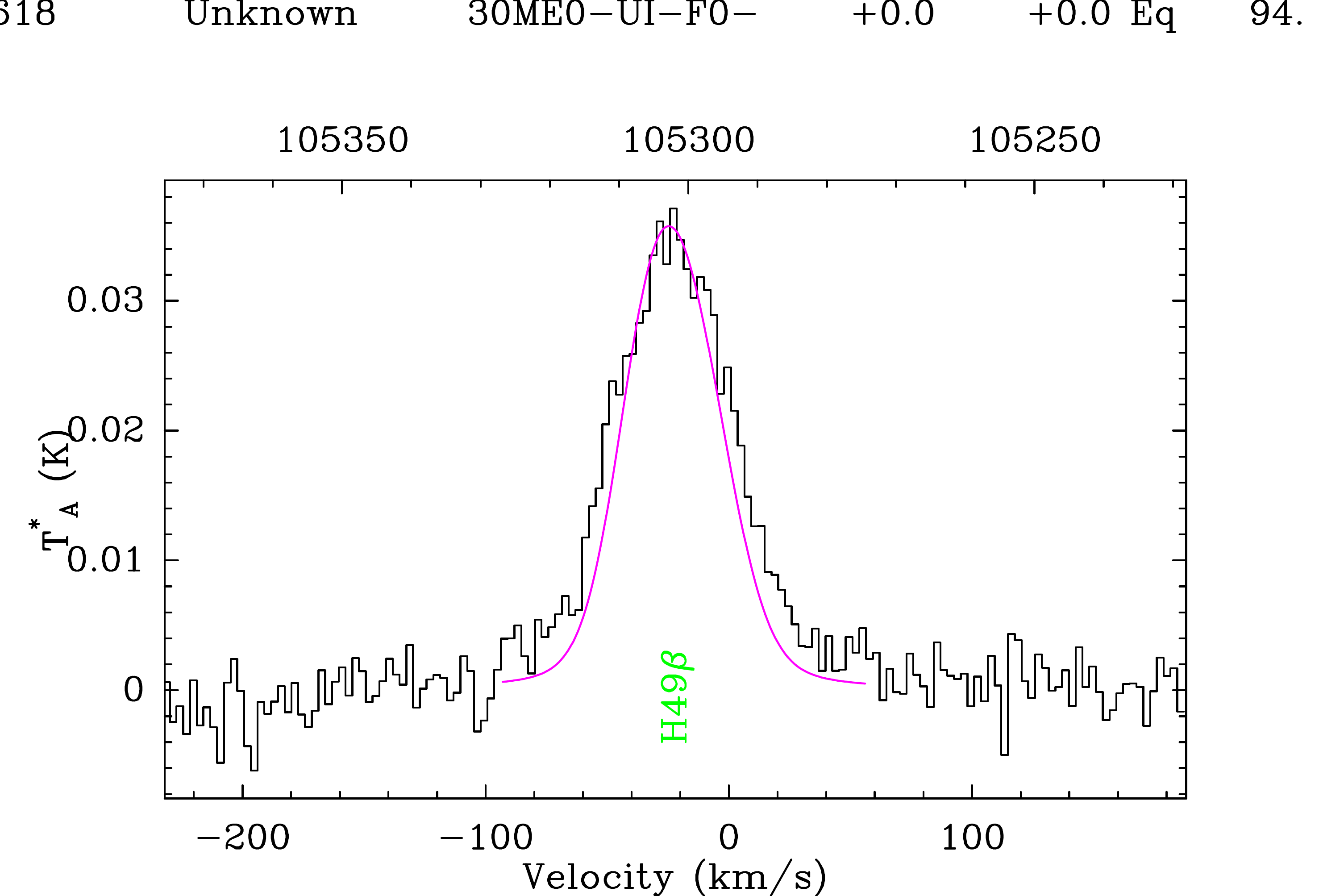}
\includegraphics*[bb=-10 0 820 520,width=0.475\textwidth]{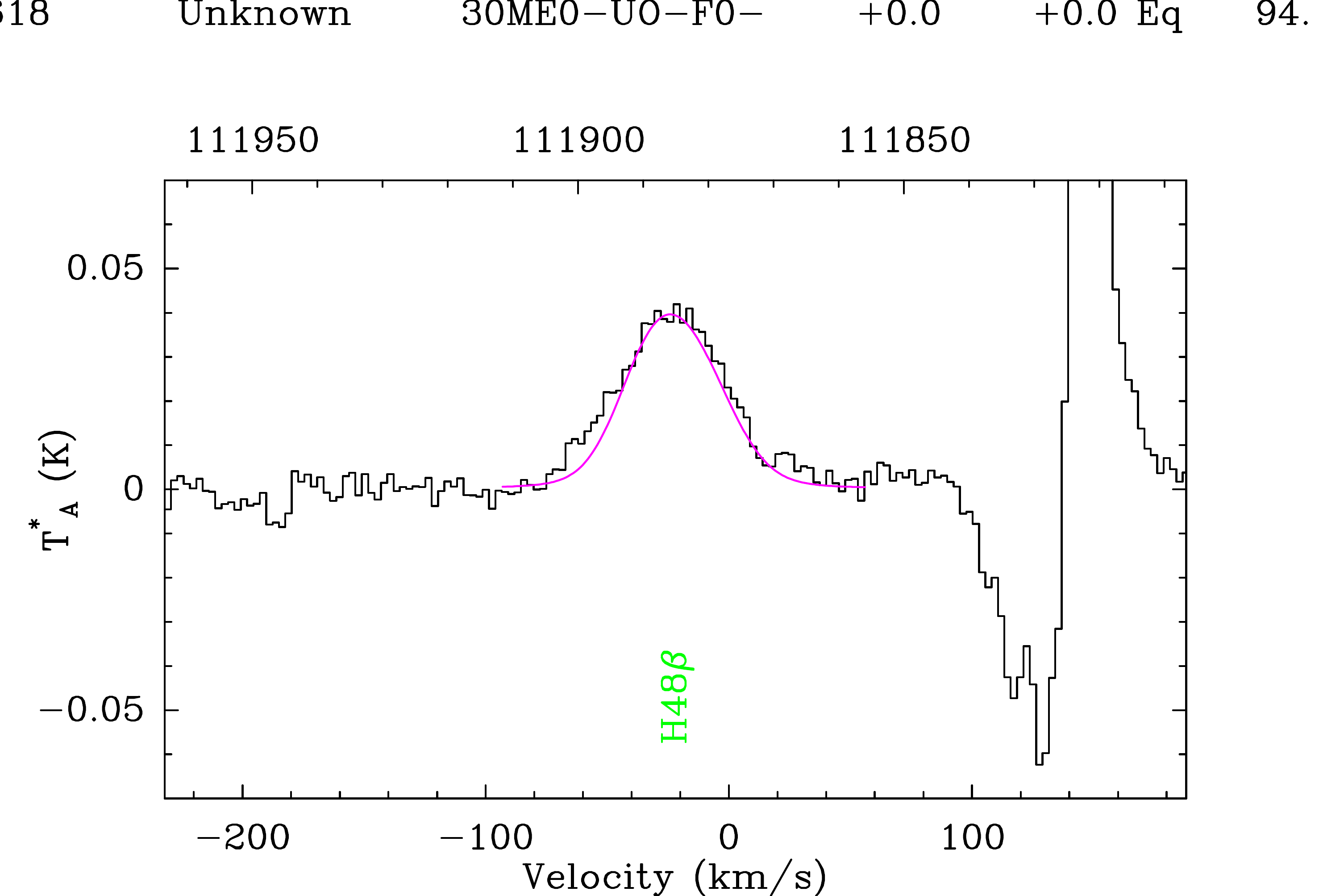}
\includegraphics*[bb=-10 0 820 520,width=0.475\textwidth]{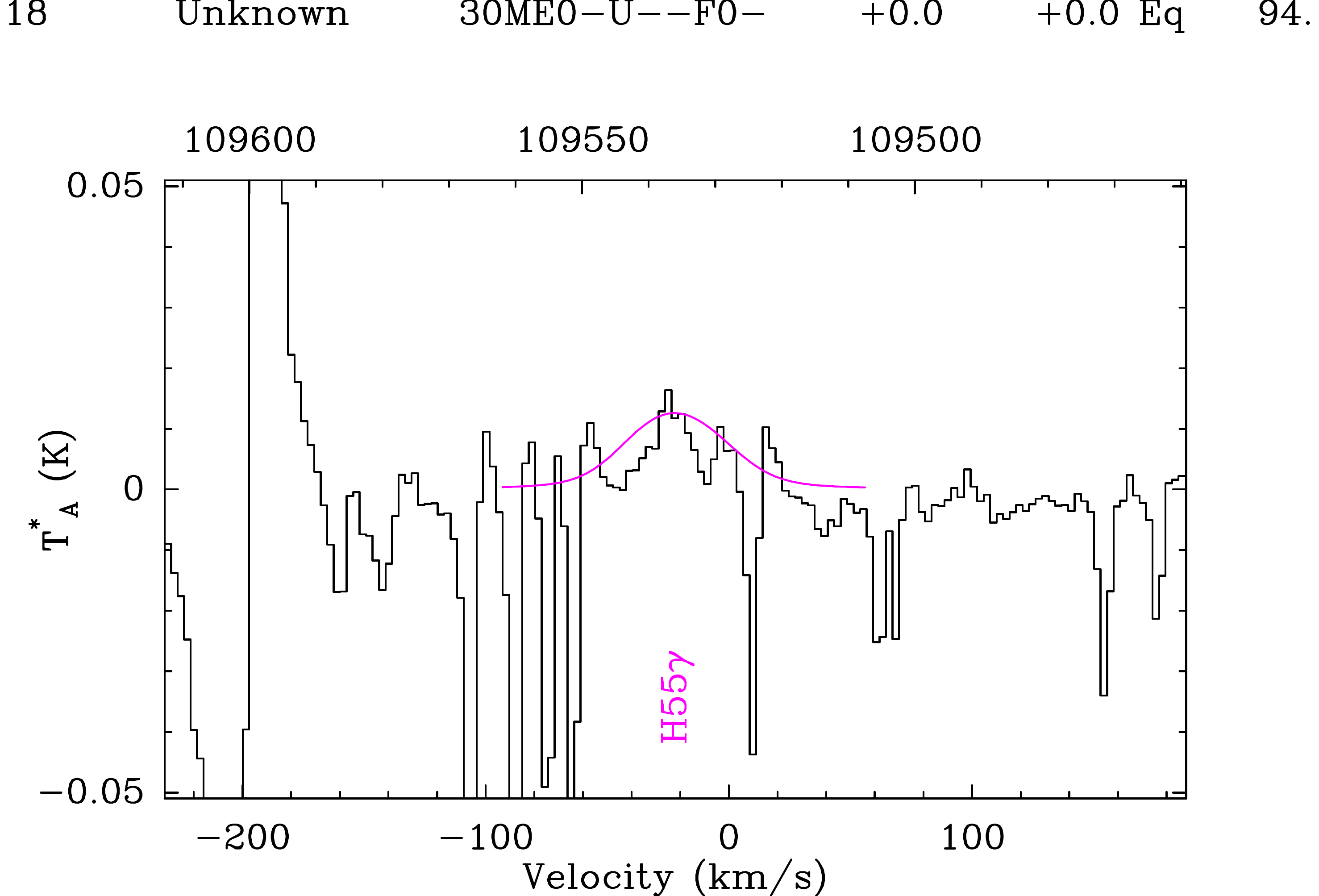}
   \caption{Recombination lines observed toward the pPN CRL\,618
     (histogram) and synthetic line profiles (pink) from our model in Table\,\ref{t-moreli} (\S\,\ref{model_c618}). The
     bottom and top X-axis represent 
     \vlsr\ (\kms) and frequency (MHz).
     The frequency of He and C $\alpha$-transitions, some
     of them detected, are also indicated. The many additional
     features observed in the spectrum, with emission and absorption
     profile components, are molecular transitions produced in the
     C-rich molecular envelope of CRL\,618. Some
     of the mm-RRLs observed are partially blended with molecular
     lines.
   }
   \label{f-crl618}
   \end{figure*}

\begin{figure*}[htbp]
   \centering 
\includegraphics*[bb=-10 0 820 520,width=0.475\textwidth]{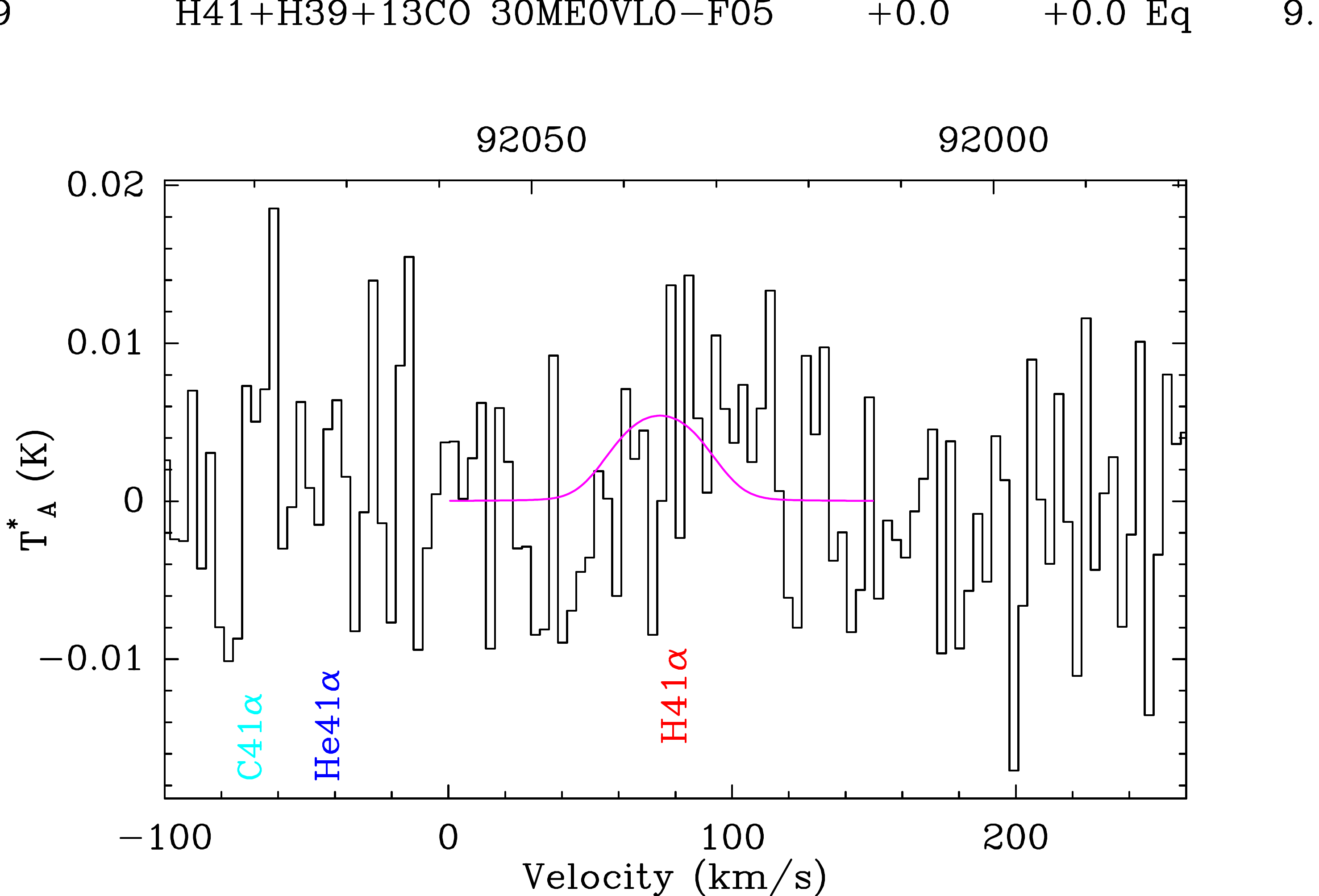}
\includegraphics*[bb=-10 0 820 520,width=0.475\textwidth]{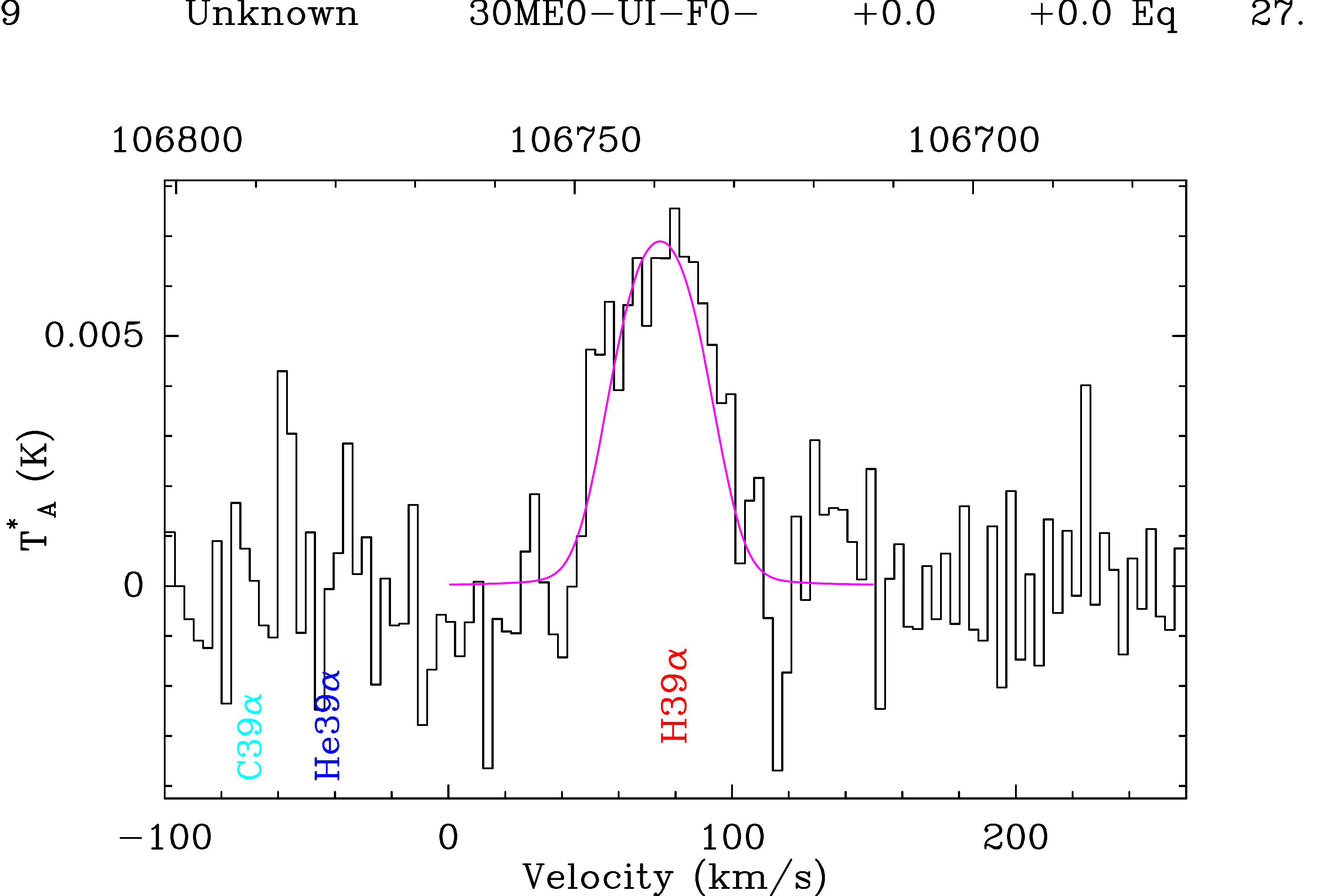}
\includegraphics*[bb=-10 0 820 520,width=0.475\textwidth]{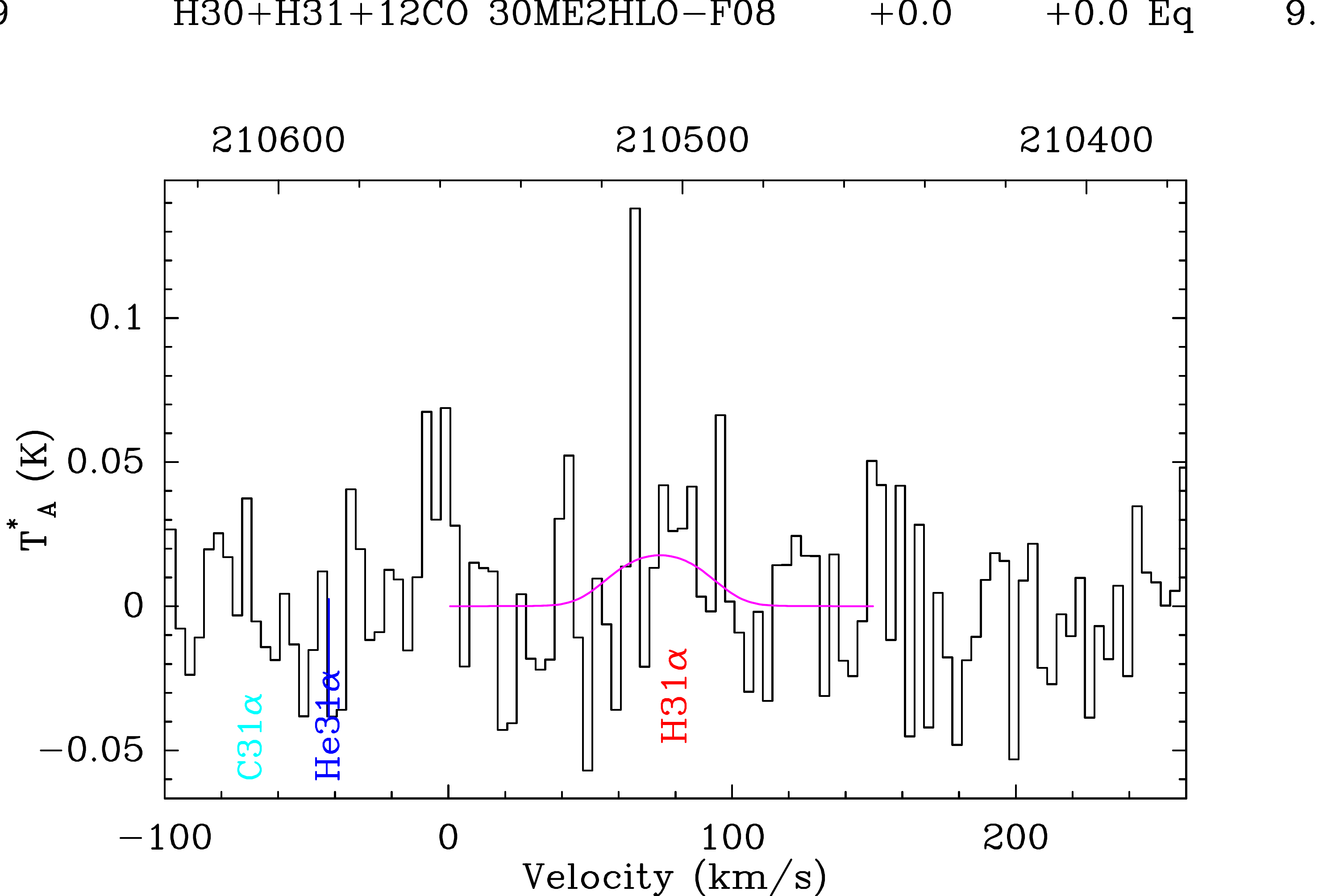}
\includegraphics*[bb=-10 0 820 520,width=0.475\textwidth]{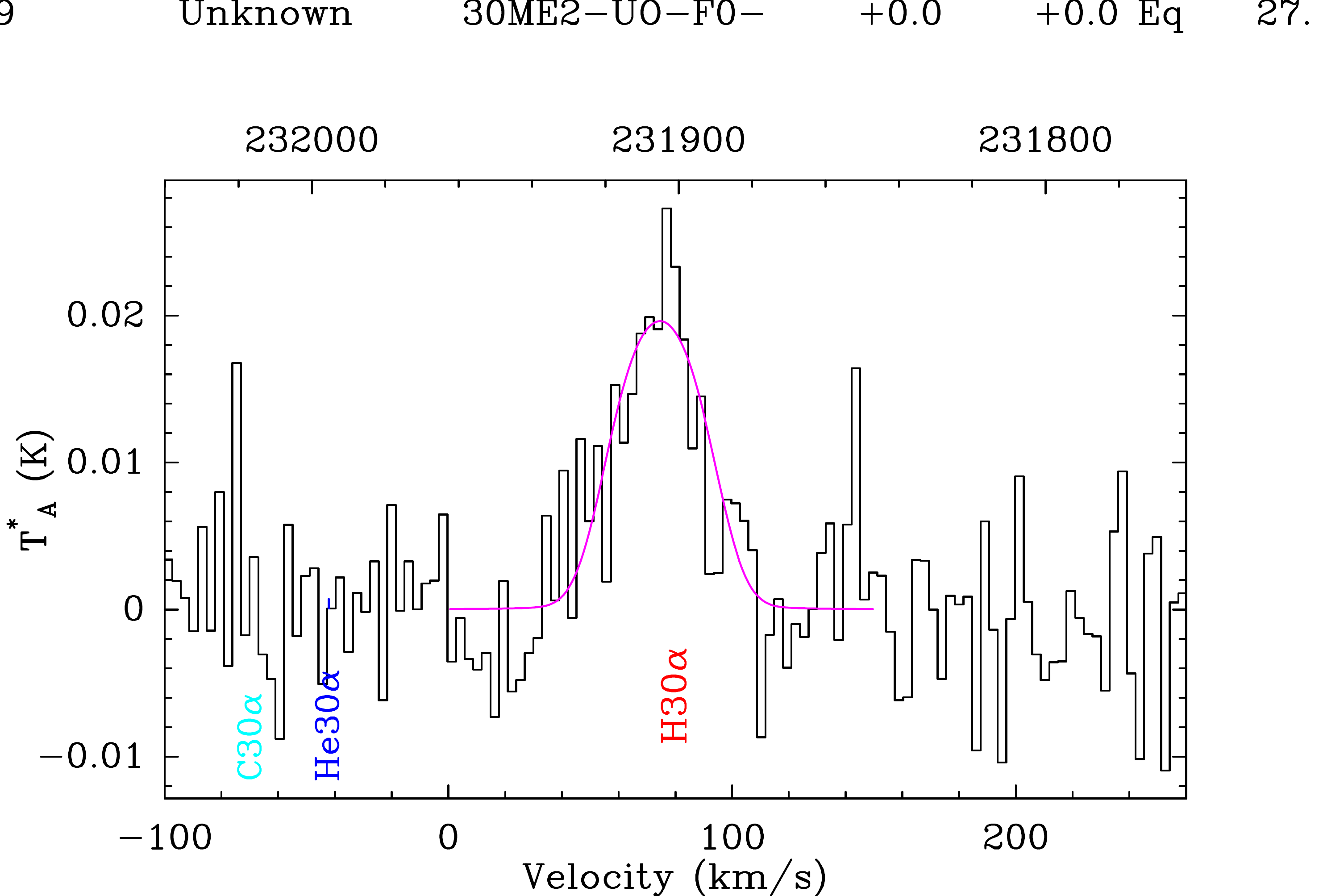} 

\vspace{1cm}
\includegraphics*[bb=-10 0 820 520,width=0.475\textwidth]{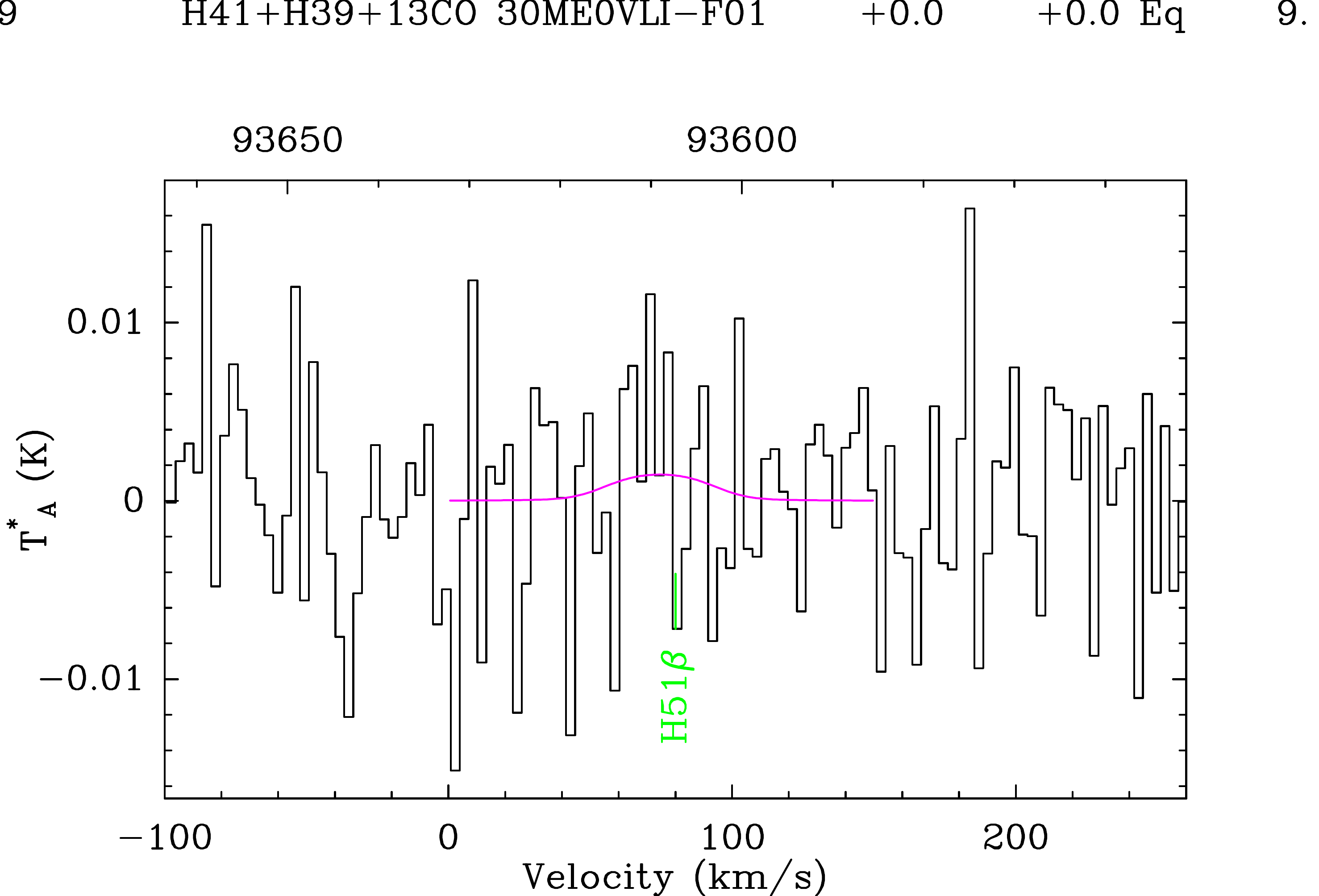}
\includegraphics*[bb=-10 0 820 520,width=0.475\textwidth]{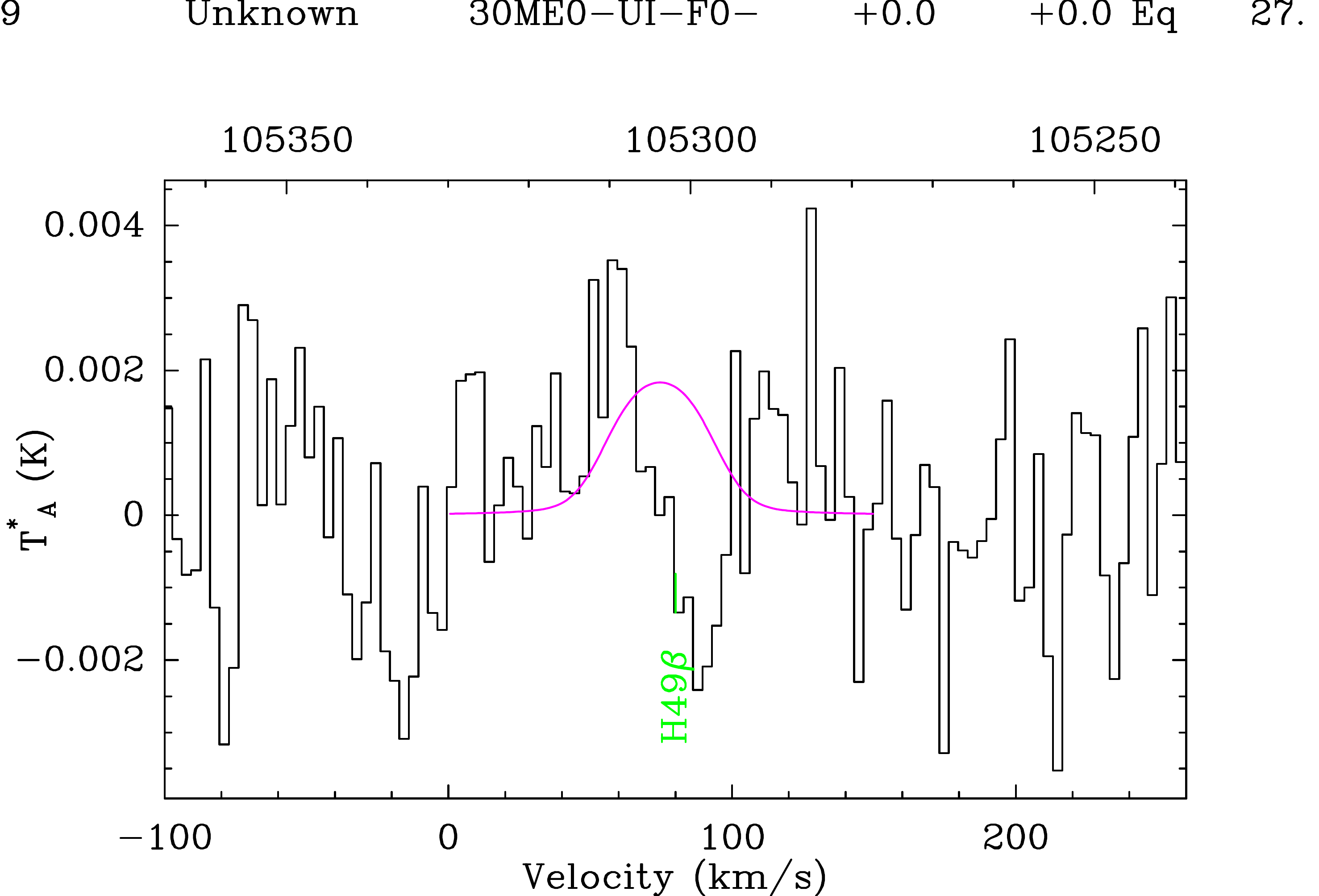}
\includegraphics*[bb=-10 0 820 520,width=0.475\textwidth]{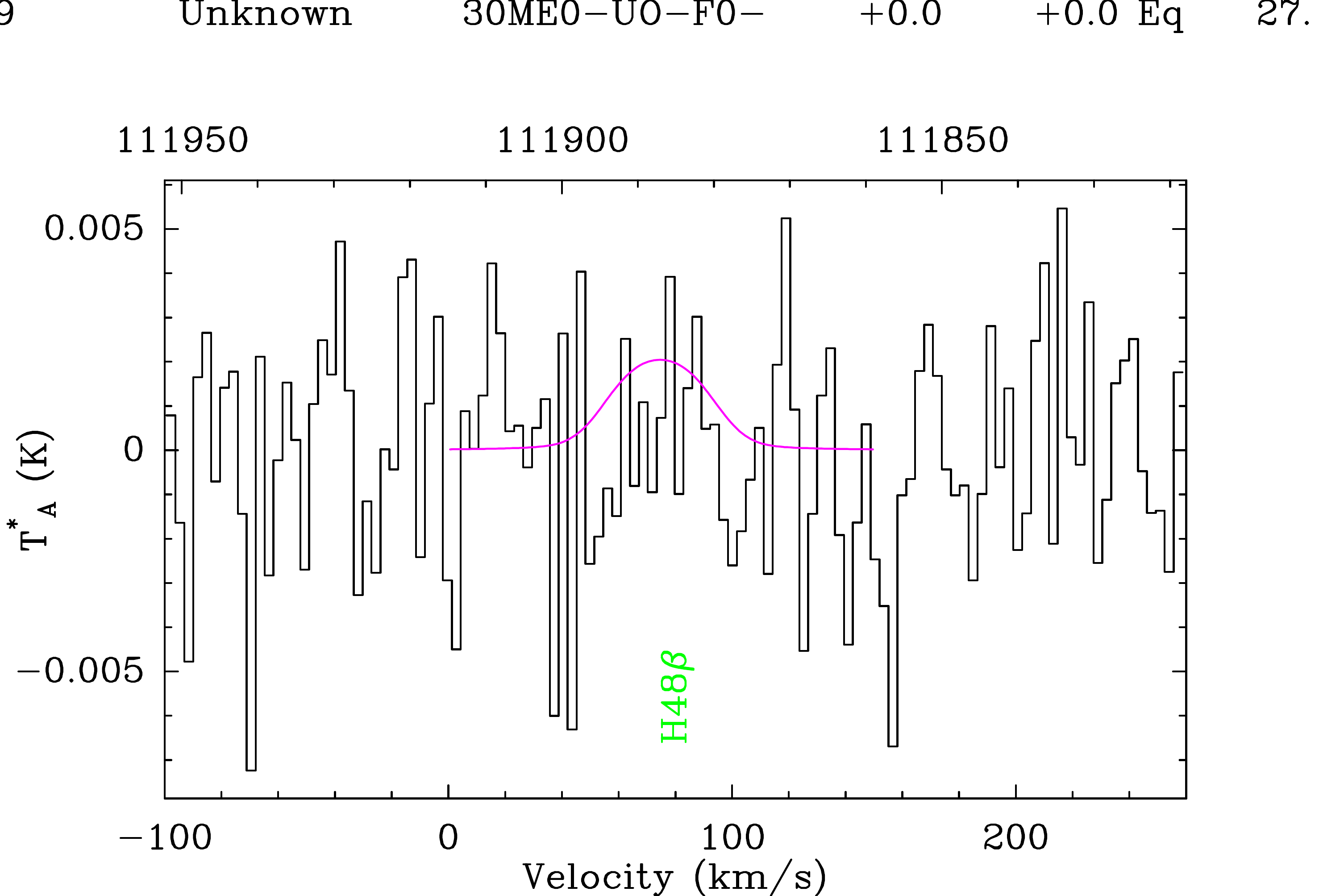}
\includegraphics*[bb=-10 0 820 520,width=0.475\textwidth]{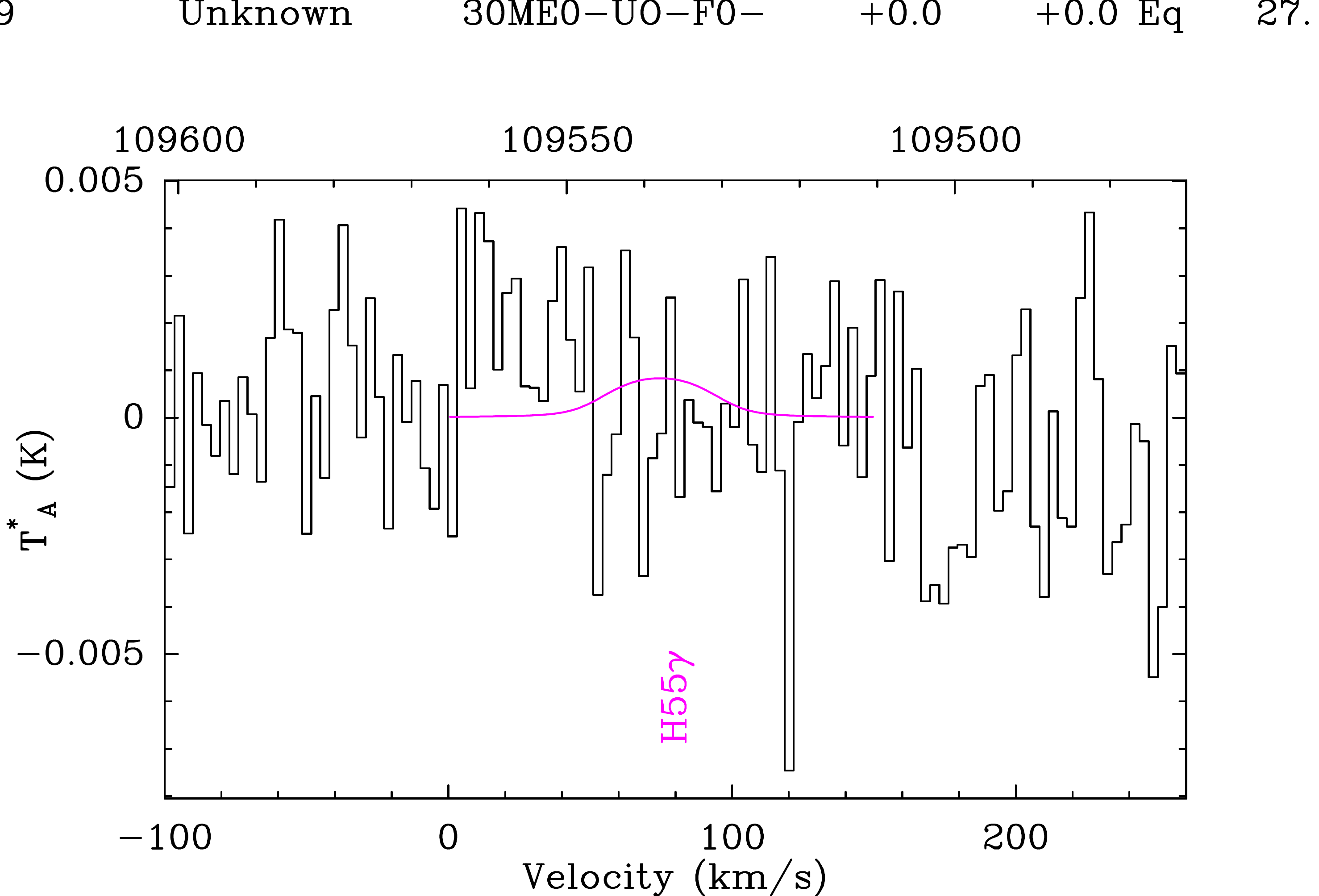}
\caption{Recombination lines observed toward M\,2-9 (histogram) and synthetic line profiles (pink) from our model in Table\,\ref{t-moreli} (\S\,\ref{model_m29}).}
   \label{f-m2-9}
   \end{figure*}

\begin{figure*}[htbp] 
   \centering 
\includegraphics*[bb=-10 0 820 520,width=0.475\textwidth]{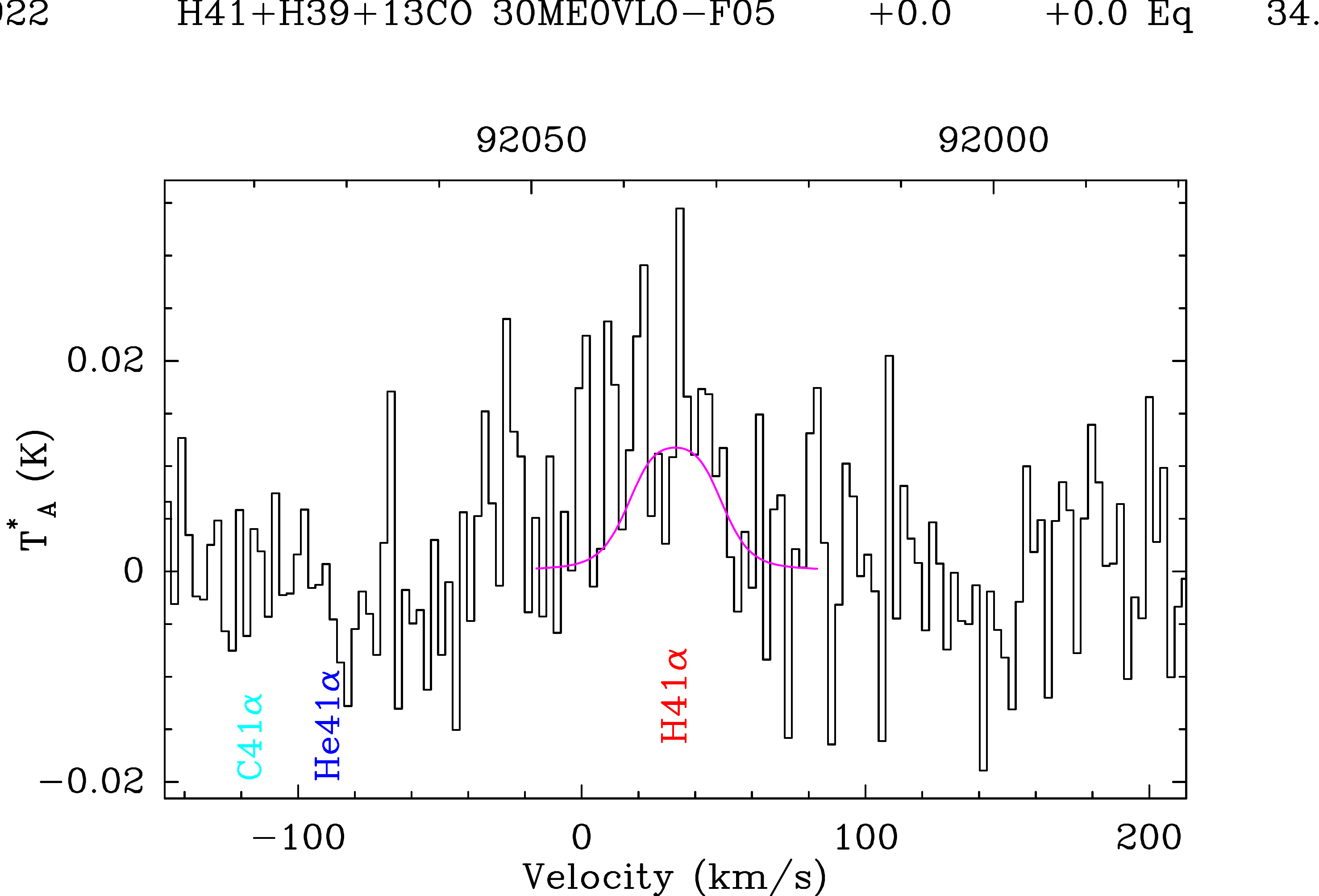}
\includegraphics*[bb=-10 0 820 520,width=0.475\textwidth]{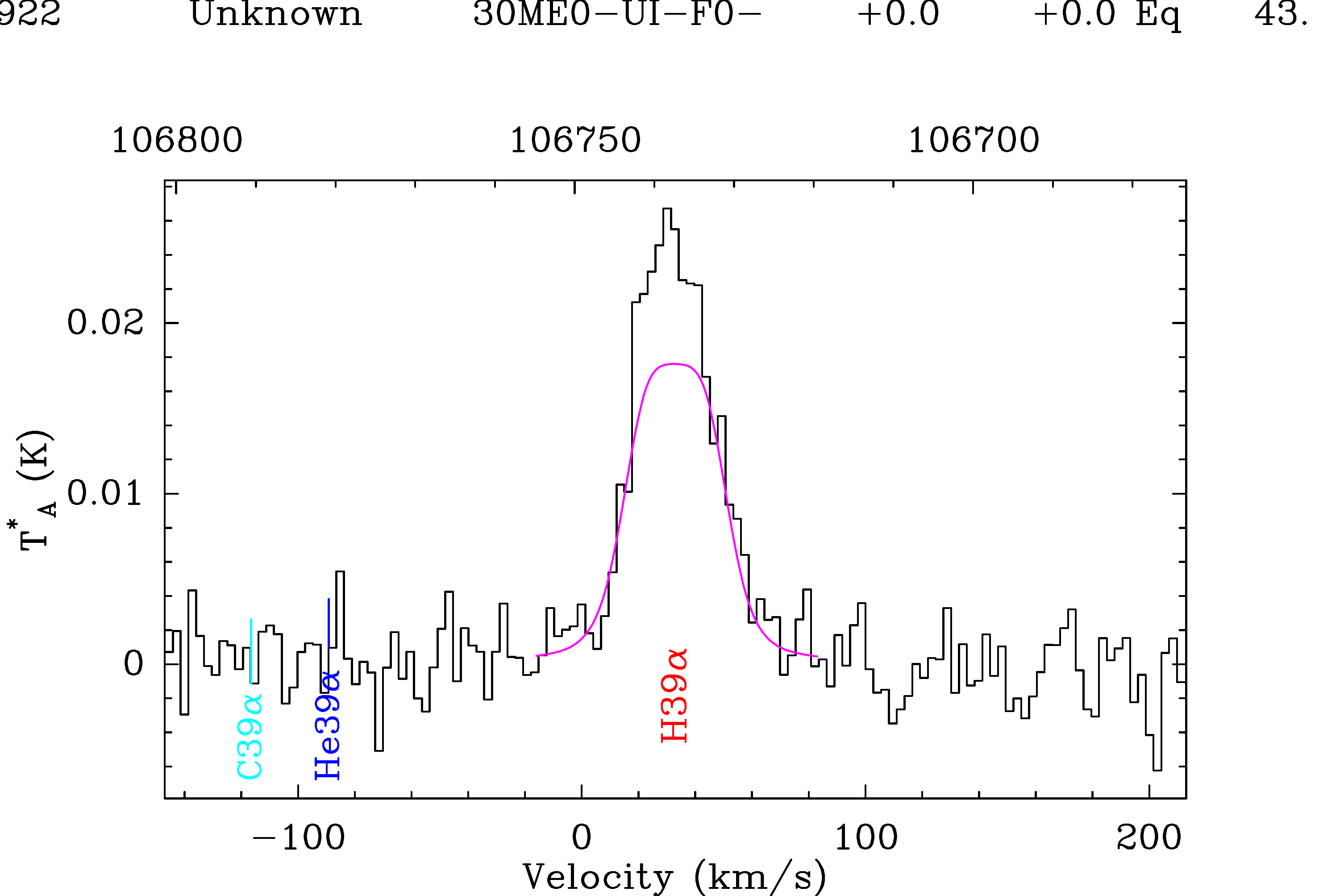}
\includegraphics*[bb=-10 0 820 520,width=0.475\textwidth]{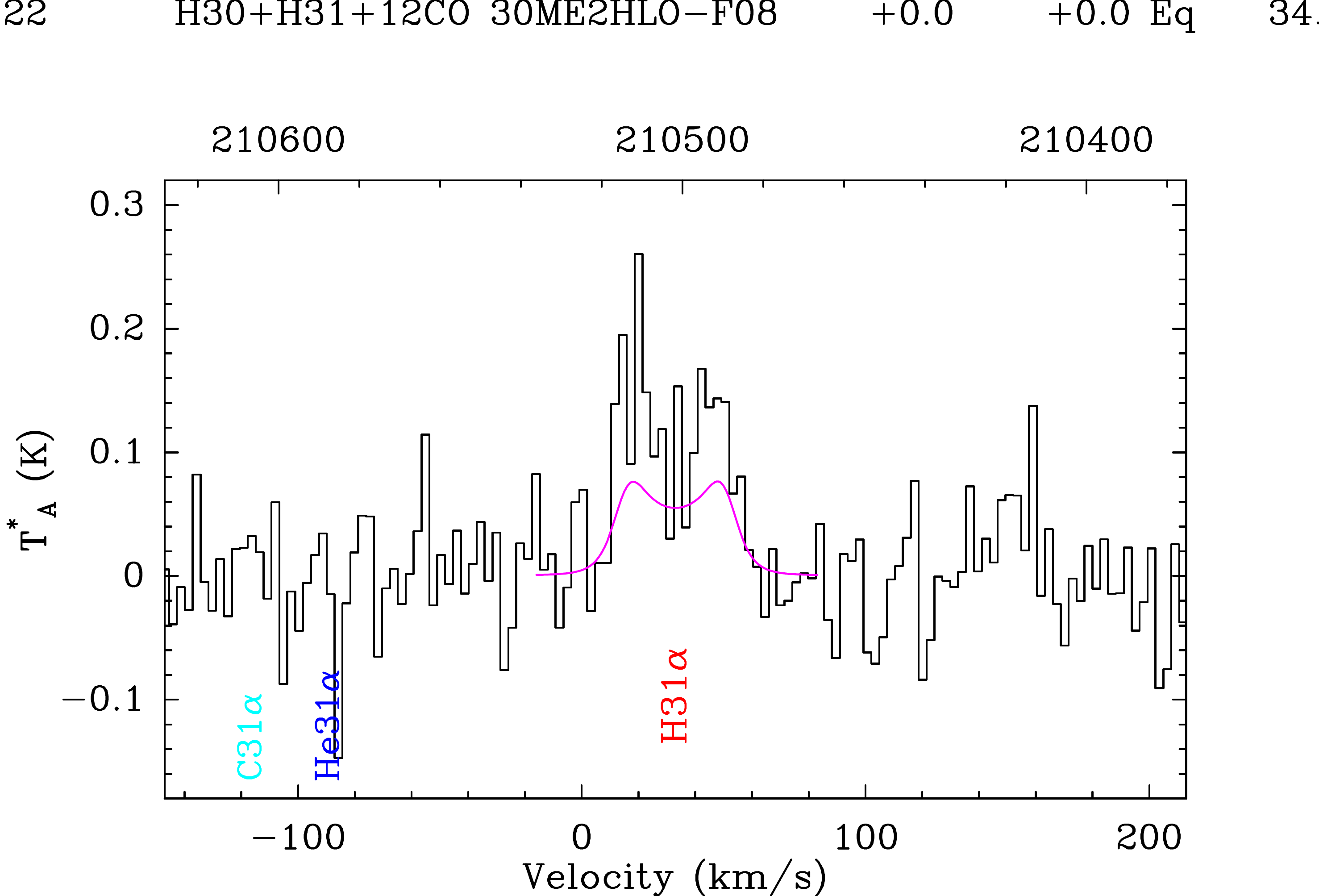}
\includegraphics*[bb=-10 0 820 520,width=0.475\textwidth]{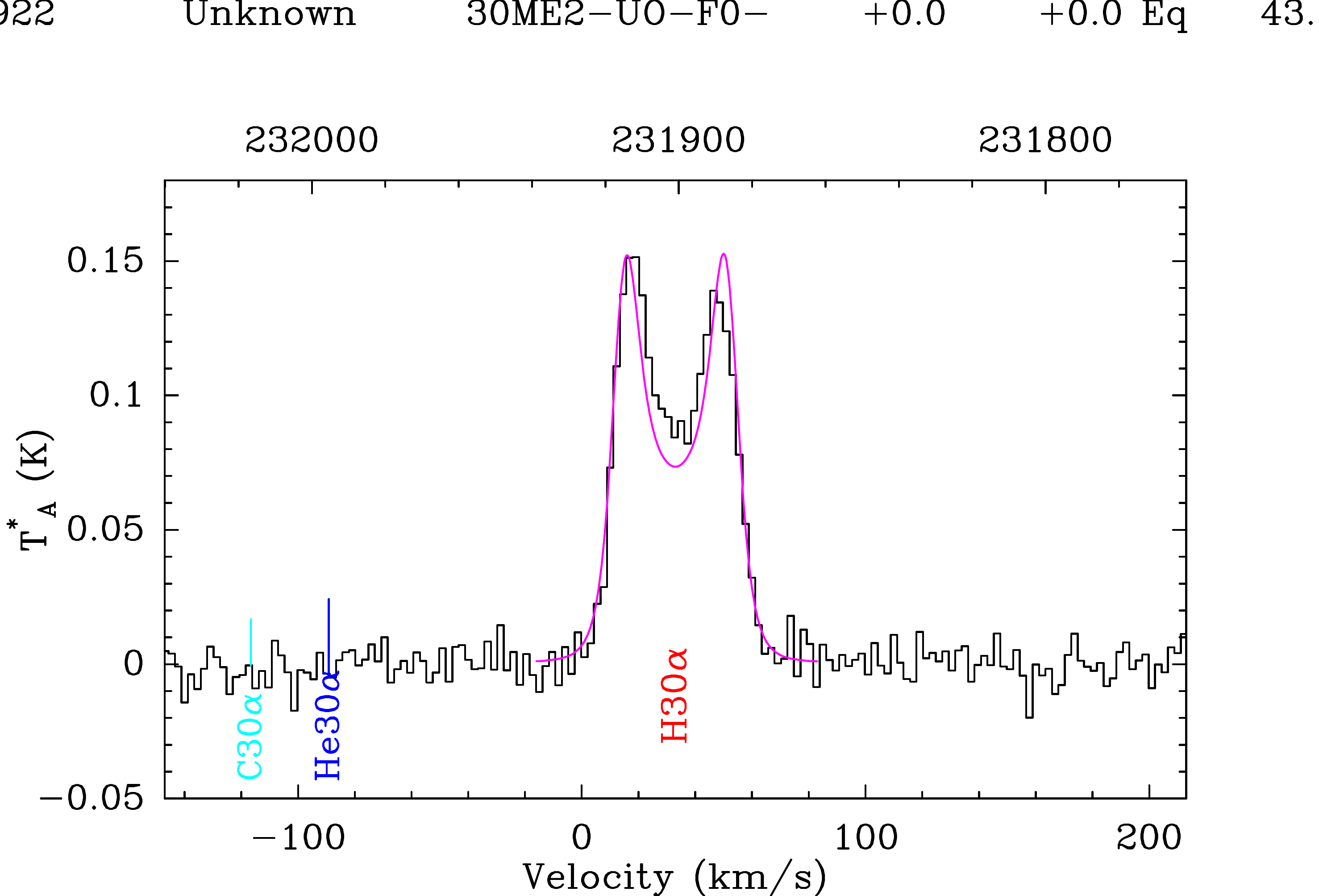} 

\vspace{1cm}
\includegraphics*[bb=-10 0 820 520,width=0.475\textwidth]{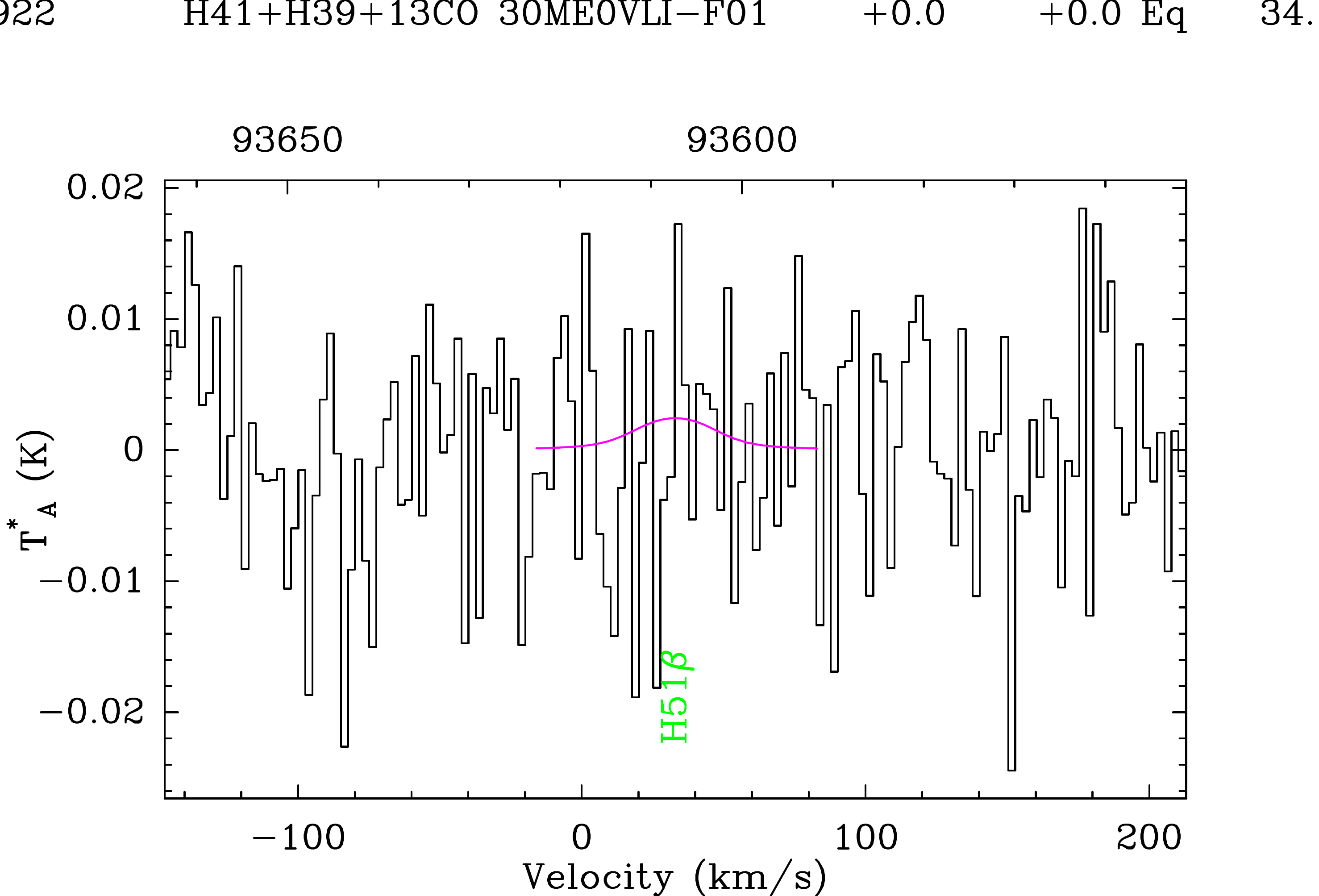}
\includegraphics*[bb=-10 0 820 520,width=0.475\textwidth]{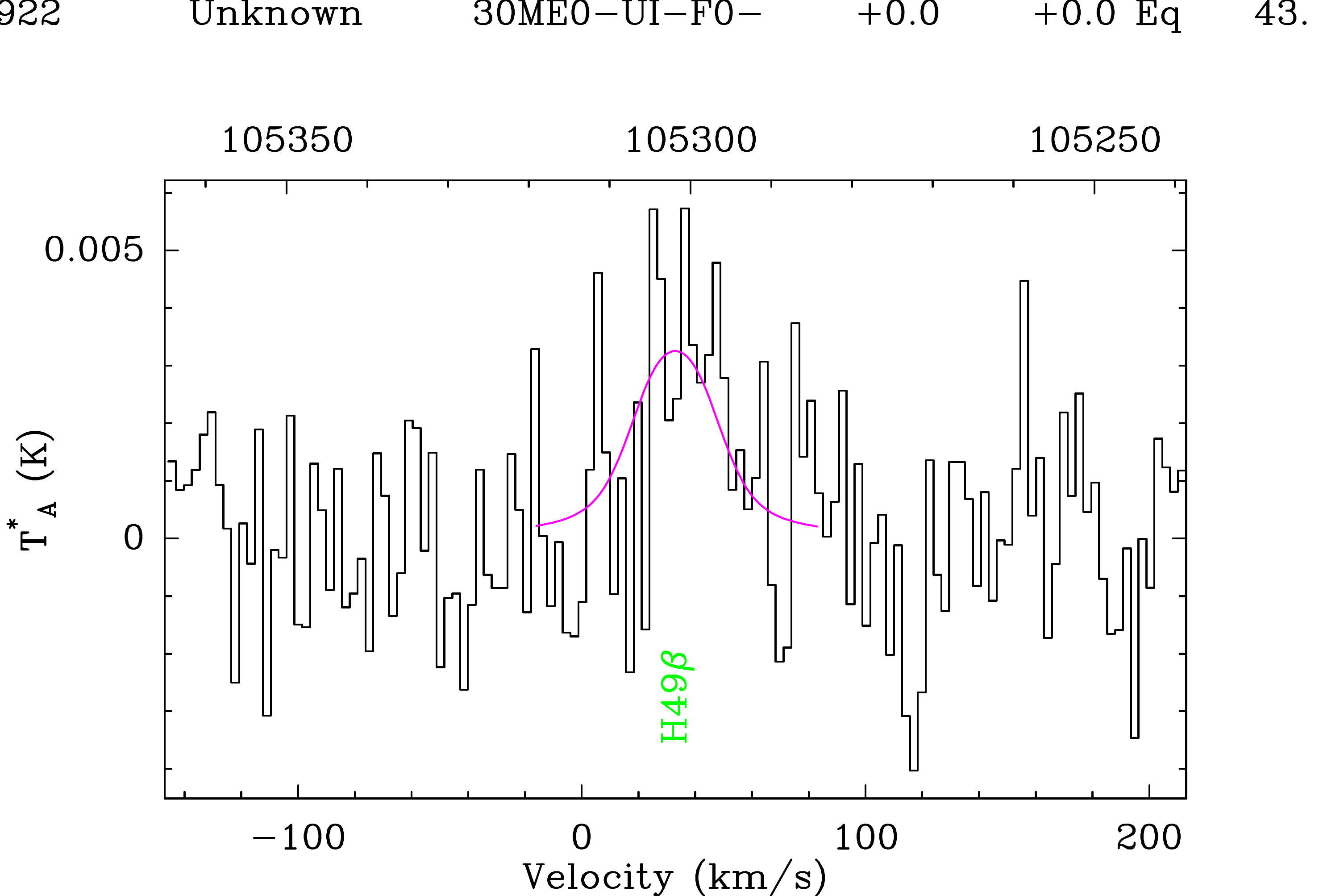}
\includegraphics*[bb=-10 0 820 520,width=0.475\textwidth]{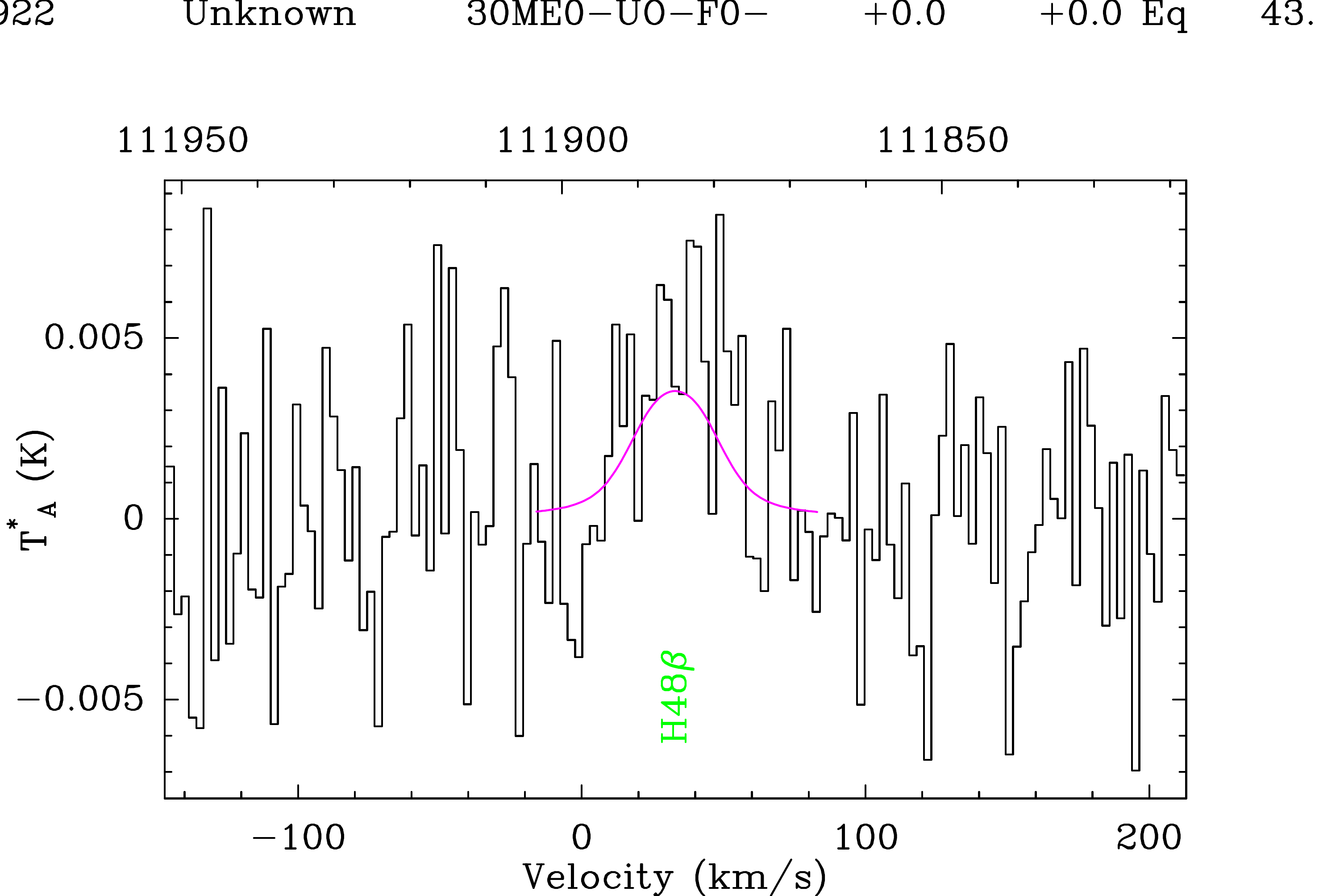}
\includegraphics*[bb=-10 0 820 520,width=0.475\textwidth]{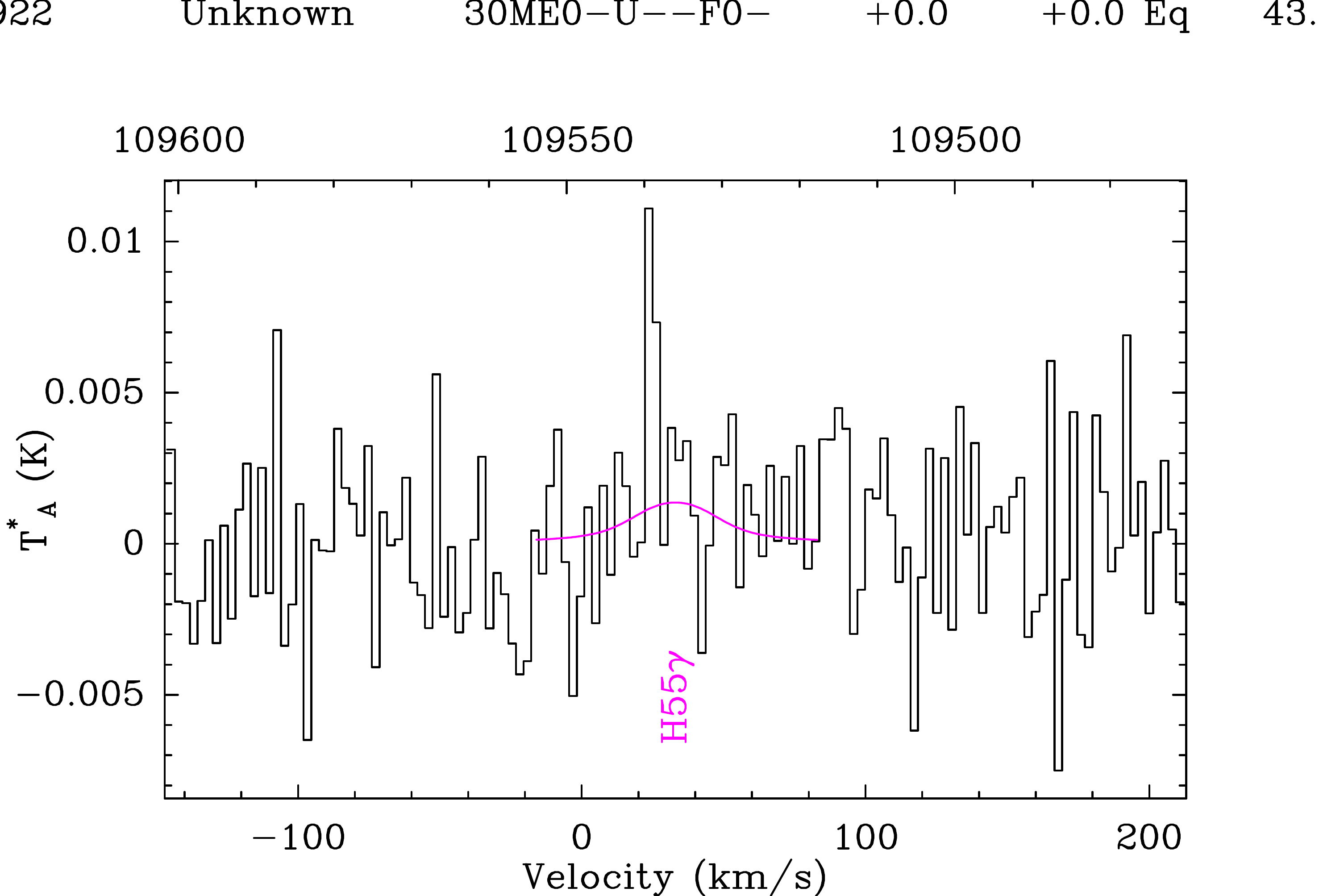}
\caption{Recombination lines observed toward MWC\,922 (histogram) and synthetic line profiles (pink) from our model in Table\,\ref{t-moreli} (\S\,\ref{model_mwc922}).
}
   \label{f-mwc922}
   \end{figure*}

\begin{figure*}[htbp]
   \centering 
\includegraphics[width=0.32\textwidth]{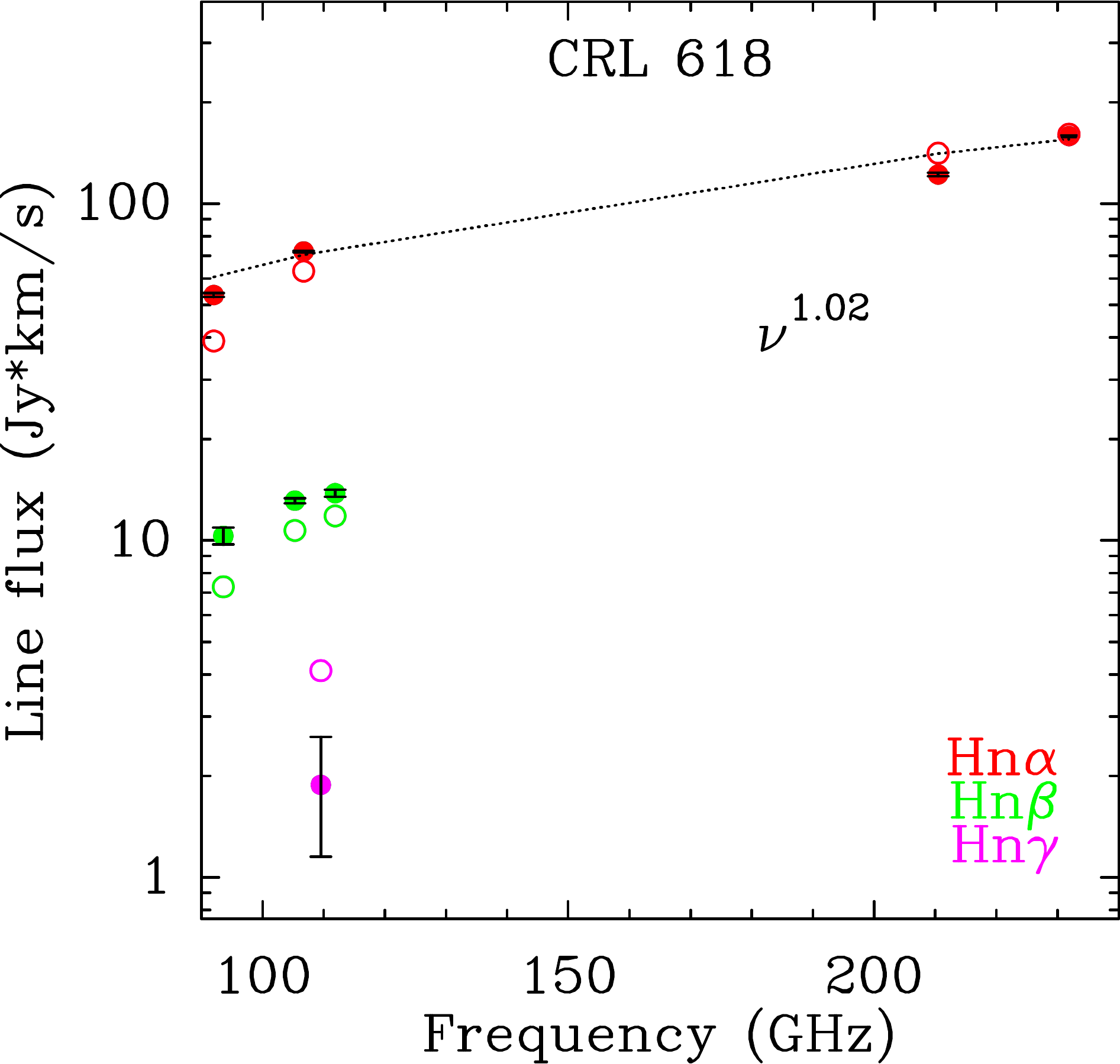}
\includegraphics[width=0.32\textwidth]{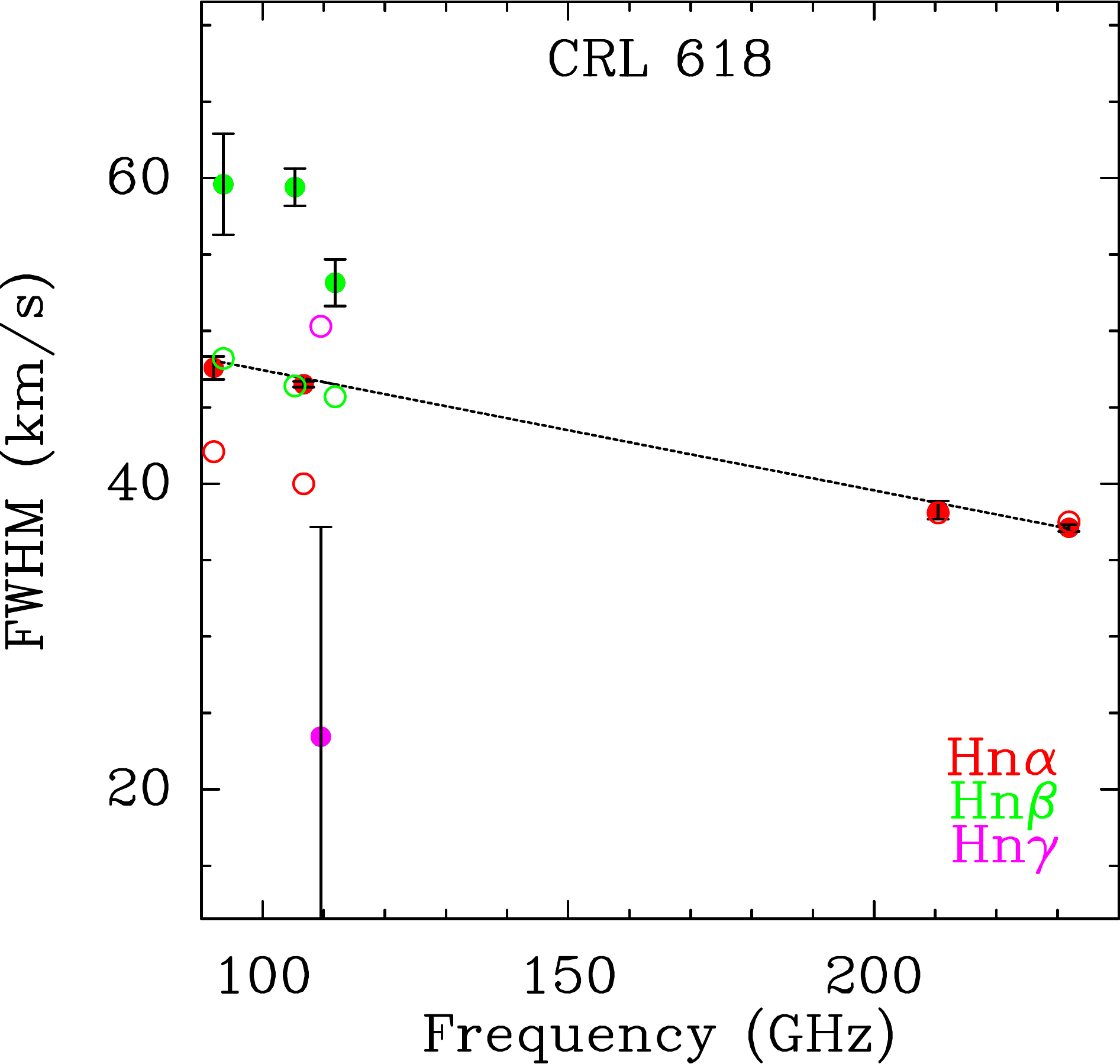}
\includegraphics[width=0.32\textwidth]{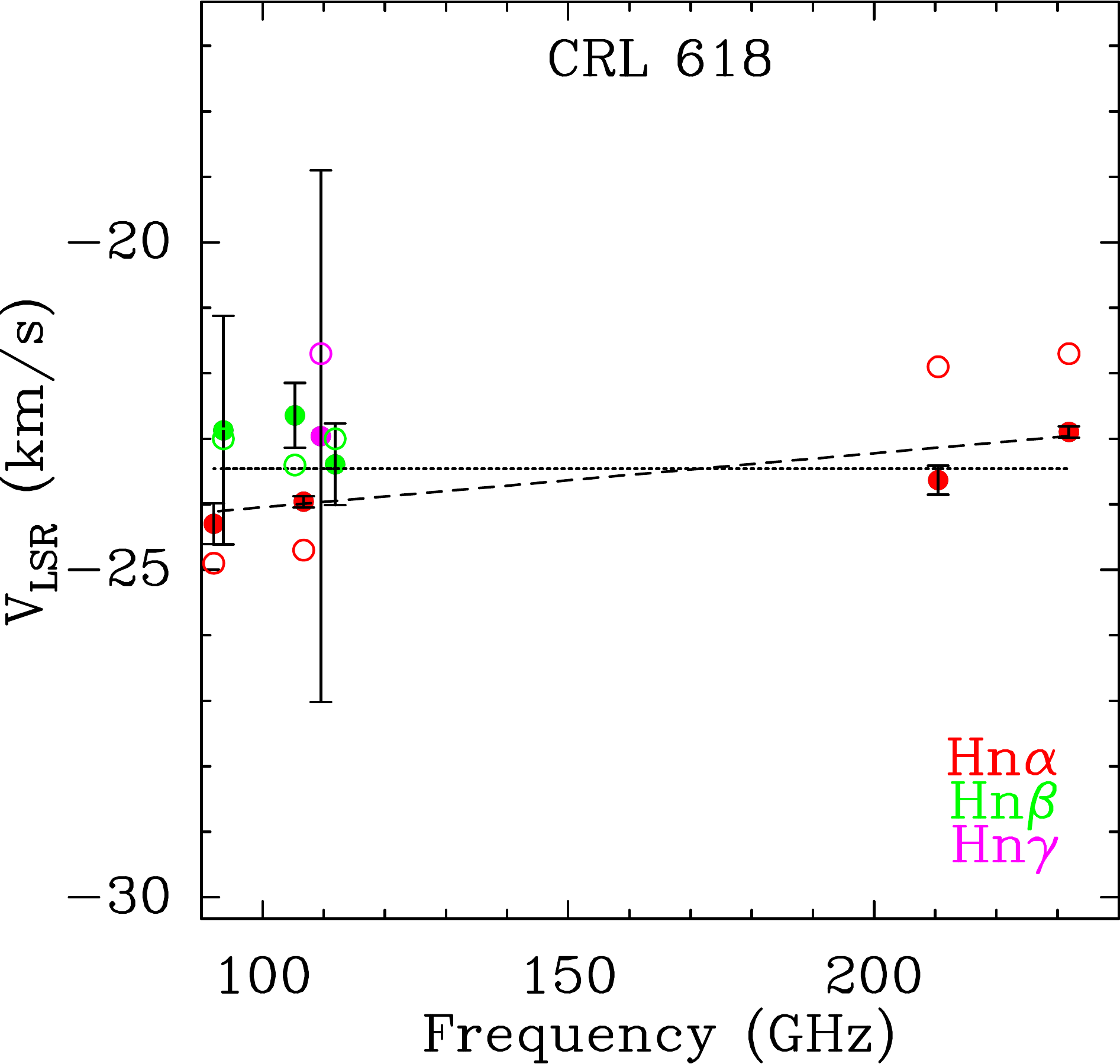} 

\vspace{1cm}
\includegraphics[width=0.32\textwidth]{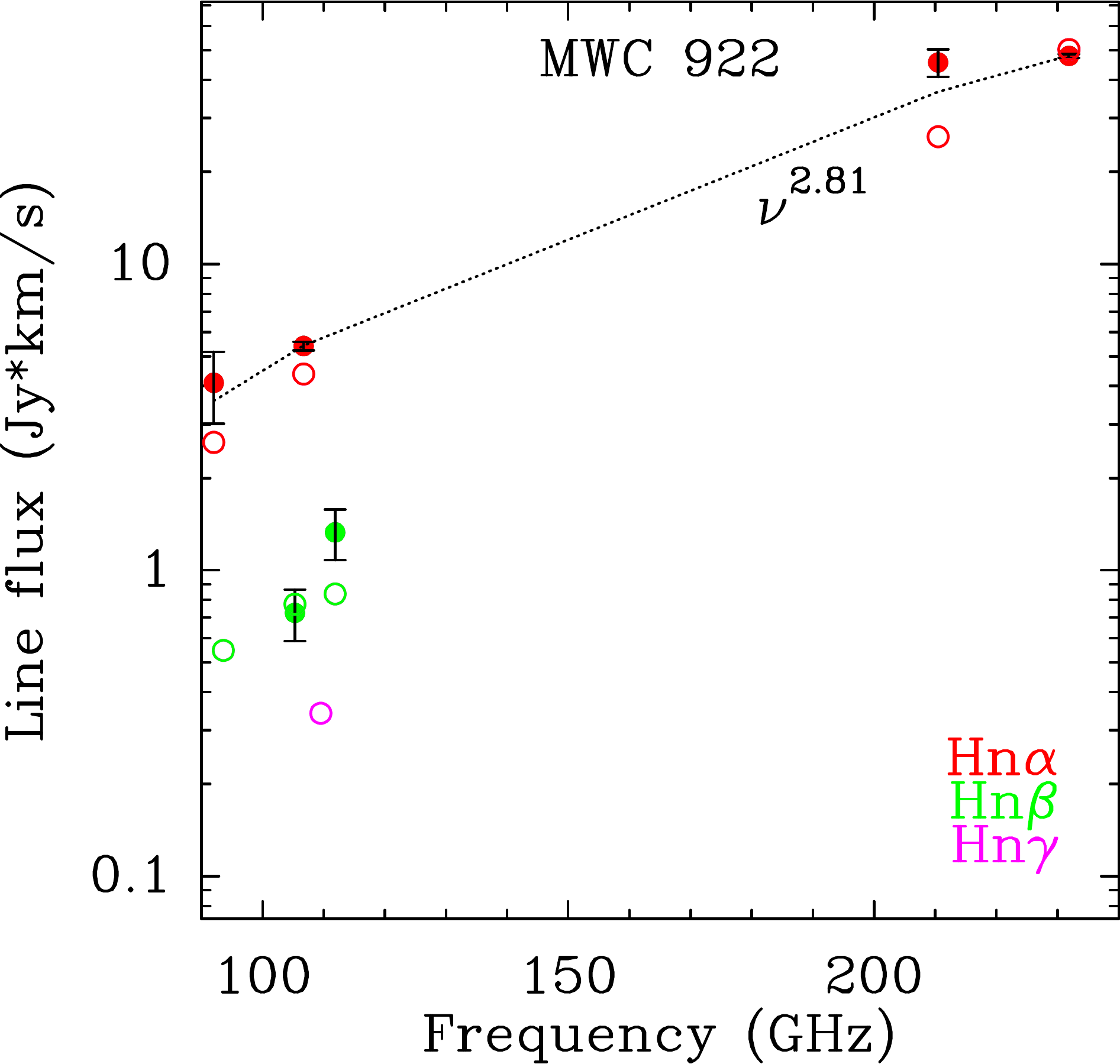}
\includegraphics[width=0.32\textwidth]{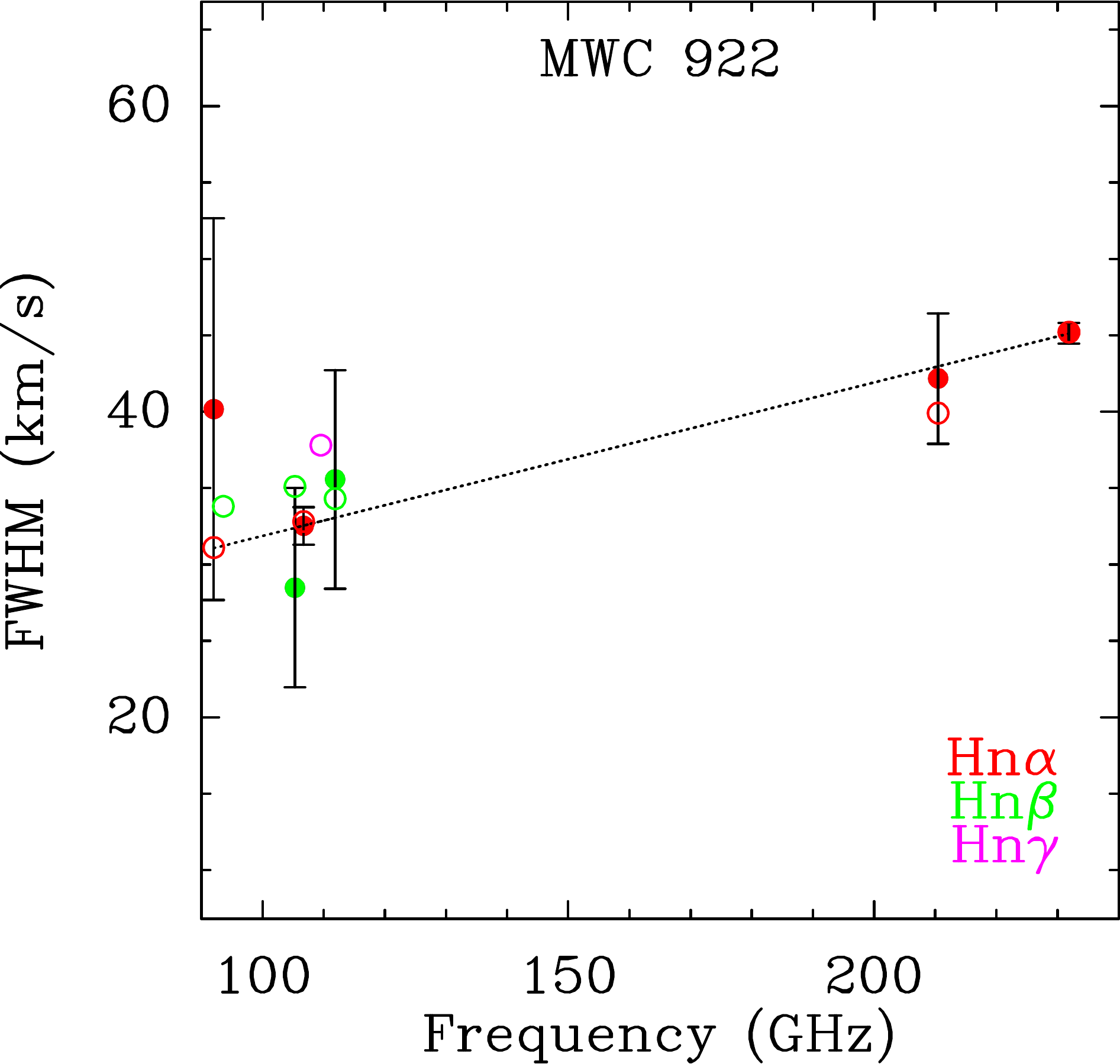}
\includegraphics[width=0.32\textwidth]{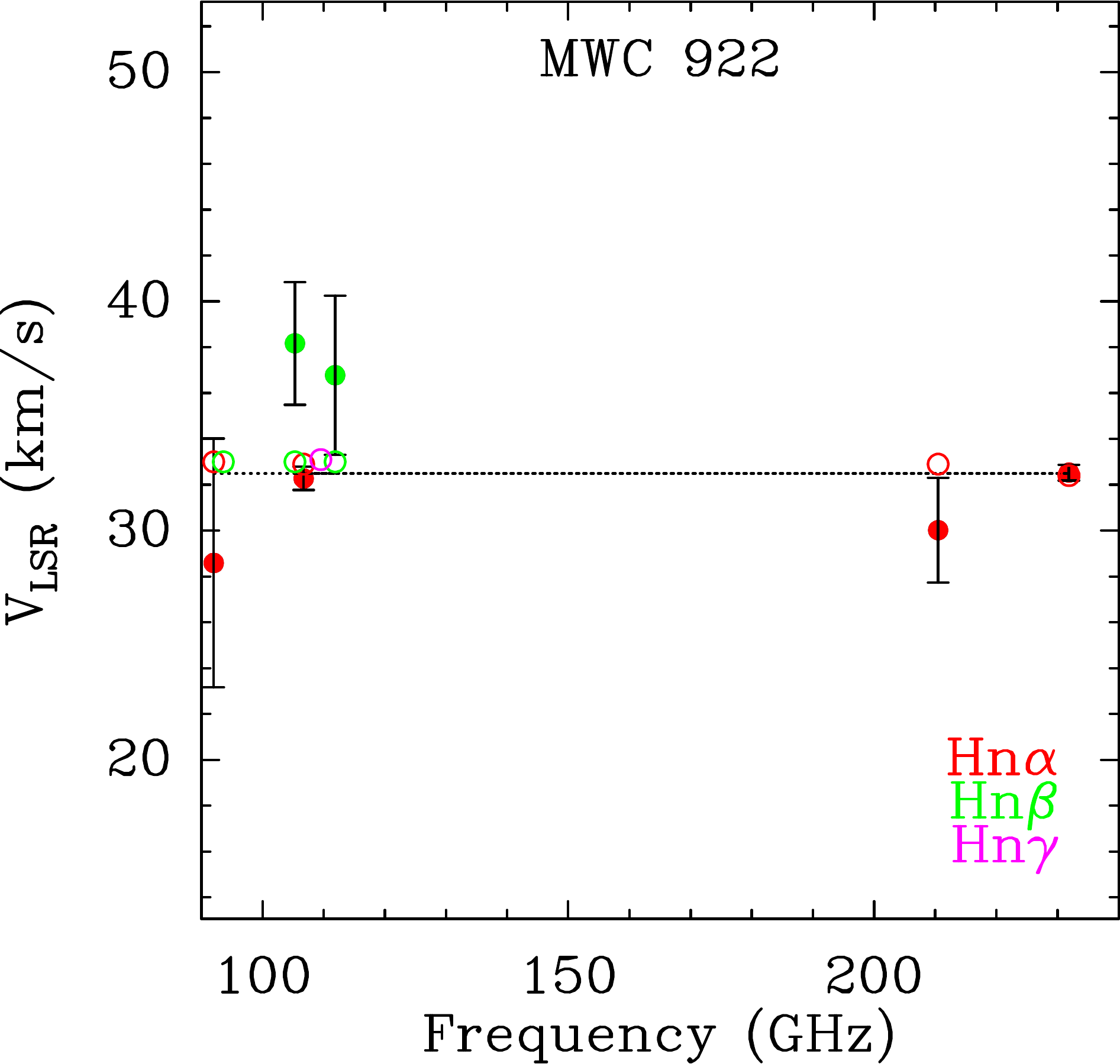} 

\vspace{1cm}
\includegraphics[width=0.32\textwidth]{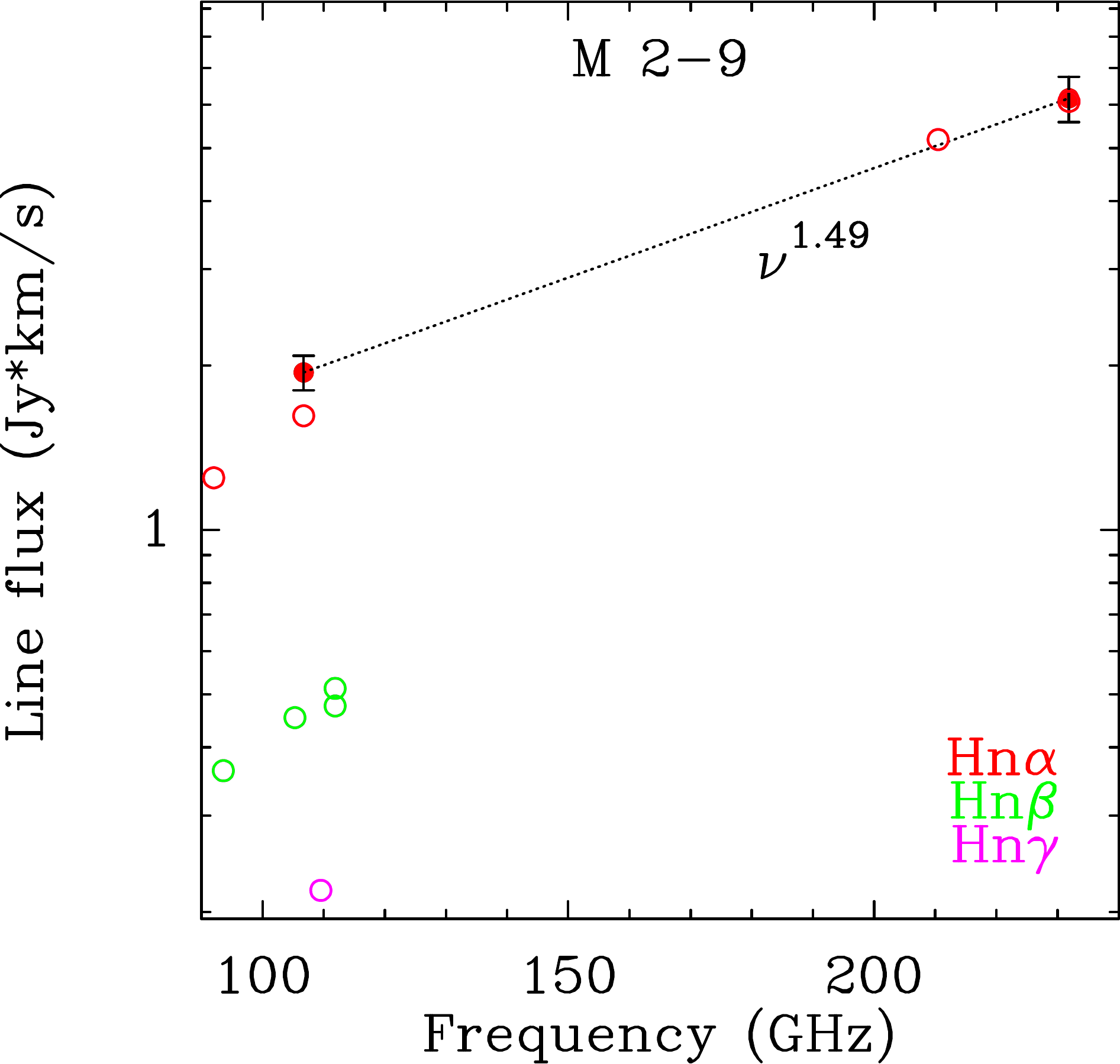}
\includegraphics[width=0.32\textwidth]{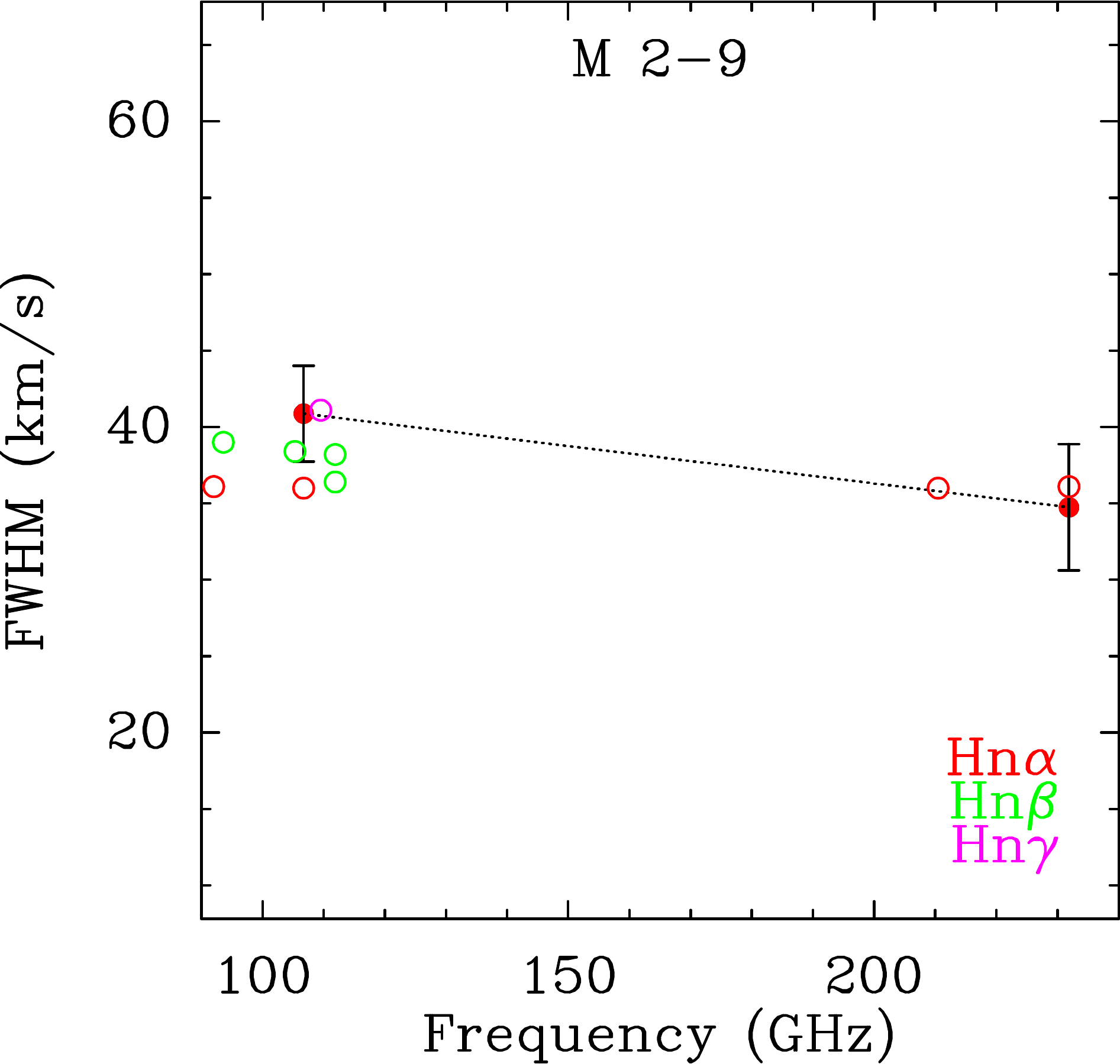}
\includegraphics[width=0.32\textwidth]{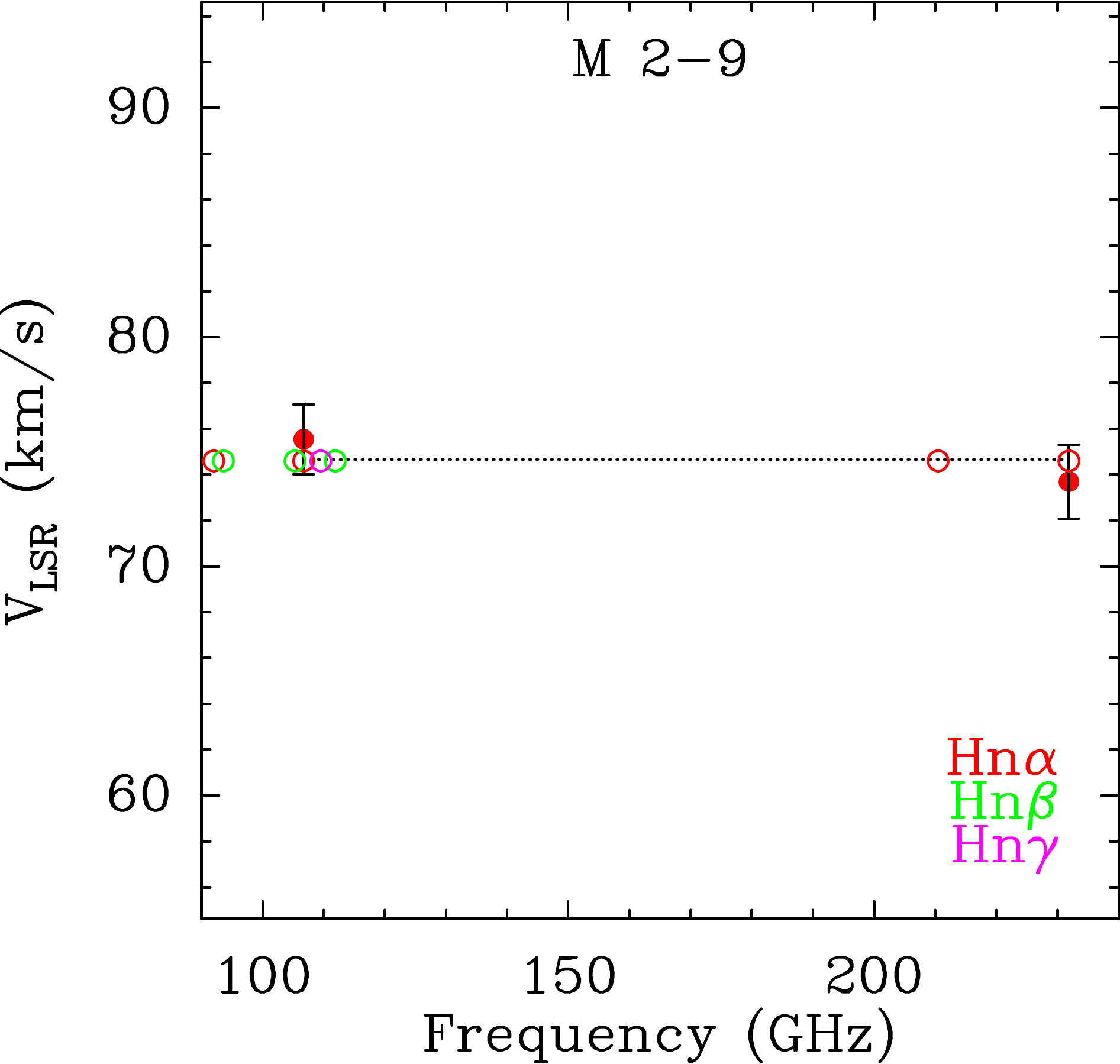}
   \caption{Line parameters of the RRLs detected in CRL\,618,
     MWC\,922, and M\,2-9 derived from the observed profiles    
       (Table\,\ref{t-parms}; filled circles) 
       and from the models in Table\,\ref{t-moreli}
       including non-detections (empty circles). Dotted lines
     are fits to the observed line parameters of the
     Hn$\alpha$ transitions (red filled circles). For CRL\,618,
     we show the mean \vlsr\ of the Hn$\alpha$ centroids (dotted line)
     and a linear fit of \vlsr\ as a function of frequency
     (dashed line). }
   \label{f-parms}
   \end{figure*}

We have detected radio recombination line emission at mm-wavelengths
for three (out of a total of eight) sources: CRL\,618, M\,2-9, and
MWC\,922 (Figs.\,\ref{f-crl618}, \ref{f-m2-9}, and \ref{f-mwc922}).
The main line parameters directly derived from the observed profiles 
are given in Table\,\ref{t-parms}. 
In Fig.\,\ref{f-parms}, we plot the flux, full width at half maximum
(FWHM), and centroids of the hydrogen RRLs detected as a function of
their rest frequencies. The rms noise of the spectra for RRL
non-detections are given in Table\,\ref{t-rms}. 
  These undetected sources are
  rather weak mm-continuum emitters, which anticipate
  low mm-RRL intensities below
  our detections limits (see \S\,\ref{model_rest}). 

%
\begin{table*}[htbp]
\caption{Line parameters from Gaussian profile fitting for RRL detections} 
\label{t-parms}      
\centering                          
\begin{tabular}{l r c c c c c }        
\hline\hline                  
Line\tablefootmark{a} & Frequency &   \inttl   &   \vlsr  & FWHM   & \tl\ & rms\tablefootmark{b} \\    
     &  (GHz)    &  (K\,\kms) &  (\kms)  & (\kms) & (mK) & (mK)  \\
\hline                        
\multicolumn{2}{c}{CRL\,618} &&&&& \\
H41$\alpha$ &  92.03443 &  9.02$\pm$0.12 & $-$24.3$\pm$0.3 &  47.6$\pm$0.8 & 178 &  4 \\ 
H39$\alpha$ & 106.73736 & 11.90$\pm$0.05 & $-$24.0$\pm$0.1 &  46.5$\pm$0.2 & 240 &  2 \\ 
H31$\alpha$ & 210.50177 & 16.4$\pm$0.2 & $-$23.6$\pm$0.2 &  38.3$\pm$0.6 & 403 & 12 \\ 
H30$\alpha$ & 231.90093 & 20.2$\pm$0.1 & $-$22.9$\pm$0.1 &  37.1$\pm$0.2 & 512 &  5 \\ 
H51$\beta$ &  93.60732 &  1.7$\pm$0.1 & $-$22.9$\pm$1.7 &  60$\pm$3 &  27 &  4 \\ 
H49$\beta$ & 105.30186 &  2.17$\pm$0.04 & $-$22.6$\pm$0.5 &  59.4$\pm$1.2 &  34 &  2 \\ 
H48$\beta$ & 111.88507 &  2.26$\pm$0.06 & $-$23.4$\pm$0.6 &  53.2$\pm$1.5 &  40 &  4 \\ 
H55$\gamma$ & 109.53600 &  0.31$\pm$0.12 & $-$23$\pm$4 &  23$\pm$14 &  12 &  6 \\ 
He41$\alpha$ &  92.07194 &  0.76$\pm$0.07 & $-$25.0$\pm$1.8 &  42$\pm$5 &  17 &  4  \\ 
He39$\alpha$\tablefootmark{*} & 106.78085 &  1.10$\pm$0.04 & $-$24.0$\pm$0.0 &  46.5$\pm$0.0 &  22 &  2  \\ 
He30$\alpha$ & 231.99543 &  2.08$\pm$0.13 & $-$23.5$\pm$1.2 &  42$\pm$4 &  46 &  5  \\ 
\multicolumn{2}{c}{M\,2-9} &&&&& \\
H41$\alpha$ &  92.03443 &  \dotfill &   \dotfill &   \dotfill &  \dotfill &  7 \\ 
H39$\alpha$ & 106.73736 &  0.32$\pm$0.02 &  75.5$\pm$1.5 &  40.9$\pm$3.1 &   7 &  1 \\ 
H31$\alpha$ & 210.50177 &  \dotfill &   \dotfill &   \dotfill &  \dotfill & 27 \\ 
H30$\alpha$ & 231.90093 &  0.79$\pm$0.07 &  73.7$\pm$1.6 &  34.7$\pm$4.1 &  21 &  5 \\ 
H51$\beta$ &  93.60732 &  \dotfill &   \dotfill &   \dotfill &  \dotfill &  6 \\ 
H49$\beta$ & 105.30186 &  \dotfill &   \dotfill &   \dotfill &  \dotfill &  2 \\ 
H48$\beta$ & 111.88507 &  \dotfill &   \dotfill &   \dotfill &  \dotfill &  3 \\ 
H55$\gamma$ & 109.53600 &  \dotfill &   \dotfill &   \dotfill &  \dotfill &  2 \\ 
\multicolumn{2}{c}{MWC\,922} &&&&& \\
H41$\alpha$ &  92.03443 &  0.69$\pm$0.18 &  29$\pm$5 &  40$\pm$13 &  16 &  8 \\ 
H39$\alpha$ & 106.73736 &  0.89$\pm$0.03 &  32.3$\pm$0.5 &  32.5$\pm$1.2 &  26 &  2 \\ 
H31$\alpha$-blue\tablefootmark{**} & 210.50177 &  3.0$\pm$0.5 &  19.2$\pm$1.2 &  15$\pm$3 & 190 & 40 \\ 
H31$\alpha$-red\tablefootmark{**} & 210.50177 &  2.8$\pm$0.5 &  45.1$\pm$1.6 &  18$\pm$4 & 150 & 40 \\ 
H30$\alpha$-blue\tablefootmark{**} & 231.90093 &  2.85$\pm$0.09 &  19.1$\pm$0.3 &  17.9$\pm$0.7 & 150 &  6 \\ 
H30$\alpha$-red\tablefootmark{**} & 231.90093 &  3.0$\pm$0.1 &  46.2$\pm$0.3 &  20.7$\pm$0.8 & 134 &  6 \\ 
H51$\beta$ &  93.60732 &  \dotfill &   \dotfill &   \dotfill &  \dotfill &  8 \\ 
H49$\beta$ & 105.30186 &  0.12$\pm$0.02 &  38$\pm$3 &  29$\pm$7 &   4 &  2 \\ 
H48$\beta$ & 111.88507 &  0.22$\pm$0.04 &  37$\pm$3 &  36$\pm$7 &   6 &  3 \\ 
H55$\gamma$ & 109.53600 &  \dotfill &   \dotfill &   \dotfill &  \dotfill &  3 \\ 
\hline                                   
\end{tabular}
\tablefoot{Continuum-subtracted line fluxes, \tl, are in units of antenna temperature. \\
\tablefoottext{a}{Standard line notation, i.e.,\, $n+\Delta n \rightarrow n$ level transitions with $\Delta n$=1, 2, or 3 for 
$\alpha$, $\beta$, and $\gamma$ lines, respectively.} \\
\tablefoottext{b}{Formal rms noise for a $\Delta \upsilonup$\,$\sim$3\,\kms\ resolution, as in Figs.\,\ref{f-crl618}-\ref{f-mwc922}. Baseline subtraction or absolute calibration uncertainties are not included.  \\ 
\tablefoottext{*}{Uncertain due to line blending; line center and width fixed to match the corresponding H39$\alpha$ values.} \\ 
\tablefoottext{**}{The blue and read intensity peaks of the two-horned profile of the H30$\alpha$ and H31$\alpha$ lines in MWC\,922 were fitted using two Gaussians.}}
}
\end{table*}
%
%
\begin{table}[t]
\caption{Rms noise (in \ta\ units) for RRL non-detections} 
\label{t-rms}      
\centering                          
\small
\begin{tabular}{l c c c c c c c}        
\hline\hline                  
Line & Freq$.$ &  \multicolumn{5}{c}{rms (mK); $\Delta \upsilonup$\,$\sim$5\,\kms } \\
     &  (GHz)    &  He\,3-1475 & M\,1-91 & M\,1-92 & 20462 & M\,2-56 \\  
\hline                        
H39$\alpha$ & 106.7     & 2.4   & 1.3   & 1.1   & 1.3   & 1.0 \\
H30$\alpha$ & 231.9     & 8.6   & 2.7   & 2.6   & 3.3   & 1.8 \\
H49$\beta$ & 105.3      & 2.8   & 1.3   & 1.2   & 1.2   & 0.9 \\
H48$\beta$ & 111.9      & 3.6   & 1.6   & 1.7   & 1.8   & 1.2 \\
H55$\gamma$ & 109.5     & 3.5   & 1.6   & 1.7   & 1.8   & 1.3 \\
\hline                                   
\end{tabular}
\end{table}
%

The strongest RRLs are observed toward CRL\,618, which also shows the
strongest mm-wavelength continuum. We detect a total of eight hydrogen
recombination lines ($\alpha$, $\beta$, and $\gamma$)
together with helium $\alpha$-transitions (Fig.\,\ref{f-crl618}).  The 
RRLs appear to be partially blended with some of the many molecular
transitions with composite emission and absorption profiles that
crowd the submm/mm spectrum of this object \cite[see, e.g.,][]{par07a}. 
The profile of the weak H55$\gamma$ line is the most
adversely affected by deep adjacent absorptions.

We find significant changes in the spectral profiles and total fluxes
of the H41$\alpha$ and H30$\alpha$ lines of CRL\,618 relative to the
observations performed in 1987 by \cite{mar88}. In particular, these
RRLs are now a factor $\sim$2-3 more intense and $\sim$20-60\%\,
broader than 29 years ago. This is not totally unexpected given the
obvious variability of the free-free continuum emission observed over
many years (\S\,\ref{intro-crl618} and Fig.\ref{f-crl6181mm}), and
suggests changes with time of the physical conditions at the inner
regions around the central source. In M\,2-9, weak emission from the
H30$\alpha$ and H39$\alpha$ lines and, tentatively, from the
H31$\alpha$ and H41$\alpha$ transitions is detected
(Fig.\,\ref{f-m2-9}).

In MWC\,922, we detect the four Hn$\alpha$ transitions observed in
this work plus the weaker H48$\beta$ and H49$\beta$ lines
(Fig.\,\ref{f-mwc922}). In contrast to CRL\,618 and M\,2-9, in this
source there is an evident drastic change from single-peaked to
double-peaked profiles in the Hn$\alpha$ lines at 3 and 1\,mm. A
similar transition from single-peaked profiles of RRLs near 3\,mm (and
at cm-wavelengths) to double-peaked profiles of the RRLs near 1\,mm
was observed for the first time toward the ultra-compact \ion{H}{ii}
region of MWC\,349A by \cite{mar89a,mar89b}. These authors showed
that, in this case, the two-horned line profiles result from maser
amplification of the emission from a rotating ionized disk. As we see in \S\,\ref{model_mwc922}, we reach a similar conclusion for
MWC\,922 that, thus, adds to the (yet short) list of sources where
mm-RRL masers have been discovered to date; to our knowledge, these include
  MWC\,349A, $\eta$ Carinae, Cepheus A HW2, and MonR2-IRS2 \citep[
see][]{mar89a,cox95,jim11,jim13,abr14}.

There are a few general observational results that can be readily seen in 
Fig.\,\ref{f-parms} and are summarized here: 

[-] The line strength of the RRLs with the same $\Delta n$ increases
with the frequency with MWC\,922 showing the steepest slope. In
particular, for the $\alpha$-transitions, the slope in this target
clearly deviates from the $\nu^{1.1}$ dependency expected in case of
optically thin LTE emission \citep[e.g.,][]{rod09}. This is in
agreement with the presence of maser amplification of the
high-frequency lines.  Non-LTE emission conditions are also inferred
in the case of CRL\,618 from the $\alpha$-to-$\beta$ line ratios,
which depart from the theoretical predictions under LTE
\cite[e.g.,][]{pet12}.

[-] The line widths of the $\alpha$-transitions range between
FWHM\,$\sim$30 and 50\,\kms\ and, thus, they are larger than expected
from thermal motions alone (e.g.,\ the sound speed is
$c_s$$\sim$13\,\kms\ for ionized gas with an average electron
temperature of \te$\approx$10,000\,K, typical of \ion{H}{ii} regions).
As discussed in \S\,\ref{moreli}, this indicates macroscopic ordered
motions of the bulk of the gas at moderate speeds of
$\approx$10-30\,\kms. We cannot rule out the presence of broad
  ($>\pm$30\,\kms) emission wings below our detection limits, i.e.,\,at
  $<$20\%\ of the line peak.

[-] The Hn$\alpha$ line width increases with increasing quantum number
$n$ in CRL\,618 and (tentatively) M\,2-9, but the opposite behavior is
found for MWC\,922.  The high-$n$ transitions trace more tenuous
regions than lower-$n$ lines (at higher frequencies and, thus, with
continuum emission optically thinner).
Therefore, assuming that the density in the ionized region falls off
with the distance to the center ($r$), the trend observed in CRL\,618
and M\,2-9 suggests that the gas velocity increases with $r$,
which is consistent with an expanding wind. In the case of MWC\,922, the
spectra seem more consistent with a rotating disk/wind, 
where one expects the rotation speed to decrease with $r$. In
CRL\,618, the widths of the $\beta$-lines are systematically higher
than the $\alpha$-lines at comparable frequencies. This is partially
due to line blending and poor baseline subtraction related
uncertainties, which happen to be on average more severe in the
$\beta$-transitions, but also due to larger pressure broadening
for the $\beta$-lines compared with the $\alpha$-lines predicted by
theory \cite[e.g.,][and \S\,\ref{moreli}]{bro72,wal90,str96,pet12}.

[-] The centroids of RRLs provide us with an estimate of the
line-of-sight velocity of the bulk of the ionized gas (and, plausibly,
of the ionizing star that lies at the center). Using the Hn$\alpha$
profiles, with the highest S/N, we find \vlsr=$-$23.5$\pm$0.2\,\kms\ for
CRL\,618, \vlsr=$+$75$\pm$2\,\kms\ for M\,2-9, and
\vlsr=$+$32.5$\pm$0.4\,\kms\ for MWC\,922.  In the case of CRL\,618,
our data are consistent with a progressive velocity shift toward the
blue at low frequencies (Table\,\ref{t-parms} and Fig.\,\ref{f-parms}).
Although tentative, the small velocity shift could be caused by the
gradual decrease of the optical depth of the lines with the frequency
\citep[e.g.,\,][]{mar88}.
In an outflowing gas, the optically thicker, lower frequency RRLs mainly trace a major part of the gas from the front (blue-shifted) of
the nebula, since part of the emission from the back side
(red-shifted) is absorbed through the line+continuum emitting
region. If this is the case, the $\sim$1\,mm lines are expected
  to be centered closer to the line-of-sight velocity of the ionizing
  star of CRL\,618 (see \S\,\ref{model_c618}). 

\section{Radiative transfer model}
\label{moreli}

In order to constrain the structure, physical conditions, and
kinematics of the emerging ionized regions of CRL\,618, MWC\,922, and
M\,2-9, we  modeled their free-free continuum and mm-RRL emission 
using the non-LTE radiative transfer code
MORELI (MOdel for REcombination LInes), which is described in detail
in \cite{bae13}. 

The MORELI code considers a given geometry (with spherical or cylindrical
symmetry) for the ionized region.  The electron density is assumed to
decrease with the distance to the central star following a power law
of the type $n_\mathrm{e}(r)\propto r^{-\alpha_n}$. Also, MORELI takes
the electron temperature, \te($r$), and a given velocity
field as inputs.  For simplicity, and since the exact gas kinematics cannot be
precisely determined from single-dish observations, in our models the
velocity field is described by radial expansion (at constant velocity
or increasing, linearly or asymptotically, with $r$) and/or Keplerian
rotation (see Sections \ref{model_c618} to \ref{model_mwc922} for more
details).  In addition to the gas macroscopic motions, thermal
broadening and electron impact (pressure) broadening from
high-density gas are included in the line profile function.
Turbulence velocity dispersion is not considered as an additional form
of line broadening since it is expected to be very small ($\sigma_{\rm
  turb}$$<$2-3\,\kms) in the envelopes of AGB and post-AGB objects
\citep[e.g.,][]{sch04,buj05,dec10} and, thus, negligible compared to
other broadening terms.

The MORELI code predicts the free-free RRLs and continuum emission by solving
the equation of radiative transfer and integrating the intensity along
the line of sight. The ionized region starts at minimum radius, \rin,
which in principle could be the stellar photospheric radius; this region
extends up to a distance that is a few times larger than the radius
that encircles a major percentage ($\ga$90\%) of the observed fluxes,
which is referred to as the effective outer radius \rout\footnote{The terms
  \rout(90\%) and \rout(99\%), which are also explicitly given in
  some cases (e.g.,\,Table\,\ref{t-ND}), refer to the radius that
  encircles $\sim$90\%\ and 99\%\ of the mm-continuum flux,
  respectively.}.  Thomson electron scattering is commonly
  neglected, to a first approximation, to solve the radiative transfer
  problem for recombination lines at millimeter wavelengths
  \citep[e.g.,][]{pet12}.  The effect of this process is, in general,
  weak and high electron concentrations of
  \dense$\approx$10$^8$-10$^{12}$\,\cm3; in electron-scattering, regions of sizes of a few$\times$10\,au
  are needed to cause a measurable effect \citep{arr03,sek12}.

The hydrogen non-LTE level populations are computed in MORELI using
departure coefficients, $b_{n}$, which are defined as the ratio
between the actual level population for the principal quantum number,
$n$, and its value in LTE
($b_{n}$$\equiv$$N_{n}$/$N_{n}^{*}$). Computations of the $b_{n}$
coefficients are available in the literature for $n$ as low as 20 and
for a broad range of electron densities and temperatures
\cite[typically, $n_e$$\sim$10-10$^{8.5}$\cm3 and
  \te$\sim$5000-15000\,K;][and references
  therein]{sal79,wal90,sto95}. The $b_n$ factors computed by different
authors for the same values of \te\ and $n_e$ do not fully agree.  The
uncertainties in the $b_n$ coefficients have a major impact on the
accuracy to which we can reproduce the intensity of RRLs where
stimulated emission appears to be present. It is particularly critical
in the case of the H30$\alpha$ and H31$\alpha$ maser lines in MWC\,922
(\S\,\ref{resRRLs}).  The $b_n$ coefficients from \cite{sto95} were
found to provide the best fit of the maser transitions in MWC\,349A
\citep{bae13} and we systematically used these for our targets.

The relative intensities of the RRLs and the
line-to-continuum flux ratios observed in M\,2-9 are consistent with
LTE conditions (or, at most, with small non-LTE effects) and,
therefore, the line profiles were modeled with MORELI in LTE
mode in this case. In MWC\,922 and CRL\,618, however, there was a need for more general
non-LTE calculations and we performed these calculations to reproduce
simultaneously the $\alpha$- and $\beta$-profiles observed at different frequencies. 

In all cases, we explored a broad range of physical conditions
until we obtained a reasonable match between the model predictions and mm-wavelength line and continuum observations. Our model
also provides reasonable predictions for radio continuum data at longer
wavelengths (whenever available), such as the total cm-to-mm free-free
continuum flux and the angular size frequency dependence. In general,
we tried to keep the number of model input parameters as low as
possible given the data available.

As shown by, for example, \cite{lee07} and \cite{bae13}, the mass-loss
rate is almost exclusively constrained by the flux of the free-free
continuum, which has a moderate observational uncertainty of
$\la$30\%\ (\S\,\ref{obs}). The other two main sources of uncertainty
in the mass-loss rate are the expansion velocity and distance to
the source.  Except for MWC\,922 (see \S\,\ref{model_mwc922}), the
expansion velocity of the bulk of the ionized wind is rather well
constrained, with a $\la$20\%\ accuracy, from the widths of the
RRLs. For the distances adopted, we find that the range of possible
mass-loss rates that can reproduce the observed continuum, and are
also consistent with the line profiles, are always within a factor
$\la$2 of uncertainty. The spatio-kinematics and electron
temperature stratification across the ionized mm-wavelength emitting
layers critically determine the shape and intensity of the RRLs, but
have a rather limited effect on the continuum flux.
The expansion velocities and temperatures are then mainly constrained by the fit of
the single-dish profiles and should be taken as representative average
values within the ionized regions under study.

The modeling results and their main uncertainties are described for
each of the modeled sources in the next subsections.

%
\begin{table*}[htp]
\small
\caption{Parameters of the emerging \ion{H}{ii} regions deduced from line and continuum radiative transfer modeling (\S\,\ref{moreli}).}
\label{t-moreli}      
\centering                          
\begin{tabular}{l c c }        %
\hline\hline                  
Parameter & CRL\,618 & M2-9 \\ 
\hline                        
Distance ($d$) & 900\,pc & 650 pc \\
LSR Systemic velocity (\vsys) & $-19$\,\kms & $+$75\,\kms \\ 
Geometry & {\sl cylindrical}\tablefootmark{\dag} &  {\sl cylindrical}\tablefootmark{\dag}  \\
Inner radius (\rin)  &45\,au & 0.05 au ($\la$2 au)  \\
Outer radius (\rout) &95\,au & 160\,au   \\    
Electron density ($n_{\rm e}$) & 2.4\ex{7}($\frac{r}{45{\rm au}}$)$^{-2.0}$\,\cm3\ &  6.4\ex{5}($\frac{r}{45{\rm au}}$)$^{-2.1}$\,\cm3\ \\
Electron temperature (\te)    & 16500\,K & 7500\,K \\
Kinematics & {\sl radial expansion} & {\sl radial expansion}  \\
Velocity ($V$)    & $10+25(1-\frac{45\,{\rm au}}{r}$)$^{0.75}$\,\kms\ & 22\,\kms\  \\
Kinematic Age ($t_{\rm k}$) & $\approx$\,15-20\,yr & \la\,15\,yr  \\ 
Inclination ($i$) & 24\degr &  17\degr  \\ 
Mass-loss rate (\mloss) & 8.4\ex{-6}($\frac{\vexp}{20}$)\,\my\ & 3.9\ex{-7}($\frac{\vexp}{20}$)\,\my\  \\ 
Ionized mass ($M_{\rm ion}$) & 2.0\ex{-4}\,\msun\ & 5.5\ex{-6}\msun\ \\ 
\hline                                   
\end{tabular} \\ 
\begin{tabular}{l c}        %
  \hline\hline                      
Parameter & MWC\,922 \\ 
\hline
Distance ($d$) & 1700 pc \\ 
LSR Systemic velocity (\vsys) & $+$33\,\kms \\ 
Geometry & {\sl double-cone}\tablefootmark{\ddag} \\
Double-cone's semi-opening angle ($\theta_{\rm a}$) & 25\degr \\
Inclination ($i$) & 0\degr \\ 
Inner radius (\rin)  & 0.05\,au ($\la$35\,au)  \\
Electron density ($n_{\rm e}$) & 9\ex{7}($\frac{r}{45{\rm au}}$)$^{-2.4}$exp($-(\theta_a-\theta)/2\degr$)\,\cm3\ \\
Electron temperature (\te) & 6700\,K \\
Opening angle of the ionized disk ($\theta_{\rm d}$) & 6.5\degr \\
Ionized disk's outer radius ($r_{\rm d}$) & 50 au \\
Ionized outflow's outer radius (\rout) & $\sim$150 au \\
Ionized outflow's velocity (\vexp)  & 5 \kms\ ({\sl radial expansion}) \\ 
Keplerian rotation velocity (\vrot) & 32.6/$\sqrt{(r/6.7 {\rm au})}$\,\kms\ \\ 
Mass-loss rate (\mloss) &  2\ex{-6}($\frac{\vexp}{5}$)\,\my\  \\ 
Ionized mass ($M_{\rm ion}$) & 3.6\ex{-5}\msun\ \\   
\hline 
\end{tabular}
\tablefoot{In all cases, $r$ is the radial distance to the center; the
  mass-loss rate (\mloss) is the equivalent isotropic mass-loss rate
  for a constant expansion velocity and \dense$\propto r^{-2}$; and
  the departure coefficients $b_{\rm n}$ used are from
  \cite{sto95}. \\ \tablefoottext{\dag}{For CRL\,618 and M\,2-9,
      we assume a cylindrical \ion{H}{ii} region with an outer radius
      \rout\ at the base (i.e.,\,along the nebula equator) and a
      central spherical cavity with inner radius \rin. The cylinder
      extends to infinity along the major axis, where $\sim$90\%\ of
      the mm-continuum is produced within $\sim$110\,au and
      $\sim$50\,au along the axis, respectively.}
  \\ \tablefoottext{\ddag}{Double-cone geometry as described in
    \S\,\ref{model_mwc922} and depicted in Fig.\,3 of \cite{bae13}.}
}
\end{table*}
%


\subsection{CRL\,618}
\label{model_c618}

Based on previous works (\S\,\ref{intro-crl618}), as a simple model of
the ionized core of CRL\,618, we adopted an isothermal stellar
wind represented by a cylinder of ionized gas with a central cavity
around the star, constant electron temperature, and an inverse-square
density profile.  This is similar to the model used by \cite{mar88} to
fit their H41$\alpha$, H35$\alpha$, and H30$\alpha$ spectra and
continuum emission observed in 1987. The presence of a central cavity,
with radius \rin, was originally proposed by these authors to explain
with this simple model the nearly flat (optically thin)
  free-free continuum at mm-wavelengths and, at the same time, the
strong continuum emission and large emission measure deduced
(EM$\approx$few\ex{10}-10$^{11}$\cm6 pc).

The outer radius of the cylinder (along PA$\sim$0\degr) was set
to \rout$\sim$95\,au, which is the size of the minor semi-axis of the
\ion{H}{ii} region expected in 2015 adopting the growth function
$\theta_{\rm min}$=\,0.03$\times$($t-1971)^{0.5}$ deduced by
\cite{taf13}.  As shown by these authors, the measured minor axis of
the cm-continuum, which remains basically constant at all frequencies,
pinpoints the locus of the ionization front along these dense
equatorial regions.
In the direction of the lobes (PA$\sim$90\degr), the ionization front was already at
distances \ga\,600\,au in 1998 and is probably near 800\,au at present.  
In our model, the cylinder extends out to infinity along its major axis.
It is not necessary to truncate the cylinder at 600-800\,au because
the ionized wind layers farther out would not contribute appreciably
($<$0.001\%) to the mm-RRLs/continuum emission.

The input parameters of the best-fit model for CRL\,618 are given in
Table\,\ref{t-moreli}.  The synthetic spectral Hn$\alpha$ profiles are
overimposed to the observed spectra in Fig.\,\ref{f-crl618}. The line
peak intensities and their FWHM are well reproduced within a
\la\,15\%. Our model recreates the progressive blue-shift of the
  line centroids as the frequency decreases, as observed
  (see \S\,\ref{resRRLs}). This is due to the moderate line optical depth
  effects in this source, which produce an overall velocity shift
  toward the blue relative to the line-of-sight velocity of the
  ionizing star, adopted to be \vsys(LSR)=$-$19\,\kms\ in our model.
The free-free continuum emission in the mm-to-cm regime predicted by
our model is also in very good agreement with the observational data
(Fig.\,\ref{f-sedmodel}).

We find a characteristic electron temperature of
\te$\sim$16,000-17,000\,K, which is somewhat larger than the value
obtained assuming LTE conditions by \cite{mar88}. The bulk of the
mm-RRLs/continuum emission is produced inside the adopted cylindrical
structure within a distance along its major axis (or semi-length) of
\rout(90\%)$\sim$110\,au. The electron densities range between
\dense\,$\sim$2\ex{7} and 4\ex{6}\cm3 in the mm-RRLs/continuum
emitting region. This is in very good agreement with the values
deduced by \cite{taf13} extrapolating their density law, which applies
to the $\sim$150-650\,au layers, to the inner wind regions. We find
that an inverse-square density profile yields satisfactory model
predictions, however, both the RRL profiles and the mm-to-cm SED can
also be reproduced with slightly different values of the power-law
index, in particular, with the steeper gradient, \dense $\propto
r^{-2.4}$ , proposed by \cite{taf13}.  Finally, since we are
able to explain our single-dish data 
suitably with a latitude-independent density law, we have not attempted any
models with equatorially enhanced density profiles. We do not
conclude that there is no equatorial overdensity present in the
ionized wind, but  increasing the model
parameter space is undesirable with the data available since it would not provide a more
accurate characterization of the density structure of the
  inner wind regions under study.

We chose to model the wind kinematics using an asymptotic
expansion velocity law similar to what is commonly used to model the
dust/wind acceleration region of mass-losing stars
(Table\,\ref{t-moreli}).  This law reproduces the broader profiles of
the $\sim$3\,mm Hn$\alpha$ lines compared to the $\sim$1\,mm
transitions.  We emphasize that, in the absence of spatially resolved
maps of the emission of mm-RRLs, it is not possible to accurately
determine the exact velocity field in the ionized core. From the
observed profiles, and taking into account that the 1\,mm and 3\,mm
lines arise in neighboring layers, probably only $\approx$10\,au
apart from each other, we deduce that there must be a steep rise of
the expansion velocity at these small spatial scales.  In particular,
the data are consistent with a minimum expansion velocity of
\vexp$\sim$7-12\,\kms\ at the innermost regions of the wind (probably
near 40-60\,au) and \vexp$\sim$20-25\,\kms\ near the outer boundary of
the modeled region ($\sim$70-90\,au).  The observed profiles can
  also be reproduced with different velocity fields (e.g.,\, with
  \vexp\,$\propto r$ or \vexp\,$\propto r^2$) as long as the full
  range of expansion velocities observed ($\approx$10-20\kms) is
  preserved. The expansive kinematics inferred imply a kinematical
age for the inner wind layers of only $t_{\rm k}$$\sim$15-20\,yr. Our
model also reproduces very well the broader profiles of the
$\beta$-transitions, which confirms that, as expected from theory,
these lines undergo pressure broadening.

The radius of the central cavity required in the model adopted is
rather uncertain and, if it exists, it can only be well determined with
spatially resolved observations at mm-wavelengths. Although the model
presented in Table\,\ref{t-moreli} provides the best fit to the data,
larger values of \rin\ are also acceptable (up to
\rin$\sim$70\,au). The larger the size of the cavity, the larger the
mass-loss rate that is required to keep the mm-continuum emission at
the observed flux level. For example, if \rin$\sim$70\,au, a mass-loss
rate of $\sim$1.1\ex{-5}\,\my\ is necessary. In the case of a bigger
cavity the velocity gradient across the 1\,mm- and 3\,mm-emitting
layers should also be larger.

Our final model also reproduces the frequency-dependence of the major
and minor axes sizes of the cm-continuum deduced by \cite{taf13} based
on their detailed analysis of VLA 1.4-23\,GHz continuum maps observed
in the period 1982-2007.  In principle, a good fit to the cm-continuum
maps observed in the past should not be a strict requirement for our
models because our mm-wavelength observations probe the
optically thin (\la\,110\,au) part of the ionized wind that was
ejected in the last $\sim$15-20 years. But the cm-wavelength
observations trace the outer regions, $\sim$150-630\,au, of an older
wind that was ejected at earlier times during the last century.  We
chose to roughly reproduce simultaneously ancillary cm-continuum
data to avoid extremely abrupt changes in the inner and outer wind
properties.

\subsection{M\,2-9}
\label{model_m29}

As explained in \S\,\ref{intro-m29}, the free-free continuum emission
of M\,2-9 at millimeter wavelengths is expected to be dominated by the
compact ionized wind at the core of the nebula, with a positive
spectral index, $\alpha$$\sim$0.7. The mm-RRLs observed are also most
likely produced in this ionized wind. The spectral distribution of the
continuum, with no apparent changes in the spectral slope over
$\sim$10-230\,GHz range, suggests that the free-free emission is
optically thick at mm-wavelengths. In fact, this emission probably becomes optically thin for frequencies $\ga$230\,GHz alone since no
transition to a flat continuum spectrum is observed at lower
frequencies. However, without angularly resolved maps
of the continuum at millimeter wavelengths, a separate, accurate determination
of the spectral index of the core and extended lobes is
not possible.

As in the case of CRL\,618, we modeled the compact ionized core of
M\,2-9 as an isothermal cylindrical wind that expands radially, which
is the simplest structure that is consistent with the elongation of the radio
core observed at cm-wavelengths. The input parameters of our best-fit
model are given in Table\,\ref{t-moreli} and the predicted RRL
profiles and free-free continuum emission from the compact ionized
wind are shown in Figures\,\ref{f-m2-9} and \ref{f-sedmodel}.

The spectral index of the compact ionized wind continuum is well
reproduced with an electron density gradient with a power index of
$\alpha_n$=2.1 and electron density of \dense$\sim$6\ex{5}\,\cm3 at
45\,au, resulting in a current mass-loss rate of
\mpagb$\approx$3.9\ex{-7}\,\my. We also derive an upper limit for the
inner radius of the \ion{H}{II} region of \rin$\la$2\,au, which is
needed to explain the turnover frequency at $\nu_{\rm t}$$>$230\,GHz,
as previously stated (Fig\,\ref{f-seds}). The bulk of the RRL and
  continuum emission at millimeter wavelengths arises within the
  $\sim$5-50\,au inner layers of the ionized wind.

We find that an LTE electron temperature of \te$\sim$7,500\,K
reproduces well the velocity-integrated line to free-free continuum
flux ratios (LTCR) observed for all $\alpha$-transitions. This implies
a value for the sound speed of $c_s$$\sim$11\,\kms. The isothermal
assumption reproduces the LTCR frequency dependence observed,
LTCR\,$\propto$$\nu^{1.1}$, which is expected under LTE conditions
(see, e.g., \citealt{mez67}).

The trend of decreasing line width  with increasing frequency guessed from
the H$\alpha$ profiles of M\,2-9 (Fig.\,\ref{f-parms}) suggests radial
expansion with a positive velocity gradient, as for
CRL\,618. 
  As a first approximation we assumed
constant radial expansion because in this case the trend is only
tentative and the inferred velocity difference between the inner and
outer layers of the ionized wind (where the H30$\alpha$ and
H39$\alpha$ emission is produced, respectively) is small
($\sim$4\,\kms). Under this assumption, the RRL profiles are
reasonably well fitted assuming \vexp$\approx$22\,\kms\ and a systemic
velocity of \vlsr=$+$75\,\kms.  Unlike for CRL\,618, the optical depth
of the mm-RRLs is very small ($<<$1) over the bulk of the emitting
region of M\,2-9 and, therefore, the line centroids are predicted to
be well centered on the adopted systemic velocity.

Finally, our simple cylindrical model is able to reproduce the RRL
profiles and free-free continuum at mm-wavelengths produced at the
compact core of M\,2-9, but our model (moderately) underestimates the continuum
flux density from the core at wavelengths \ga3\,cm. This is because a
non-negligible fraction of the cm-continuum flux arises in the outer
($>$160\,au) layers of the ionized wind, which are being truncated,
and thus excluded, in the cylindrical model along the equatorial
direction. We used a cylindrical geometry based on the elongated
shape of the cm-continuum brightness distribution observed by
\cite{kwok85} and \cite{lim03}. However, the model presented here mainly aims to constrain the properties of the inner regions of
the ionized wind ($\sim$5-50\,au), where the free-free (RRL and
continuum) emission at mm-wavelengths reported in this work arises and
which are satisfactorily reproduced. Detailed modeling of the
properties of the cm-continuum emission produced in the outermost
regions of the ionized wind is beyond the scope of this paper.

We stress again that, as for the other targets, the simple model
presented here is most likely not unique and that more complex
geometries, kinematics, density, and temperature profiles
(e.g.,\ latitude-dependent) cannot be ruled out. In the particular case
of M\,2-9, we checked that a double-cone geometry, similar to
that used for MWC\,922 (see next section), could also reproduce the
mm-wavelength data. In this case, to fit the widths of the mm-RRLs, the expansion velocity of the wind
would have to be larger 
and the mass-loss rate moderately smaller (for example,
\mloss$\sim$1.4\ex{-7}\,\my\ for a semi-opening angle of
$\sim$20\degr\ and \vexp=25\,\kms). High-angular resolution maps of
the mm-continuum emission are needed to accurately determine the
spatio-kinematic structure and physical conditions of the inner
ionized wind layers of our targets.

\subsection{MWC\,922}
\label{model_mwc922}

The ionized core of MWC\,922 has been represented by a double-cone
geometry like that used for modeling the ultra-compact \ion{H}{II}
region MWC\,349\,A in \cite{mar11} and, more recently, \cite{bae13}.  This is based on the resemblance
of the maser and non-maser RRL profiles of both sources and it is also
consistent with the X-like morphology of the optical/NIR nebula surrounding
MWC\,922 (\S\,\ref{intro-mwc922}).
The ionized emission is thus assumed to arise in two opposing
conical structures. Each of these structures is formed by two nested
cones co-axial with, and inscribed in, an extended neutral rotating
disk. A sketch of this geometry is given by \cite{bae13} in their
Fig.\,3.  The outermost surface of the bicone (in contact with the
neutral disk) has a semi-opening angle $\theta_{\rm a}$ measured from
its revolution axis to the midplane. 
The outer (shell-like) bicone has an angular width $\theta_{\rm d}$.
The regions that are inside
of (surrounded by) the innermost surface of the outer bicone, at colatitude
$\theta < \theta_{\rm a} - \theta_{\rm d}$, form the inner
bicone. Regions of colatitude $\theta > \theta_{\rm a}$ reside in the
neutral disk, which is adopted to be edge-on.

The inner and outer ionized bicones can have different kinematics. The
outer bicone is assumed to rotate in a Keplerian fashion, presumably
sharing the same kinematics with the neutral disk in which this bicone 
is hypothetically inscribed. The outer bicone
would then simply represent the thin conical surface of the rotating
disk that is illuminated and photoionized by the central source.  If
the density above the disk is lower than that in the equatorial plane,
the central star irradiates/ionizes the upper thin surfaces of the
disk. We note that the disk material at low latitudes could remain
largely neutral if the ionizing radiation is additionally blocked
along the midplane by dust, for example, at the inner edge of a
compact circumstellar disk.  The inner bicone is modeled as a rotating
and radially expanding wind; the same Keplerian rotation field adopted
in the disk is vectorially added to a radial expansion field. The
inner bicone could represent the stellar wind and/or gas
photoevaporating from the ionized surface of the rotating disk. We
refer to the inner and the surrounding outer bicone as the ``ionized
outflow'' and the ``ionized disk component'', respectively.

The gas in the rotating disk outside a certain radius $r_{\rm d}$ is
considered to be part of the outflow, that is, it is expanding and
rotating. The value of $r_{\rm d}$ is always chosen to be smaller than
or equal to the disk gravitational radius, which is defined as the
point where the Keplerian velocity ($\sim$ escape velocity) and the
sound speed are equal; the disk region
outside the gravitational radius is no longer gravitationally bound to the central mass
and it flows away freely \citep[e.g.,][]{hol94,owe12}.

The electron density is assumed to vary throughout the disk and outflow as a function of the radial distance, following a power-law
distribution and also as a function of the latitude, decreasing
exponentially from the midplane toward the poles.
In principle, our model enables the electron temperature to be different in the disk and outflow (but see below).

In Table\,\ref{t-moreli} we give the input parameters of a model that
successfully reproduces the mm-RRL profiles and the free-free
continuum emission of MWC\,922 (Figs.\,\ref{f-mwc922} and
\ref{f-sedmodel}). This model predicts the observed transition from
single-peak profiles at $\sim$3\,mm to double-peak profiles at
$\sim$1\,mm, including the velocity peak-to-peak separation of the
latter, and the steep increase of the Hn$\alpha$-line FWHM and
intensity as a function of frequency; the latter
is indicative of non-LTE conditions and, in particular,
of H30$\alpha$ maser emission in this object.  It also recreates the
spectral continuum emission distribution from the mm- to the
cm-wavelength range and the angular diameter ($\sim$0\farcs2=340\,au
at 1700\,pc) of the 3.6\,cm-continuum emitting region
(\S\,\ref{intro-mwc922}).

Before discussing this model in detail, we stress that in the case of
MWC\,922, the geometry, kinematics, and electron temperature deduced
are particularly uncertain. This is because of the extremely high
sensitivity of the maser line intensity not only to small changes in
the local physical conditions but also to the {\sl uncertain} $b_n$
departure coefficients (\S\,\ref{moreli}). Our final model presented in
Table\,\ref{t-moreli} uses the coefficients by \cite{sto95}, however,
we also explored a large range of model parameters using the
$b_n$ coefficients \cite{wal90}. Since the $b_n$ coefficients computed
by these two groups differ significantly, the physical model needed to
reproduce the observed line intensities are different when using one
set of $b_n$ or the other. Irrespective of the $b_n$ set adopted, the
biggest challenge is always to reproduce simultaneously the maser and
non-maser RRL profiles observed.

Most of the mm-wavelength free-free emission observed in MWC\,922 arises in a
rather compact region within $r$$\la$150\,au.
In particular, 90\%\ of the RRL and continuum emission at 1 and 3\,mm arise in the inner 
($\sim$10-80\,au) and outer ($\sim$20-150\,au) parts of the
central ionized core, respectively. 
The full extent of the \ion{H}{II} region is probably larger as
suggested by the 3.6\,cm-continuum maps. The inner radius of the
ionized core of MWC\,922 is poorly determined from these data since
the deepest ionized layers at $\la$10-15\,au are very optically thick at
mm-wavelengths and thus are not probed by our mm-RRL profiles.  We
chose a nominal inner radius in our model of \rin=0.05\,au,
which is equivalent to \rin$\sim$1\,\rs, adopting the values of
\teff\ and luminosity given in Table\,\ref{t-buj}.
As for M\,2-9, we derive an upper limit to the radius of an
hypothetical central cavity, which could be devoid of ionized gas, of
\rin$\la$35\,au; this is the largest cavity radius that would still
reproduce the uniform spectral index of the free-free continuum as
observed, which indicates optically thick emission from cm wavelengths
up to $\sim$230\,GHz (at least).

The electron temperature distribution in the ionized core can only be
poorly constrained from single-dish data. Initially, the temperature
\te\ has been enabled to be different in the disk and in the outflow,
however, for the geometry and opening angle adopted in our final
model, the temperature in the disk and outflow have to be comparable to
reproduce simultaneously the observed maser and non-maser
profiles, \te$\sim$6500-7000\,K.

The electron density rapidly declines from the ionized surface/layer
of the rotating disk toward the poles. Along
$\theta$=$\theta_a$+$\theta_d/2$, i.e., within the thin ionized layer
of the disk, the density ranges from \dense$\approx$10$^7$ to
$\approx$10$^5$\cm3, following a $r^{-2.4}$ power law, across the
$r\sim$15-150\,au layers that emit the bulk of the continuum and emission of RRLs at mm-wavelengths. The density falls off by a factor of
$\sim$20-30 across the full width of the ionized disk, $\theta_d$, and
continues to decrease rather steeply across the ionized outflow,
dropping by 5 dex at the revolution axis.

As discussed in \S\,\ref{res}, the gradual increase of the width of
the Hn$\alpha$ profiles as a function of frequency is consistent with
Keplerian rotation. To explain the velocity peak-to-peak separation
observed in the H30$\alpha$ maser line, we considered a central
mass of $\sim$8\,\msun, which is equivalent to the rotation velocity
field given in Table\,\ref{t-moreli}; we find that a range of
$\sim$5-10\,\msun\ gives also acceptable predictions.  The maser
spike emission arises mainly in the Keplerian disk, at radial
distances $\sim$35-55\,au, because this region has the geometry (a
high coherence length) and the densities, of
\dense$\approx$10$^6$-10$^7$\,\cm3, required to produce strong maser
amplification of the H30$\alpha$ line.
The double-peaked profile of the H31$\alpha$ line, which is also
indicative of stimulated non-LTE amplification, is well reproduced by our
model. However,
the intensity of the H31$\alpha$ line relative to the rest of the RRLs
detected is always notably lower than observed for
all models regardless of the physical conditions and kinematics
adopted. For this reason, and given the extremely high uncertainties
of the $b_n$ coefficients, in our final model, we used the same
departure $b_{30}$-coefficient for the H31$\alpha$ and H30$\alpha$
transition, which results in a much better model-data agreement. 

In the ionized outflow, we adopted a constant expansion velocity of
\vexp$\sim$5\,\kms, which is consistent with the profile of the
non-maser H39$\alpha$ and H41$\alpha$ lines, and, in particular, with
the lack of emission wings.  Given the dimensions of the outermost
regions of the ionized wind probed by our observations, $\sim$140\,au,
we derive a kinematical age for the outflow of $t_{\rm  k}$$\sim$150\,yr.
This is a lower limit, since the outer boundaries
of the ionized core of MWC\,922 are not observationally constrained
and could reach larger distances.  The innermost layers of the
outflow, for example,\ at \rin=15\,au, would have been ejected only
$\approx$10\,yr ago. The age of the rotating disk cannot be estimated from
these data but it, in principle, it could be a much older long-lived
structure.

Finally, we already mentioned the large uncertainties of the
model input parameters in the case of MWC\,922. From our multiple
model runs, we conclude that it is also possible to reproduce the data
used in this work adopting a larger opening angle for the ionized
disk+outflow structure. For example, using $\theta_a$$\sim$53\degr, we
are able to find a set of physical parameters that satisfactorily
explain the observations. One important difference with respect to the model in
Table\,\ref{t-moreli} is that, for a larger opening angle,
the temperature in the ionized disk
must be notably larger than in the outflow. 
In our opinion, the presence of a hot ionized thin disk
sandwiched between two cooler structures (namely, the outflow and the neutral disk) is improbable.\footnote{For example, based on
  our theoretical and observational current understanding of photoevaporating disks; see, for example, \cite{hol94,yor96,bae13}.  }

As for the other targets, we estimate an equivalent isotropic
mass-loss rate for the outflow. For a constant expansion velocity of
\vexp=5\,\kms, and considering that we are adopting a non-isotropic
(bipolar) outflow with a latitude-dependent density, we find
\mloss$\sim$1.8\ex{-5}$\times$(1$-$cos(25\degr))$\sim$2\ex{-6}\,\my\ for
the value of $\theta_a$=25\degr\ used in our model.  As expected, the
equivalent isotropic mass-loss rate does not notably change when using
a different opening-angle, since it is mainly constrained by the
free-free continuum flux. In particular, for $\theta_a$$\sim$53\degr,we find
\mloss$\sim$5.2\ex{-6}$\times$(1$-$cos(53\degr))$\sim$2.0\ex{-6}\,\my.
However, the nature of the outflow in MWC\,922 is
unclear (stellar wind and/or photoevaporating gas from the ionized
disk) and, therefore, the meaning/interpretation of the rate derived is
not straightforward.

\subsection{Remaining targets: mm-RRL non-detections}
\label{model_rest}

We modeled the free-free continuum emission of the sample
targets with no mm-RRL detections. This mainly aims at obtaining
an estimate of the mass-loss rate for these objects as well. In the
simplified model used in this case, we assume that the free-free
emission arises in a spherical isothermal wind expanding at constant
velocity.  In particular, we adopt \vexp=15\,\kms, \te=10,000\,K, and
an electron density power-law of the type
\dense($r$)$\propto$$r^{-\alpha_n}$. In this case, the free-free
continuum is expected to vary with frequency as
\snu{\frac{(6.2-4\alpha_n)}{(1-2\alpha_n)}} and, therefore, the
$\alpha_n$ parameter can be constrained from the spectral index of the
free-free continuum \citep[see, e.g.,][]{rod09}.

\begin{figure*}[htbp]
     \centering 
     \includegraphics[width=0.45\textwidth]{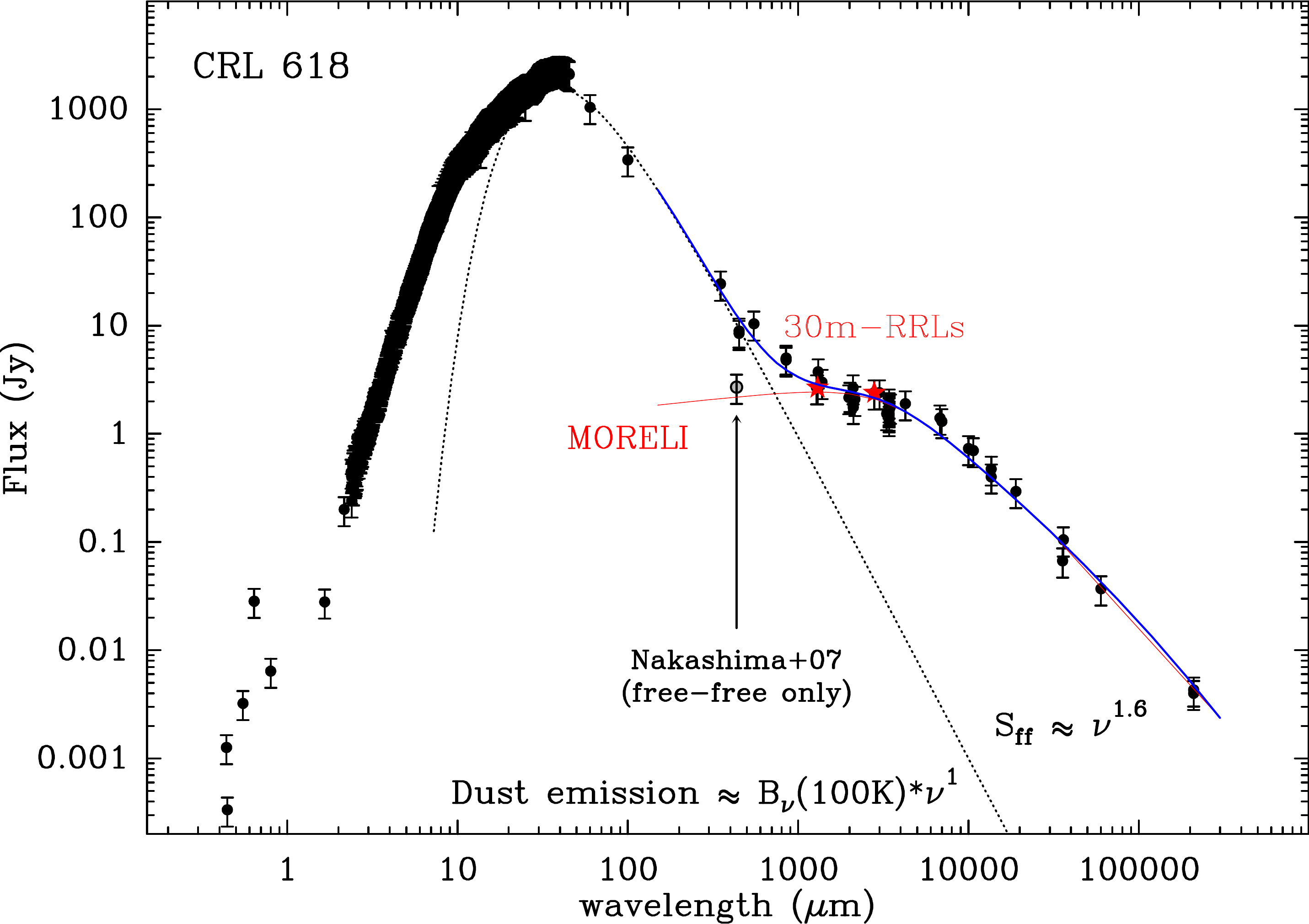}  
     \includegraphics[width=0.45\textwidth]{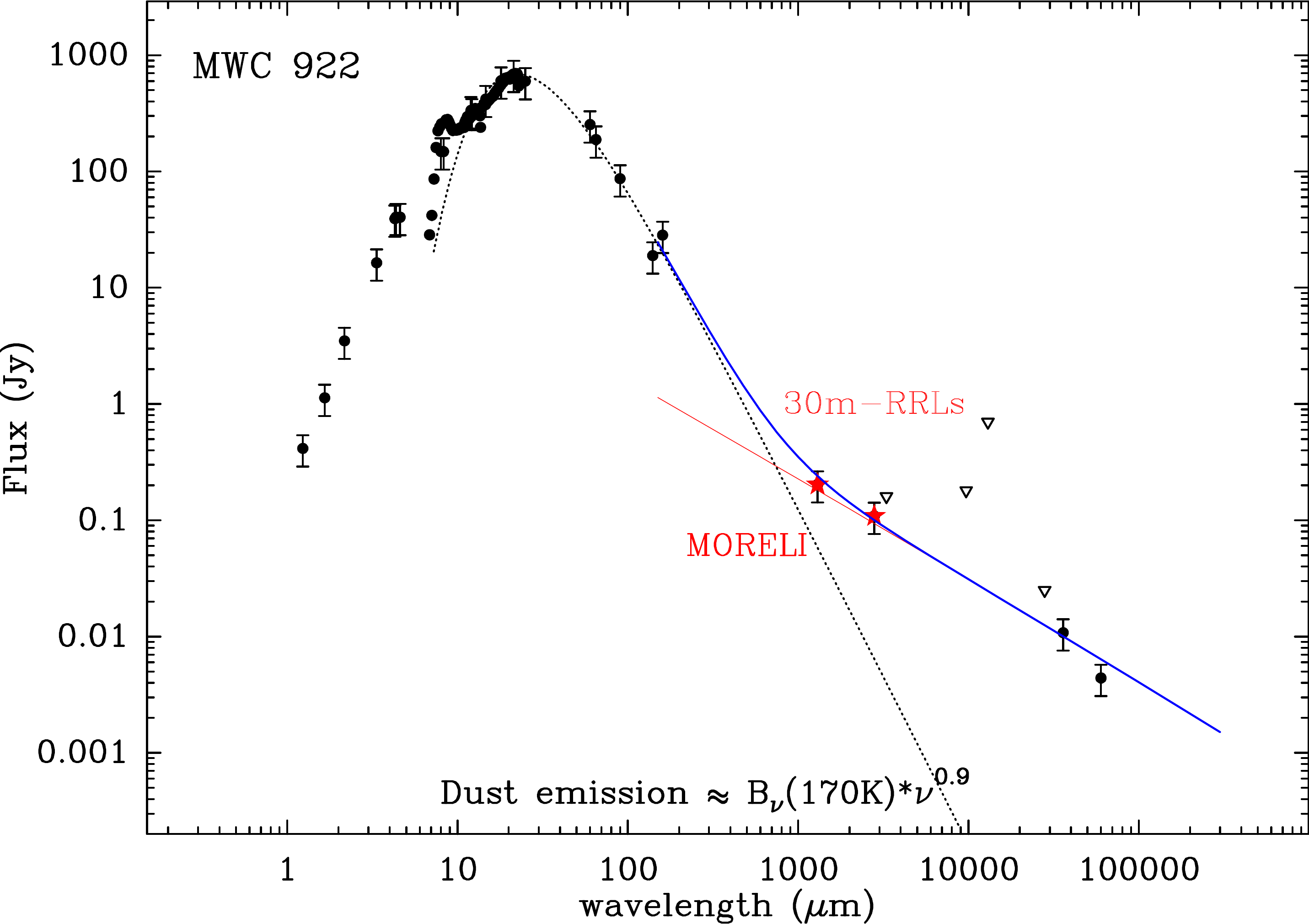} 

     \includegraphics[width=0.45\textwidth]{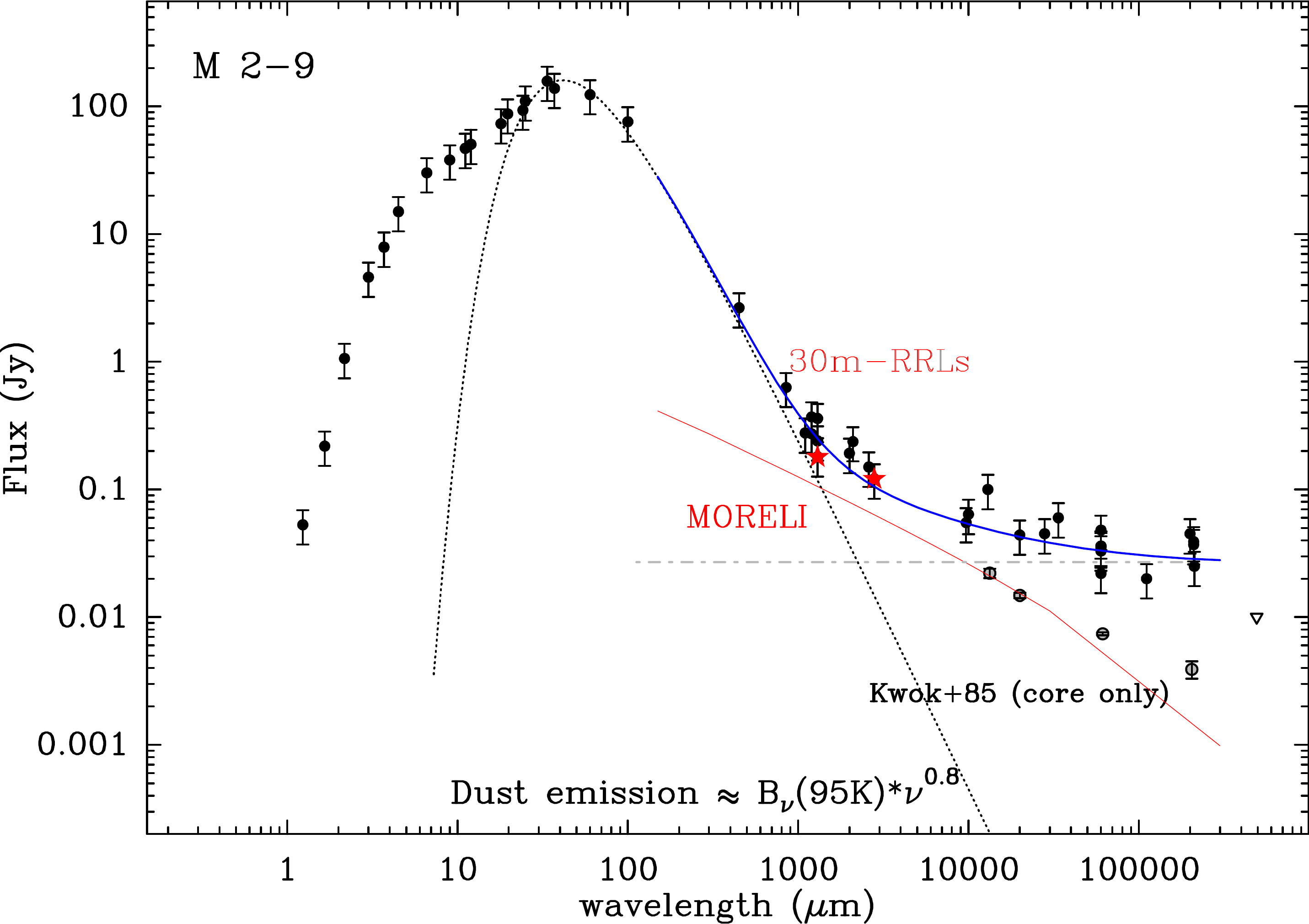} 
     \includegraphics[width=0.45\textwidth]{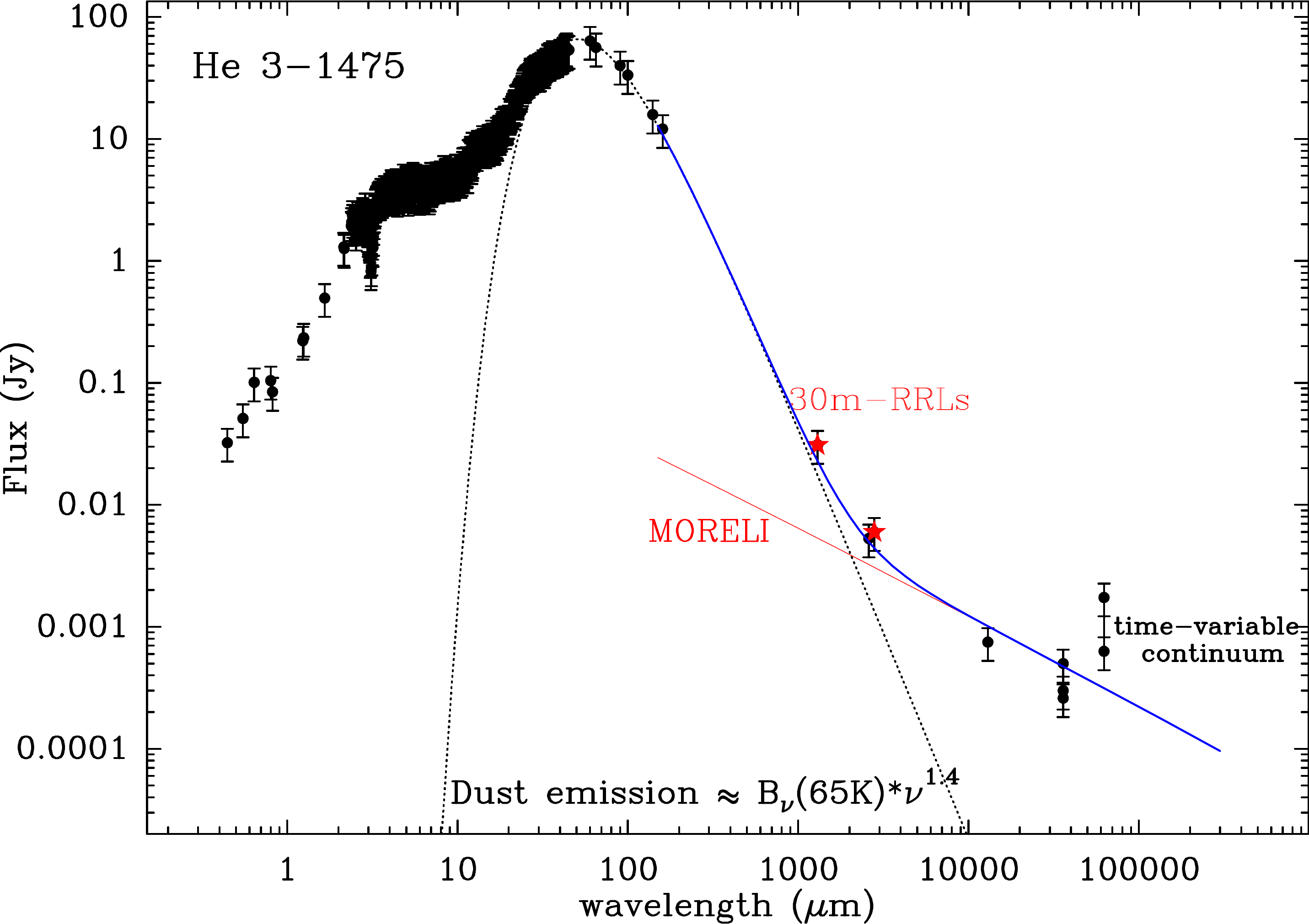}  

     \includegraphics[width=0.45\textwidth]{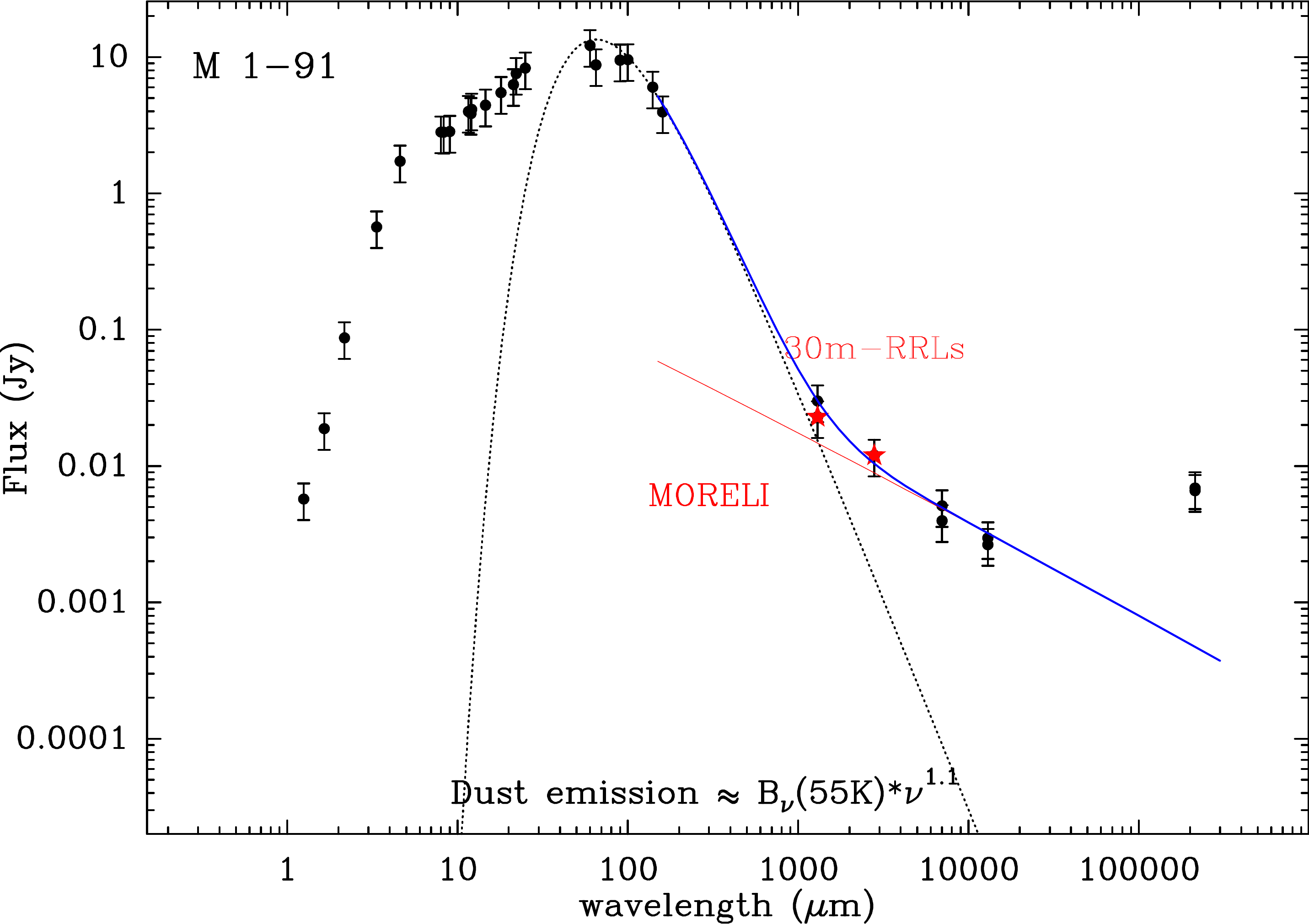}
     \includegraphics[width=0.45\textwidth]{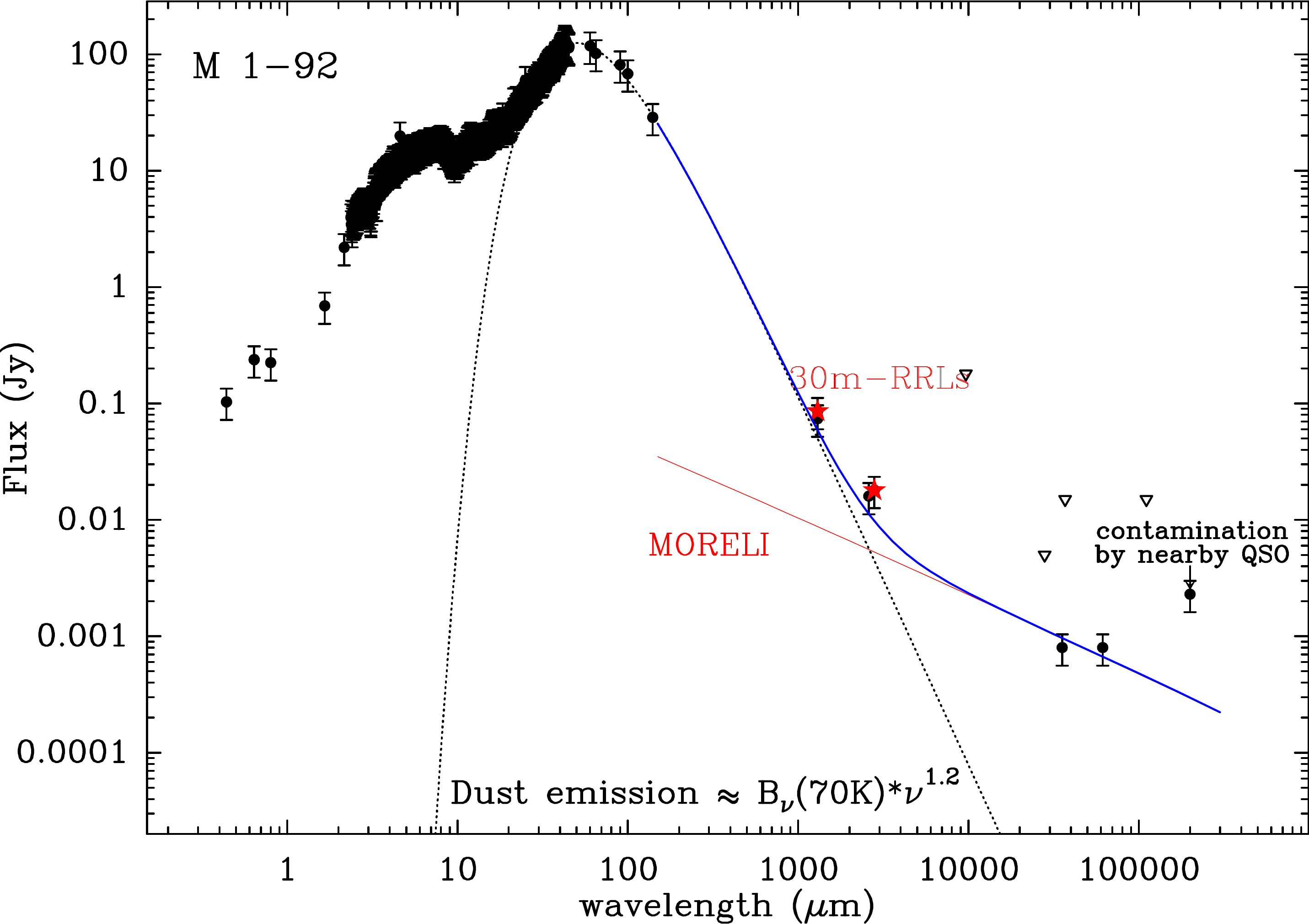} 

     \includegraphics[width=0.45\textwidth]{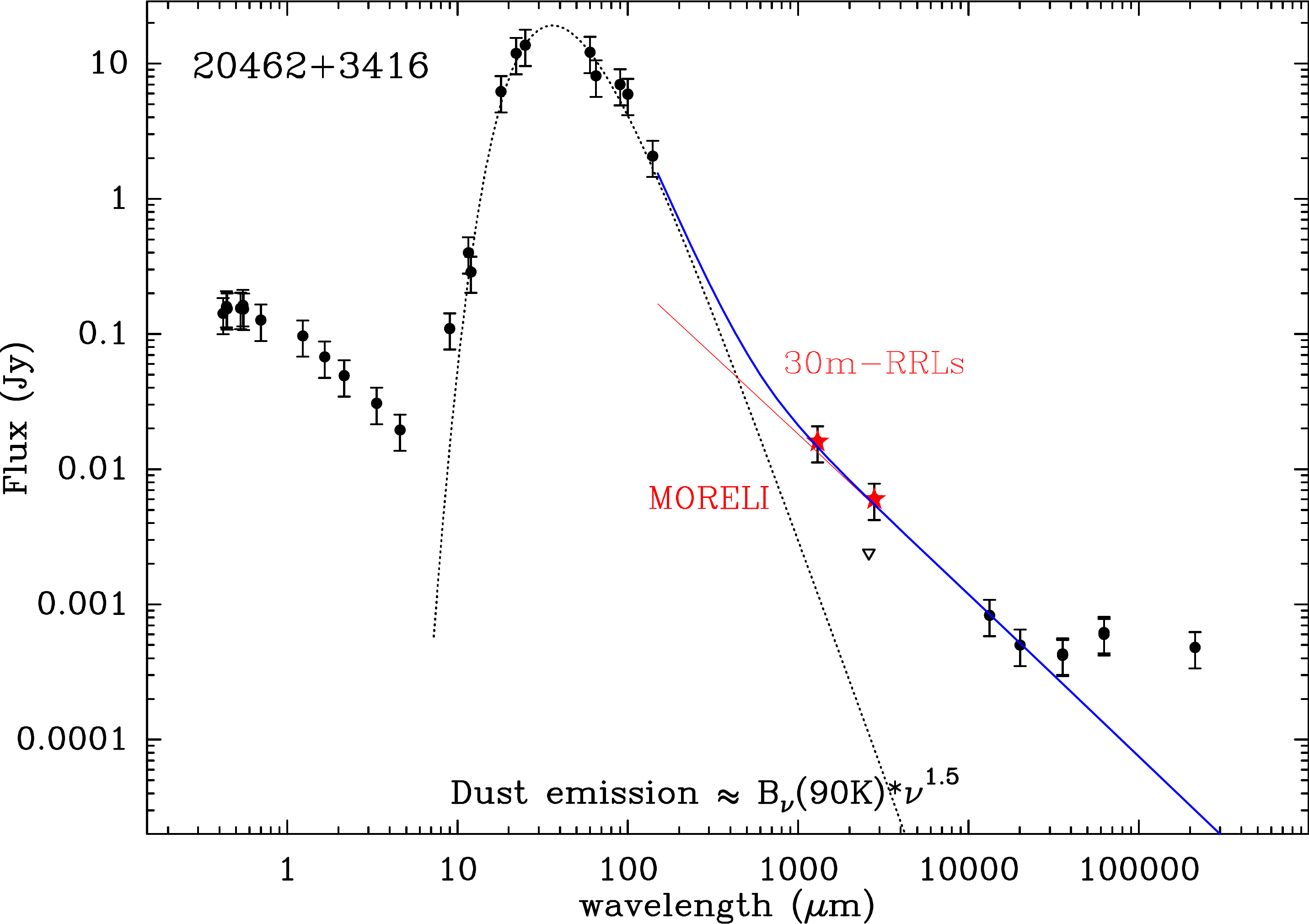}
     \includegraphics[width=0.45\textwidth]{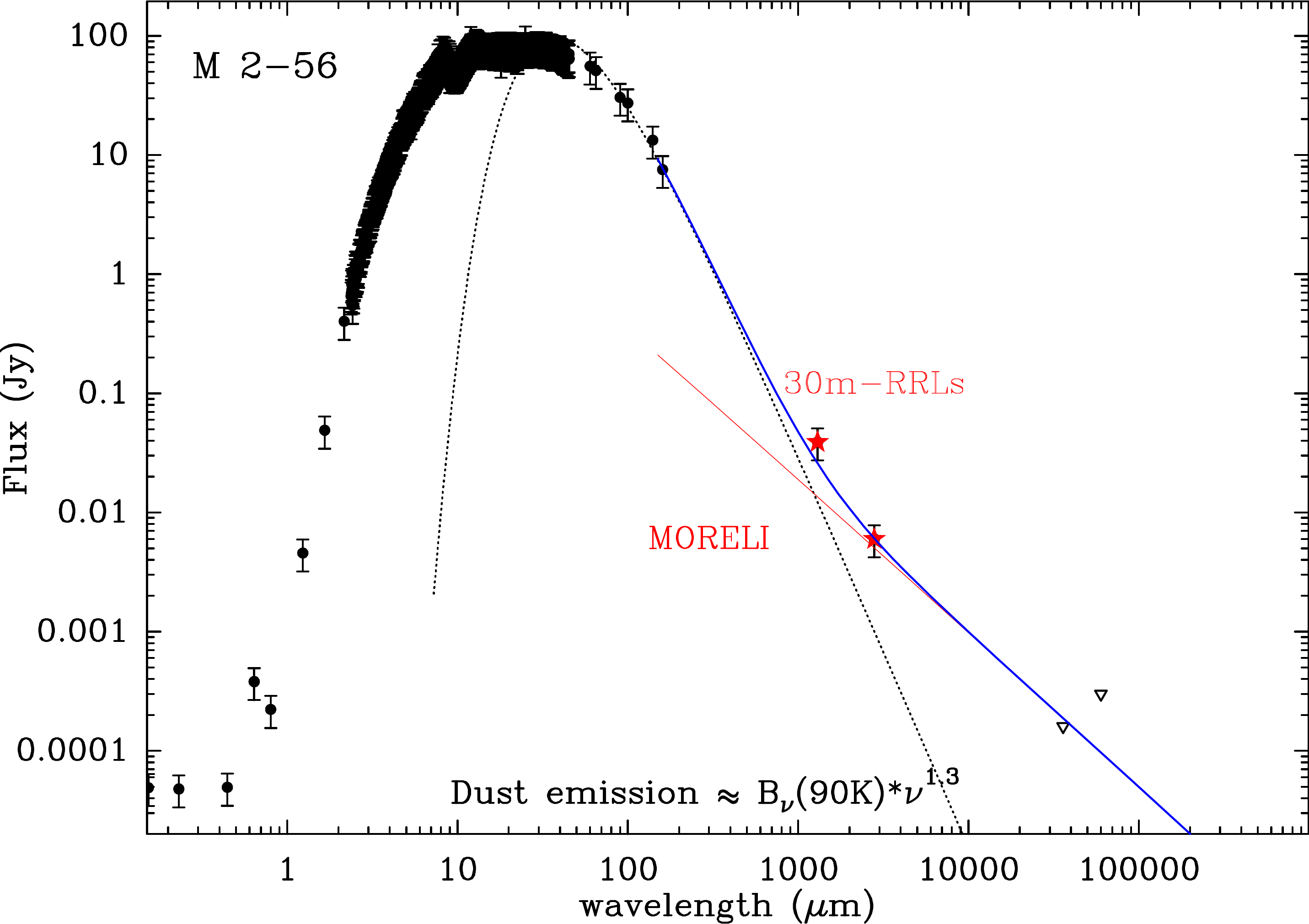}  
     \caption{Same as in Fig.\,\ref{f-seds} with the free-free continuum emission  predicted by our model (red line).
       Model parameters are given
       in Table\,\ref{t-moreli} for CRL\,618, MWC\,922, and M\,2-9, and in Table\,\ref{t-ND} for the rest of the targets.}
   \label{f-sedmodel}
   \end{figure*}

The main input parameters and synthetic free-free continuum emission
of the best models are given in Table\,\ref{t-ND} and in
Fig.\,\ref{f-sedmodel}. In all cases, we checked that MORELI
predicts mm-RRLs with intensities below our detection limits.  As we
can see, most of the free-free continuum emission at mm-wavelengths
arises at the very inner layers of the wind around the star, typically
within \rout(99\%)\la\,130\,au and, in some cases such as M\,2-56, in
regions as compact as \rout(99\%)\la30\,au.  The pPNe/yPNe with the
most extended \ion{H}{II} region is He\,3-1475 (with
\rout(99\%)$\sim$800\,au). In all but one case (He\,3-1475), these
inner layers represent very recent mass ejections that happened less than
$\sim$15$\times$($\frac{15 {\rm kms^{-1}}}{\vexp}$)\,years ago.

We find spectral indexes roughly consistent with inverse-square
density laws except for \mbox{M\,2-56} and IRAS\,20462+3416. In these
two cases, a steeper radial density fall, as \dense$\propto
r^{[-3.1:-3.5]}$, is inferred, which may suggest a gradual
increase of the mass-loss rate in recent times.

We derive characteristic mass-loss rates in the range
\mpagb$\sim$[0.3-4]\ex{-6}\,\my. We do not observe a clear trend
between \mpagb\ and the bolometric luminosity; there are rather
luminous objects, namely, M\,1-91 and M\,2-56, with relatively low
mass-loss rates of \mpagb$\sim$[2-3]\ex{-7}\,\my, as well as dim
sources, namely, IRAS\,20462+3416, with relatively high mass-loss
rates of \mpagb$\sim$[1-2]\ex{-6}\,\my.


%
\begin{table*}[htp]
\small
\caption{Results from modeling of the
  free-free continuum for mm-RRL non-detections (\S\,\ref{model_rest}).}
\label{t-ND}      
\centering                          
\begin{tabular}{l c c c c c c c c}        %
  \hline\hline
  Source & \rout(90\%) & \rout(99\%) & \dense(6.68\,au) & $\alpha_n$ & \mpagb\ & $M_{\rm ion}$(90\%) & $t_k$(90\%) \\
  name   & (au) & (au) & (\cm3) & & (\my) & (\msun) & (yr) \\ 
  \hline\hline
He\,3$-$1475 & 185 & 840 & 6.2\ex{8} & 2.18 & 4.4\ex{-6} & 1.7\ex{-6} & 60 \\ 
M\,1$-$91 & 44 & 144 & 4.7\ex{7} & 2.10 & 3.4\ex{-7} & 4.2\ex{-6} & 14 \\ 
M\,1$-$92 & 40 & 132 & 4.1\ex{7} & 2.10 & 3.0\ex{-7} & 3.5\ex{-6} & 13 \\ 
IRAS\,20462 & 22 & 67 & 2.4\ex{8} & 3.13 & 1.7\ex{-6} & 2.9\ex{-5} &  7 \\ 
M\,2$-$56 & 12 & 24 & 7.0\ex{7} & 3.50 & 5.0\ex{-7} & 2.3\ex{-5} &  4 \\ 
\hline                                   
\end{tabular} \\
\tablefoot{In all cases, the central ionized regions were  modeled assuming spherical isothermal winds; we
  adopt \vexp=15\,\kms, \te=10,000K, \dense$\propto$$r^{-\alpha_n}$,
  and inner radius \rin=0.05\,au. The mass and kinematic age of the
  ionized gas layers within \rout(90\%), i.e., the radius that
  encircles 90\% of the $\sim$3\,mm continuum flux, are given in the
  last two columns.  \\
}
\end{table*}
%

\section{Mass-loss history and evolutionary status} 
\label{dis1}

The emission from the mm-RRLs reported and modeled here allowed
us to characterize 
the relatively dense, \dense$\sim$10$^{6-8}$\cm3, inner
($\approx$\,10-100\,au) ionized regions of three of our program
objects: CRL\,618, M\,2-9, and MWC\,922.  Considering their expansion
velocities and small extent (Table\,\ref{t-moreli} and
\S\,\ref{moreli}), these inner regions probe very recent mass
ejections events that happened less than $\sim$15-20\,yr. This is probably also
true for our sources with no mm-RRL detections, since there is
no reason to expect remarkably different bulk gas motion speeds  in
these regions.

The mass-loss history in more ancient epochs (back to few thousand
years ago) is relatively well known for CRL\,618 and M\,2-9 from
previous works but, unfortunately, not for MWC\,922 (\S\,\ref{intro}).
In the next subsections, we discuss our
findings in the context of these former mass-loss episodes, the nature and evolutionary status of these objects.  

\subsection{CRL\,618} 
\label{h-crl618}
In the case of CRL\,618, we found a significant difference between
the H30$\alpha$ and H41$\alpha$ line profiles (widths and peak
intensity) and those observed in 1987 by \cite{mar88}. These changes
are not totally surprising since our observations do not trace the same wind material as observed in the 1980s.
This is because the expanding wind layers
probed by the mm-RRLs observed $\sim$30 years ago by \cite{mar88} have
traveled
$\sim$125\,au outward during this time (at
$<$\vexp$>$$\approx$20\,\kms) starting from their position in 1987
(at \rin$\sim$70\,au). Therefore, they are now beyond
$\sim$195\,au and outside the mm-RRL emitting regions. 
The changes in the line profiles observed are then indicative of wind
variability, i.e., evolution with time of the wind properties.\footnote{Incidentally, in the case of unperturbed wind expansion, the layers
  studied by \cite{mar88} would have now entered the optically thick
  part of the ionized wind (encompassing $\sim$150-650\,au; see
  Fig.\,5 in Tafoya et al.\,2013) and, thus, should be best traced at
  present by RRLs and continuum observations {\sl at cm-wavelengths.}}

The kinematic age of the layers of the post-AGB wind of CRL\,618
observed by us is $\sim$10-20\,yr (Table\,\ref {t-moreli}).  The
mass-loss rate deduced from our mm-RRL observations,
\mpagb\,$\sim$8.4\ex{-6}\,\my, is that of a post-AGB wind that was
ejected in the $\sim$10-20\,years previous to 2015. Similarly, the
mass-loss rate deduced by \cite{mar88}, \mpagb\,$\sim$7.6\ex{-6}\,\my,
probably corresponds to a wind ejected in the $\sim$10-20\,years
previous to 1987.  The slightly larger mass-loss rate estimated by us
in more recent epochs is then in agreement with the trend deduced by
\cite{taf13}, who proposed that the mass-loss rate had increased (from
\mpagb$\sim$4\ex{-6} to $\sim$6\ex{-6}\,\my) in the $\sim$130 years
previous to 1998.  In this context, our result is consistent with a
mass-loss rate that has continued to rise and, indeed, at nearly the
same rate contemplated by \cite{taf13}.

\cite{mar88} considered a central cavity with radius \rin$\sim$70\,au
in their LTE model of the ionized core of CRL\,618. As explained in
\S\,\ref{moreli}, the size of this cavity, which could result from a
short period of diminished mass-loss rate, is rather uncertain because of the degeneracy of the turnover frequency with \rin\ and \mpagb.
However, we would like to mention that the presence of a cavity with a
radius of \rin$\sim$70\,au in 1987 is not inconsistent with our data
and model. This is because, as explained in the first paragraph of
this section, the cavity (as the rest of the layers of the ionized
wind) would have moved forward by $\sim$125\,au in the last
$\sim$30\,years. Therefore, if the cavity hypothesized by \cite{mar88}
was real and its size in 1987 roughly correct, 30 years later, it
should constitute a low-density shell at $\sim$195-265\,au,
i.e.,\ beyond the ionized wind layers traced by our mm-wavelength
observations.  The fact that a cavity is also needed to explain the
observations in 2015 could indicate another (periodically recurrent?)
recent short intermission of the mass loss. In any case, we caution
that the cavity is needed in the context of the simple model used here
(Table\,\ref{t-moreli}). The cavity may not be a requirement if one
adopts a more sophisticated model with a sufficiently complex
geometry, density, temperature, and velocity stratification (for
example, an inhomogeous wind with dense clumps).

Of course, the comparison of the mass-loss rates (and other
parameters) derived from different works has to be carried out with extreme
caution since these are affected by hardly quantifiable uncertainties
that are inherent to the different analysis methods and assumptions
adopted. Therefore, our conclusions about the overall increase of
\mpagb\ and intermittent reduced mass-loss rate intervals should be
considered tentative.

The mass-loss history of CRL\,618 in earlier epochs is written in its
nebular architecture (\S\,\ref{intro-crl618}). From detailed
observations and modeling of its molecular envelope, two major
large-scale, mass-loss episodes in the form of slow winds
(\vexp\,\la\,17\kms) have been identified. The first took place
\ga\,2500\,yr ago at a rate of \mloss\la\,10$^{-5}$\,\my\ and generated
a tenuous extended halo. The second mass-loss event started
$\sim$\,400\,yr ago at very high rate of
\mloss$\sim$few\ex{-4}-10$^{-3}$\,\my\ resulting in the formation of a
dense central core \citep{san04b,lee13a,sor13}.
More recently, $\sim$200\,yr ago, the interaction between fast
bullet-like ejections and the pre-existing AGB circumstellar envelope
probably shaped the optical lobes \citep{bal13,vel14}. The compact, fast ($\approx$\,100\kms) molecular outflows
of CRL\,618, with a kinematical age of $\sim$30-80\,yr, could have
been partially shaped by the mentioned bullet-like ejections or by a distinct, more
recent (\la\,80\,yr) collimated fast wind \citep{san04b,lee13b,hua16}.

The mass-loss rate at which the post-AGB wind was ejected over the
last $\sim$150\,years in CRL\,618  is
\mpagb$\sim$[4-8]\ex{-6}\,\my\ . Considering the whole range of values
deduced including our work (see above), this mass-loss rate is then only {\sl moderately
  lower} than the mass-loss rates of the slow AGB winds that led to
the molecular extended halo and dense core. Given their low expansion
velocity and high mass-loss rates, the halo and core were most
likely ejected when the central star was ascending the AGB toward its
tip and immediately before leaving this phase, respectively; this is
when the most intense winds are believed to be blown out
\citep[e.g.,][]{blo95,vas93}. Therefore, in the case of CRL\,618, the drop of
the mass-loss rate, which is believed to mark the end of the AGB phase
(after the so-called ``superwind'') and the beginning of the AGB-to-PN
evolution, is rather modest, in particular, only about one order of
magnitude lower than that of the core-wind, and comparable to that of
the halo-wind.

\subsection{M\,2-9}
\label{h-m29}

Our mm-RRL observations trace the inner layers of a dense ionized wind at the core of M\,2-9 that has
been ejected at an average rate of \mpagb$\sim$3.5\ex{-7}\,\my\ and
expansion velocity \vexp$\sim$22\,\kms\ over the last \la\,15\,yr
(Table\,\ref{t-moreli}). The mass-loss rate estimated by us is
consistent (within uncertainties) with that obtained by \cite{kwok85}
for the outer layers of the ionized core from their analysis of the
cm-continuum emission maps (1.4-22\,GHz, see also
\S\,\ref{intro-m29}). These authors find
\mpagb=3.3\ex{-5}(\vexp/1600\kms)($d$/kpc)$^{3/2}$\,\my, which is
equivalent to \mpagb=2.4\ex{-7}\,\my\ after scaling it to the values
of the distance and the expansion velocity in Table\,\ref{t-moreli}.

In 1982, the outer boundary of the ionized core of M\,2-9 along the
nebula axis was probably near 0\farcs3 (\rout$\sim$200\,au); beyond
this radius most of the material remains neutral, as deduced from the
cm-continuum maps presented by \cite{kwok85}. Adopting the same
expansion velocity measured by us, \vexp$\sim$22\,\kms, the wind
layers traced by the cm-continuum observed by these authors must have
been ejected in the $\la$40-50\,years previous to 1982 (the observing
date), that is, sometime after 1932. Therefore, the current wind observed
by us could have been ongoing with a constant or, at least, not
markedly variable mass-loss rate of $\sim$3\ex{-7}\,\my\ since it
began at least $\sim$50\,yr ago.

Prior to this, two other major mass-loss episodes have been identified
in this object leading to two ring-shaped, eccentric molecular
structures at the nebular equator mapped in CO emission by CC12; see
also \S\,\ref{intro-m29}. Recent 0\farcs18-resolution CO emission maps
obtained with ALMA confirm the spatio-kinematics of the two rings
\citep{cc17}. These authors conclude that these
two expanding rings were formed during two short mass-loss episodes
(of duration $\sim$40\,yr) produced when the mass-losing star was at
different positions in the orbit of the central binary.
This scenario is also
supported by the different systemic velocities found for the two
rings, \vlsr=80.0 and 80.6\,\kms\ ($\pm$0.1\,\kms) for the outer and
inner ring, respectively, which is readily explained if the
mass-losing star changed its velocity due to its orbital motion
within a binary (or multiple) system. CC12 proposed that the
outer ring, with \rout$\sim$\,2340\,au and \vexp=7.8$\pm$0.1\,\kms,
was ejected at a rate of $\sim$9\ex{-5}\,\my\ about 1400 years ago,
whereas the inner ring, with \rout$\sim$\,780\,au and
\vexp=3.9$\pm$0.1\,\kms, would have occurred $\sim$900\,yr ago at a
slightly smaller rate.  It is unknown whether the mass-losing central
star of M\,2-9 was in the late AGB or early post-AGB phase when the
CO-rings were ejected, but the relatively high mass-loss rates (and
low expansion velocities) observed are most consistent with a late-AGB
status. 
The mass-loss rate of the younger ionized core-wind studied here
(Table\,\ref{t-moreli}) is about two orders of magnitude smaller than
that of the preceding mass outburst that led to the inner CO ring, and
thus most consistent with a post-AGB evolutionary stage at present.



\subsubsection{The central binary} We found a significant difference
between the centroids of the mm-RRLs,
\vlsr=75$\pm$2\,\kms\ (\S\,\ref{res}), and those of the CO lines.
This \vlsr\ difference is consistent with the scenario proposed by
CC12 in which the mass-losing star is moving along its orbit; the
centroid of the mm-RRLs would then correspond to the line-of-sight
velocity of the star averaged during the last \la\,10-20\,years (when the
inner layers of the ionized wind were ejected). The
  kinematical age of the inner wind layers probed by our mm-RRL
  observations only represents a small fraction ($\la$1/5) of the
  orbital period.
We can obtain a lower limit to the orbital velocity of the
mass-losing (primary) star of
\mbox{\vorbu\ga(80.6-75)/(2*cos(17\degr))$\approx$2.9\,\kms}, which would imply
an average distance to the center of mass of $a_1$\ga\,8.7\,au for an
orbital period of $P$$\sim$90\,yr.
With this limited knowledge of the orbital elements for only one of
the stars ($m_1$, the mass-losing star), we cannot derive the mass
($m_2$) and orbit of the secondary with certainty, but we can
obtain some constraints by applying the well-known relationship
between the binary mass function, $f(m_1,m_2)$, and the
observables $a_1$ and $P$. Assuming a circular orbit, the mass
function, which is derived from the third Kepler's law, is given by

\begin{equation}
f(m_1,m_2) = \frac{m_2^3}{(m_1+m_2)^2} = m_1 \frac{q^3}{(1+q^2)} = \frac{a_1^3}{P^2} 
,\end{equation}

\noindent 
where $m_2$ is the mass of the companion, $q$ is the dimensionless
secondary-to-primary mass ratio $q$=$m_2$/$m_1$; units are
given in \msun, au, and years for the masses (including $f(m_1,m_2)$),
semi-major axis, and period, respectively. 

Using the lower limit to $a_1$ deduced above, we can also obtain a
lower limit to the left hand of this equation,
$f(m_1,m_2)$\ga\,0.08\,\msun\ and, thus, a lower limit to the mass of
the companion relative to the primary.  The mass of the primary star
($m_1$), which is the stellar remnant after most of the envelope was ejected in the earlier red giant and AGB phases, is expected to
be within $\sim$0.5-1\,\msun\ range \citep{sch83,blo95,vas94}. In
particular, the low luminosity of M\,2-9 ($\sim$700-3000\,\ls\ at
$d$=650-1700\,pc, Table\,\ref{t-buj}) is most consistent with a
low-mass progenitor (with an initial mass of \la\,1\,\msun) and, thus,
with a low-mass stellar post-AGB core of about 0.5\,\msun\ (see
\S\,\ref{dis2} and Fig.\,\ref{f-tracks}).  Adopting $m_1$=0.5\,\msun,
the limit to $f(m_1,m_2)$ implies
that the companion mass must be $m_2$\ga0.4\,\msun. The
uncertainty in the systemic velocity of the RRLs,
\vsys=75$\pm$2\,\kms, translate into a relative error of 50\% in the mass of the
companion, which could then be as low as $m_2$$\sim$0.2\,\msun\ as derived from these data.

We now analyze the possible evolutionary status of the secondary
for different values of the mass ratio. A main sequence companion
would be consistent
with a mass in the range $m_2$\ga0.2 and \la\,1\,\msun.  Larger
  masses of the secondary (assuming a dwarf star) would be in conflict
  with its slower evolution relative to the low-mass (\la\,1\,\msun)
  primary.
In this case, the relative orbital separation would be 
$a=a_1+a_2$\,$\sim$\,18-23\,au.  
If the companion is a white dwarf (WD), the upper limit to its mass is $m_2\sim$1.4\,\msun, 
which would imply a slightly larger orbital separation of
$a$\,$\sim$25\,au.  If this is the case, then 
M\,2-9 could have been a classic symbiotic system (AGB+WD) until recently, when the 
primary mass-losing star would have left the AGB; at present, the whole system could be dying a second death as as 
post-symbiotic (`post-AGB'+WD) system.   

Values of $q$ larger than those discussed above are, in principle,
possible based on our observables and interpretation
($a_1$\ga\,8.7\,au and $P$$\sim$90\,yr), but it would imply that the
companion is a dark, compact stellar remnant such as a neutron star
(NS, with a typical mass in the range $m_2 \sim$\,1.1-3.2\,\msun) or a
black hole (BH, if $m_2>$3\,\msun). If this is the case, the orbital
separation of the system would be larger than for the dwarf or WD
companion scenario (for example, $a$\,$\sim$44\,au for
$m_2$$\sim$10\,\msun). If the mass-losing star is accompanied by such
a NS/BH then, given the relatively large orbital separation, it would
be most likely in a ``silent'' mode; that is, this star would not be undergoing
significant mass-transfer that, if present, should be accompanied by
high-energetic phenomena. This is indeed consistent with the non-detection of X-ray emission toward M\,2-9 \citep{rui14}.

The mass of the companion deduced here ($m_2$$\ga$0.2\,\msun) is
somewhat larger than that estimated by CC12, who proposed a low-mass companion of $m_2$$\la$0.1-0.2\,\msun\ orbiting around a
$m_1$$\sim$1\,\msun\ mass-losing star when the CO rings were ejected.
The low secondary-to-primary mass ratio proposed by these authors
results from the low orbital velocity of the primary used in their
calculations, \vorbu$\sim$1\,\kms, which is smaller than the value
inferred from the separation between the centroids of the
mm-RRLs and the CO line (\vorbu$\ga$2.9$\pm$1\,\kms).
We note that CC12 interpret the difference between the CO line centroids
($\sim$0.6\,\kms) and velocity gradients observed in the two rings as
the result of two short mass-ejection events that took place when the
mass-losing star was at two specific positions in a circular
orbit. In particular, they proposed that the large CO ring was ejected
when the system was near conjunction
(with the stars moving perpendicularly to the line of
sight) and the small ring was blown out $\sim$500\,yr later, when the
stars were at quadrature (and, thus moving in
a direction very close to the line of sight); see red and green
marks in their Fig.\,4.

The low radial velocity shifts deduced from the CO emission from the
two rings are not inconsistent with a moderately larger orbital
velocity, as deduced here, if the rings were expelled at positions
slightly different from those adopted by CC12 or, most importantly, if
the orbit is not circular. For elliptical orbits, the radial velocity
curve exhibit a marked skew symmetry with different maximum and
minimum absolute radial velocities. In fact, for eccentric orbits the
observed radial velocity can be much smaller than the semi-amplitude
of the curve during most of the orbital phase. If this is the case,
the relatively low radial velocities derived from the
spatio-kinematics of the CO rings would not be incompatible with a
larger radial velocity measured at a third epoch from our RRLs,
probably nearer to periastron.

Finally, one can also consider the possibility that the ionized gas
traced by our mm-RRLs is not centered around the primary mass-losing
star but around the companion. In this case, the difference between
the centroids of the RRLs and CO transitions would represent the
orbital velocity of the less massive secondary (moving faster than the
primary about the mass center). If the companion is a very low-mass
star, $m_2$\la0.2, an external source of ionization, other than the
companion itself, would be needed. This is because a main sequence
star of this mass or lower would have a spectral type later than
$\sim$M\,4, i.e.,\,\teff$\la$3200\,K, and would not be able to ionize
the gas in its vicinity.  High-speed shocks in a putative accretion
disk around and/or within fast winds launched by the secondary
(presumably, after wind mass transfer from the primary) may constitute
an alternative ionization agent.  If this is correct, at present, the
companion must be undergoing active accretion from the wind of the
primary (given the short lifetime, $\la$10-20\,yr, of the ionized
core) and broader RRL profiles from the hot shocked emitting gas
would be expected.  In the case of high-speed shocks, one would also
expect partial ionization of the slow dense wind around the
mass-losing star by the UV radiation generated by the shocks.

With the observables available to date and, in particular, in the
absence of direct empirical determination of its orbital or stellar
parameters, an accurate characterization of the system at the core of
M\,2-9 is not possible.

\subsection{MWC\,922}
\label{h-mwc922}

The B\,[e] star MWC\,922 and its surrounding nebulosity remain poorly
characterized to date and, therefore, the long-term history of the
mass loss and nebular shaping cannot be reconstructed as for CRL\,618
and M\,2-9. Our mm-RRL
observations are consistent with the presence of an ionized outflow
and a Keplerian rotating disk within a radius of $\sim$150\,au from
the center. The origin of these two structures is unknown and cannot be
determined from our data.

The rotating kinematics of the disk, if it is Keplerian as adopted in
our model, provides an estimate of the mass of the central
object. This parameter is poorly constrained, mainly owing to the large
uncertainties of the departure coefficients $b_{\rm n}$ on which the
maser line profiles critically depend, but values of $\sim$5-10\,\msun\ are
consistent with our observations. This large mass could represent 
the total mass of a binary or, more generally, multiple system.  

The location of MWC\,922 in the HR diagram is uncertain, mainly due to
the unknown distance. The temperature of the central star is also not
well established, although it probably lies within the range
\teff$\sim$20,000-30,000\,K (\S\,\ref{intro-mwc922}). As discussed by
\cite{tut07}, the presence of gas and dust around
MWC\,922 suggests either a pre-main sequence star still partially
embedded in its natal cloud or a post-main sequence star surrounded
by matter ejected in its late evolutionary stages.

The range of \teff\ and luminosities given in Table\,\ref{t-buj} are
consistent, in principle, with MWC\,922 being a post-AGB object
evolved from a massive progenitor (Fig.\,\ref{f-tracks}). Considering
the post-AGB evolutionary tracks by \cite{blo95}, which include stars
with initial masses of up to 7\,\msun, MWC\,922 could be the
evolutionary product of a $\sim$5\,\msun\ main sequence star [adopting
  $d$=1.7\,kpc and $\log{(L_*/\ls)}$=4.25] or even more massive,
$>$7\,\msun\ [if $d$=3.0\,kpc and $\log{(L_*/\ls)}$=4.77]. The mass of
the remnant post-AGB core of MWC\,922 would be around 1\msun\ at
present.  Since the total mass at the center of the rotating disk of
MWC\,922 deduced from our analysis is significantly larger,
$\sim$5-10\,\msun, either a massive companion ($m_2$$\sim$ 4-9\,\msun)
or multiple lower mass companions must exist.  As explained earlier,
the mass of the central object in uncertain. If we take the lowest
value to be consistent with the observations, which is 5\,\msun, then the
secondary could be a $\sim$4\,\msun\ star still on the main sequence
[with $\log{(L_*/\ls)}$$\sim$2.2 and spectral type $\sim$B\,6-B7],
while the more massive primary (with an initial mass $\geq$5\,\msun)
would have evolved off the main sequence already (in less than
$\approx$10-100\,Myr).

If this scenario is correct and MWC\,922 is an evolved/post-main
sequence star, its age would still be sufficiently low to be
associated with other young less massive stars born in the same natal
cloud, which could also remain not fully dispersed.  This is not
inconsistent with the strong ISM contamination of the CO spectra
toward this target (see Appendix\,\ref{app-CO}), which could be
completely masking out the emission from the presumptive
  mass-ejecta from the star.
In any case, given its location close to the
Galactic plane, strong ISM contamination is expected toward MWC\,922.

If the total mass of the system at the core of MWC\,922 is
$\geq$8-10\,\msun, and $m_1$$\sim$1\,\msun, then the companion (or
companions) should account for the remaining stellar mass of
$\ga$7-9\,\msun. In this case, and if there is only one companion, a
pre-main sequence nature would be probably favored.
This is the case unless, of course, the initial mass of MWC\,922 was even larger,
i.e., $\geq$10\,\msun, and it is now in a pre-oxygen-neon WD 
or pre-supernova stage; this stage is expected to be
extremely short, which makes this scenario less probable.



\section{Post-AGB mass-loss rates and post-AGB evolution}
\label{dis2}

The post-AGB phase is, beyond question, one of the least understood
phases of the evolution of low-to-intermediate mass stars. Mass loss
is the dominant mechanism impelling these stars across the HR diagram
on their way to the CSPN phase, before fading along the white dwarf
cooling sequence.  The total AGB-to-PN transition time depends on the
heating rate of the central star, which is, indeed, critically
dictated by the mass-loss rate, especially in the early post-AGB
stages: the more intense the post-AGB wind, the faster the evolution
to the final (hottest) point of the CSPN phase. In spite of its
decisive role in stellar evolution, empirical data on post-AGB
mass-loss rates are notably lacking and, therefore, evolutionary
models are bound to adopt completely unconstrained mass-loss
prescriptions, for example, interpolating between those theoretically
or semi-empirically determined for the preceding AGB and the following
CSPN phases \citep[e.g.,][]{sch83,blo95,vas93,ber16}.

\begin{figure}[htbp]
   \centering 
\includegraphics[width=0.475\textwidth]{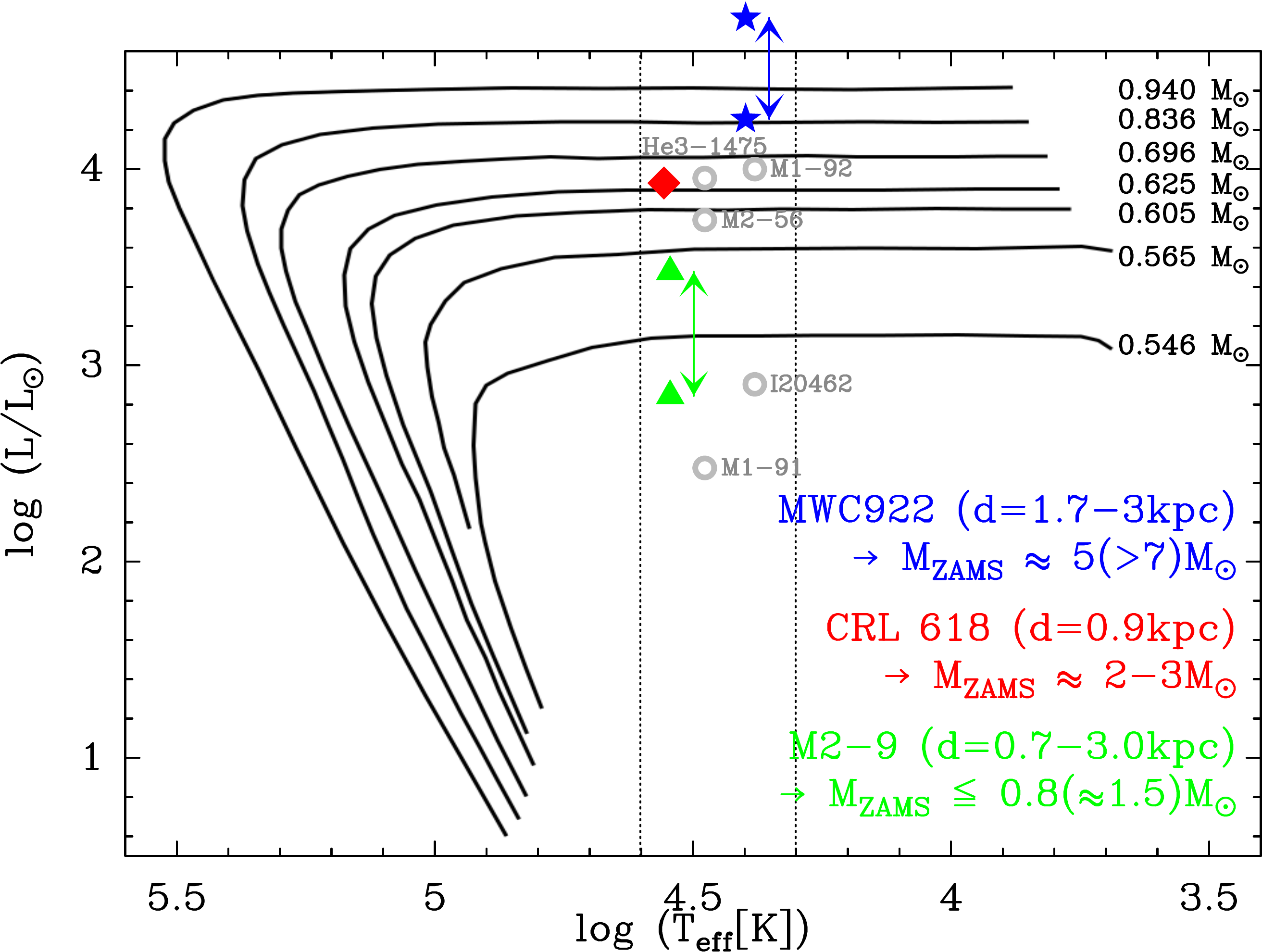} 

\vspace{0.25cm}
\includegraphics[width=0.485\textwidth]{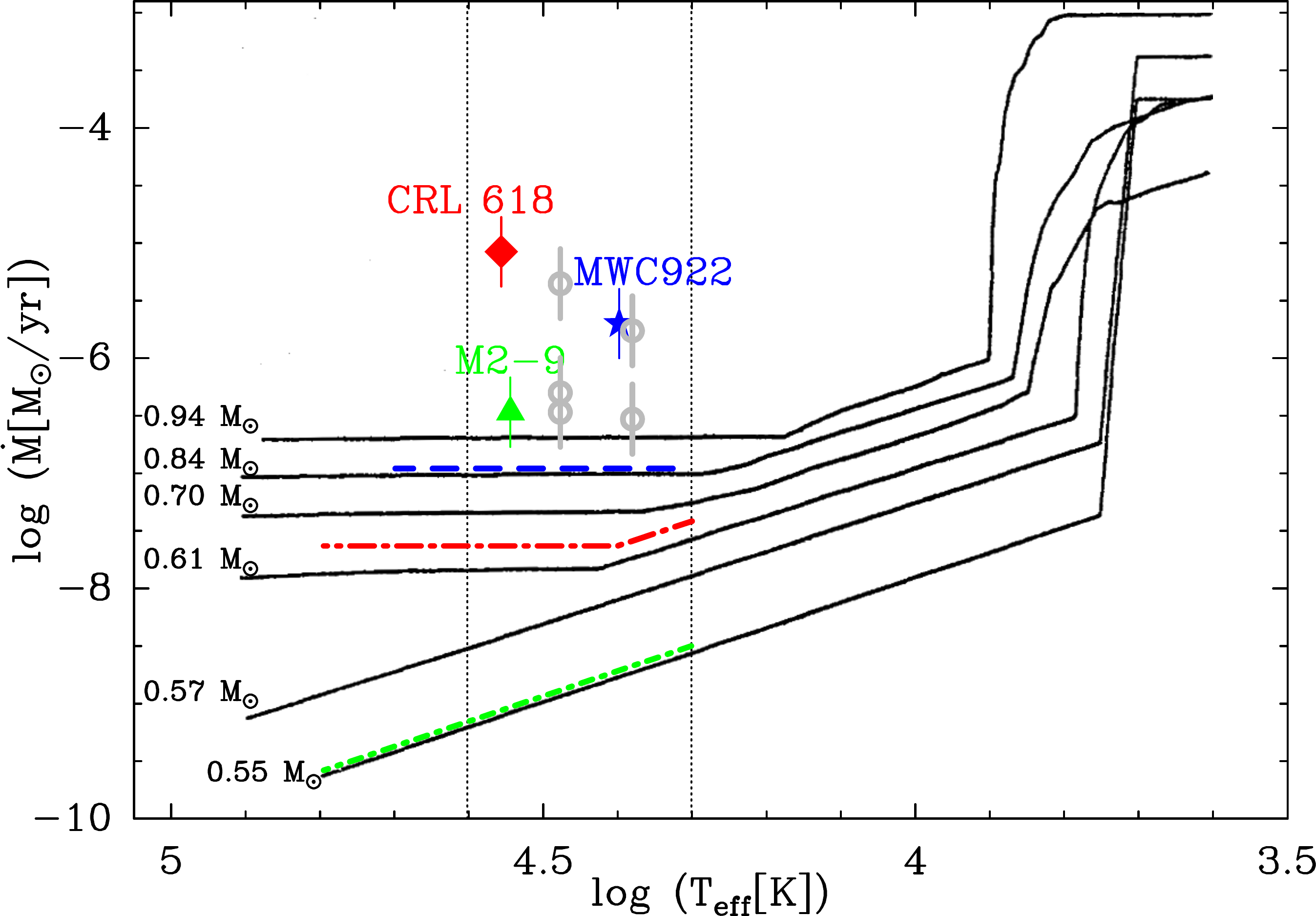}
\caption{Adapted from Figs.\,12 and 5 of \cite{blo95}. {\bf Top)}
  Post-AGB evolutionary tracks with observational data for RRLs and
  non-RRL detections (filled and empty symbols, respectively --
  Table\,\ref{t-buj}). For MWC\,922 and M\,2-9, the uncertainty of
  the luminosity due to the uncertainty of the distance is indicated
  by the arrows.  The stellar mass of the post-AGB remnant core of
  each track is indicated to its right.  Vertical dotted lines
  delimit the ionization onset region (\teff$\sim$20,000-40,000\,K)
  where our targets lie. {\bf Bottom)} Post-AGB mass-loss vs.
  effective temperatures adopted by post-AGB models of 0.605, 0.696,
  0.836, and 0.894\,\msun\ \citep{blo95} and of 0.546\msun\ and
  0.565\msun\ \citep[from][]{sch83}. As in the top panel, the symbols
  indicate the location of our targets using the values of
  \mpagb\ deduced in this work 
  (Tables\,\ref{t-moreli} and \ref{t-ND}; vertical error bars correspond to an
  uncertainty factor of 2).  The thick dash- and dot-dashed lines
  indicate
  the mass-loss rate typically assumed by evolutionary models for our three objects with RRL detections.}
   \label{f-tracks}
   \end{figure}

Commonly, the post-AGB mass-loss is described by
a sudden abrupt drop immediately after the departure from the AGB
phase (where very high rates of $\approx$10$^{-3}$-10$^{-4}$\,\my\ are
applied) followed by a less steep, linear
decline, and final flattening
in the CSPN regime (see Fig.\,\ref{f-tracks}, bottom panel, and
references above). The mass-loss rates derived for CRL\,618, MWC\,922,
and M\,2-9 in this work (Table\,\ref{t-moreli}) are well above the
typical values adopted in post-AGB evolutionary models, which vary
between \mpagb$\sim$1\ex{-9} and 2\ex{-7}\my\ for the range of
\teff\ considered here.

The largest discrepancy, by three orders of magnitude, is found for
the well-known pPN CRL\,618. The location of this object in the HR
diagram (Fig.\,\ref{f-tracks}, upper panel) is consistent with a
remnant core of $\sim$0.625\msun, which is the descendant of a star with an initial
mass of $\sim$3\msun\ on the main sequence, for which
\mpagb$\sim$1\ex{-8}\my\ would be assumed by evolutionary
models currently in use (Fig.\,\ref{f-tracks}, bottom panel).

The mass-loss rate deduced for M\,2-9 is also much higher than the
values adopted by models, although the progenitor mass in this case is
less certain because the distance to this object is not well
known. Adopting the largest distance proposed in previous works,
$d$=1.7\,kpc, the position of M\,2-9 in the HR diagram would be
consistent with a remnant mass of $\la$0.565\msun, which is the descendant of a
$\la$1\,\msun\ star, for which \mpagb$\sim$10$^{-9}$\,\my\ or less is
normally assumed (Fig.\,\ref{f-tracks}). The lowest luminosity point
in the HR diagram ($\log{(L_*/\ls)}$$\sim$2.8, adopting $d$=650\,pc)
falls below the evolutionary tracks of the least massive stars
($\sim$0.8\,\msun), which are expected to evolve off the main sequence
in less than a Hubble time.

In the case of MWC\,922, the mass-loss rate is also above the model
prescriptions assuming that it is the evolutionary product of a
$\geq$5\,\msun\ star (with a $\geq$0.836\msun\ remnant core).
However, not only the post-AGB nature of this object is questionable
(\S\,\ref{h-mwc922}); the origin of the expanding outflow (+rotating
disk), which is consistent with the observed mm-RRL line profiles, is
not clear since it could represent a stellar wind or gas
photoevaporating from the ionized surface of a circumstellar disk, or
a mixture of both.

Although the mass-loss rates inferred for the five objects in our
sample with no mm-RRL detections are uncertain (given that the
spatio-kinematics of their ionized cores is unconstrained), for all of
these objects we obtain relatively large values of
\mpagb$\approx$10$^{-6}$-10$^{-7}$\,\my\ (Table\,\ref{t-ND}) that
clearly exceed
the rates adopted by models.  For example, the most luminous pPNe/yPNe
in our sample with no mm-RRL detections, namely, He3-1475, M\,1-92,
and M\,2-56, would have similar $\sim$2-3\,\msun\ progenitors to
CRL\,618 for which models would use \mpagb$\approx$10$^{-8}$\,\my.
IRAS\,20462+3416 and M\,2-56 have luminosities that are comparable to M\,2-9
and, therefore, they also probably descend from a low-mass ($\la$1\,\msun)
progenitor.  For such a low-mass progenitor, the post-AGB
mass-loss rates used by models are 3 orders of magnitude below the
rates inferred by us.




As mentioned in the beginning of this section, the mass-loss rates
applied by models during the transition between the AGB and hot
CSPN stage are expected to influence the HR crossing
time dramatically.  Some models, for example,\, those by \cite{sch83} for
$\sim$0.8-1\msun\ stars, assume that most of the envelope mass is
removed at very high rates of
$\sim$10$^{-3}$-10$^{-4}$\,\my in a short
``superwind'' phase that happens only in the vicinity of the AGB,
i.e.\,for \teff$\sim$4000-6000\,K. After that point, which is
arbitrarily defined by most theorists as the beginning of the post-AGB
phase, the adopted mass-loss rates fall well below the growth rate of
the core due to hydrogen burning, and therefore, they are considered
to influence only moderately the transition of the star to the CSPN
region.  Other evolutionary models, such as those by \cite{blo95} for
stars with initial masses between 1 and 7\,\msun, however, indicate
that the transition time from the AGB to the CSPN region depends
strongly on the treatment of mass loss {\sl beyond} the AGB phase,
although rather low post-AGB mass-loss rates (inferred by
radiation-driven wind theories) are still applied in these
computations.

We have shown that the mass-loss rate in the post-AGB phase, when the
star has reached \teff$\approx$20,000\,K and the ionization has begun,
can be orders of magnitude larger than the values assumed by
evolutionary models.  If this is found to be a common property of most
post-AGB stars (which needs to be investigated using larger samples), much faster evolutionary speeds to the PN region may be expected. 

A similar result was obtained by \cite{tra89} from the analysis of
H$\alpha$ P-Cygni profiles in four candidate post-AGB stars with
\teff$\sim$6500-10,000\,K, which yielded order-of-magnitude estimates
of the mass-loss rate of $\sim$10$^{-8}$ to 10$^{-7}$\my. These
authors discussed the drastic consequences of high mass-loss rates for
the formation of PNe, especially for objects with low-mass
progenitors. For example, for a low-luminosity object like M\,2-9,
with a $\sim$0.55\msun\ core, the time needed to evolve from
\teff$\sim$5000\,K to its current temperature is $\sim$100,000 yr
when adopting the model mass-loss prescriptions, $\la$2\ex{-9}\my, but less
than $\sim$5000 yr when adopting the observed rate (see
Fig.\ref{f-tracks}). The latter transition time leads to a much better
agreement with the dynamical lifetime of the nebulosity in this and 
other similar objects.

\section{Conclusions}
\label{summ}

We have detected RRL emission at mm-wavelengths from the central
ionized regions of CRL\,618, M\,2-9, and MWC\,922 at a moment when
these \ion{H}{ii} region are in a very early stage of development after
the ionization onset.  The RRLs at mm-wavelengths trace the inner regions
of these targets at scales $\approx$10-100\,au, indicating bulk motion
speeds of $\sim$10-30\,\kms.

Our analysis of the RRL spectra and free-free continuum data included
detailed modeling of CRL\,618, M\,2-9, and MWC\,922 using the non-LTE
radiative transfer code MORELI. This has enabled us to set constraints
of the density and temperature profiles and morpho-kinematics in the
ionized cores of these objects. Typical densities are found to range
between \dense$\approx$10$^6$ and $\approx$10$^8$\,\cm3, following
radial power-law distributions
\dense($r$)\,$\propto$\,$r^{[-2.1:-2.4]}$. The total mass of ionized
gas at the core of these objects is
$\approx$10$^{-6}$-10$^{-4}$\,\msun, which is currently being ejected
at rates of $\approx$10$^{-6}$-10$^{-7}$\,\my.  Simple models of the
free-free continuum emission have also been performed for the
remaining five targets with no mm-RRL detections under the assumption
that their ionized cores are spherical isothermal winds. The post-AGB
mass-loss rates derived in these cases range between
\mpagb$\sim$3\ex{-7} and 4\ex{-6}\,\my.  These rates are significantly
higher than those adopted by current post-AGB evolutionary models. New
evolutionary tracks should be made allowing post-AGB stars to lose
mass at higher rates, to study their
influence on the star heating rate, AGB-to-PN transition times,
initial-final mass relation, etc.

\subsection{Future prospects}
Our models are necessarily simplified representations of the
presumably more complex spatio-kinematical and physical structure of
the dense stellar surroundings in these objects.  We need RRL emission maps
with \la\,0\farcs05-angular resolution to better constrain
the properties of these very inner winds, in particular, to test our
conjectures on nebular morphology and kinematics.

For example, these maps are needed to check the rotating `disk+wind' scenario
in MWC\,922; if confirmed, it will be possible to obtain from the disk
kinematics a more reliable estimate of the mass of the central
source(s) and, thus, to clarify its nature. Disentangling the
composite nebular structure of MWC\,922 and, in particular, isolating
the wind from the disk is needed for an accurate interpretation of the
mass loss in this case. In parallel, measuring the outflow
rotation and studying the disk dynamics can provide fundamental
information regarding the origin of the outflow and the role of the disk in
the wind launching/collimation process.  High-angular resolution maps
of RRLs are also needed to confirm or not the expansive kinematics and
the suspected velocity gradient in the \la\,100\,au inner regions of
CRL\,618 and M\,2-9: Does it follow a \vexp$\propto r$ law as observed
at much larger linear scales in these, and most, pPNe? \ Also, these maps are necessary to unveil
the presence of {\sl jet}-like features, of which {\sl indirect}
observational indicia exist but whose direct characterization remains
elusive. High angular-resolution of the ionized core of M\,2-9 is also
critical for a truthful judgement of the binary system at its nucleus
and to unravel the different ionized gas (and dust) components that
contribute to the observed mm-continuum.

With its unprecedented capabilities, ALMA is the only facility in the
world that is able to map with sufficient sensitivity and angular
resolution the dense inner winds of young PNe at the precise moment
when these come into existence and begin to be ionized.


\begin{acknowledgements} 
We thank the referee for his/her comments and very valuable
suggestions.  The data presented in this paper were reduced using
the packages available in the GILDAS software ({\tt
  http://www.iram.fr/IRAMFR/GILDAS)}.  This work has been partially
supported by the Spanish MINECO through grants CSD2009-00038,
AYA2009-07304, AYA2010-2169-C04-01, AYA2012-32032, AYA2016-78994-P,
FIS2012-39162-C06-01, ESP2013-47809-C03-01 and ESP2015-65597-C4-1.
A.\ B\'aez-Rubio acknowledges support from DGAPA postdoctoral grant
(year 2015) to UNAM.  This research has made use of the SIMBAD
database and VizieR catalog access tool (CDS, Strasbourg, France),
the NASA Astrophysics Data System, and Aladin.


\end{acknowledgements}

%
%

\newpage 
%

%
%

%


%

\newpage

\begin{appendix}

\section{Spectra of the bonus \docetto\ and \trece\ lines.}
\label{app-CO}
\begin{figure*}
   \centering 
   \includegraphics*[bb=-10 0 820 470,width=0.45\textwidth]{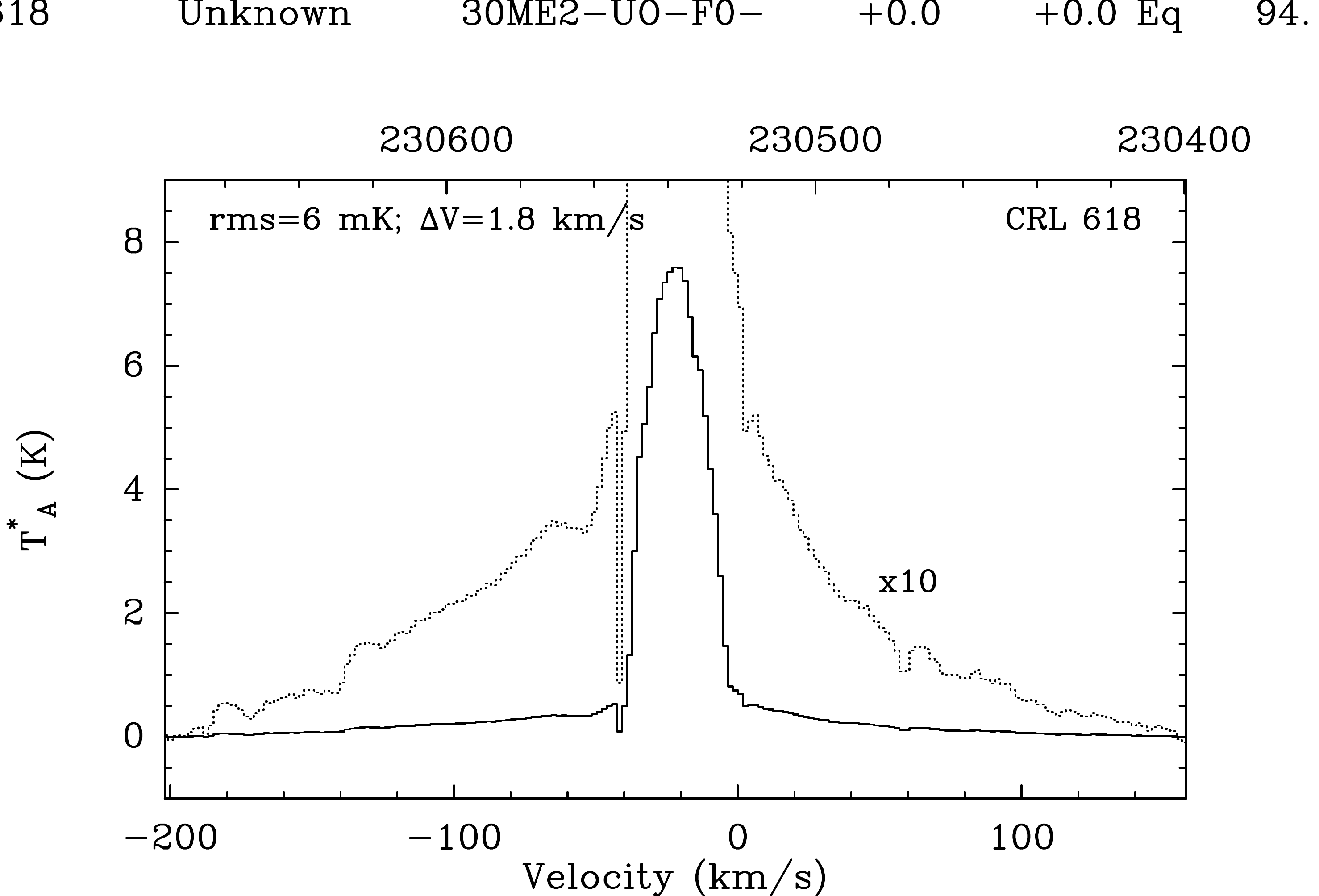}
   \includegraphics*[bb=-10 0 820 470,width=0.45\textwidth]{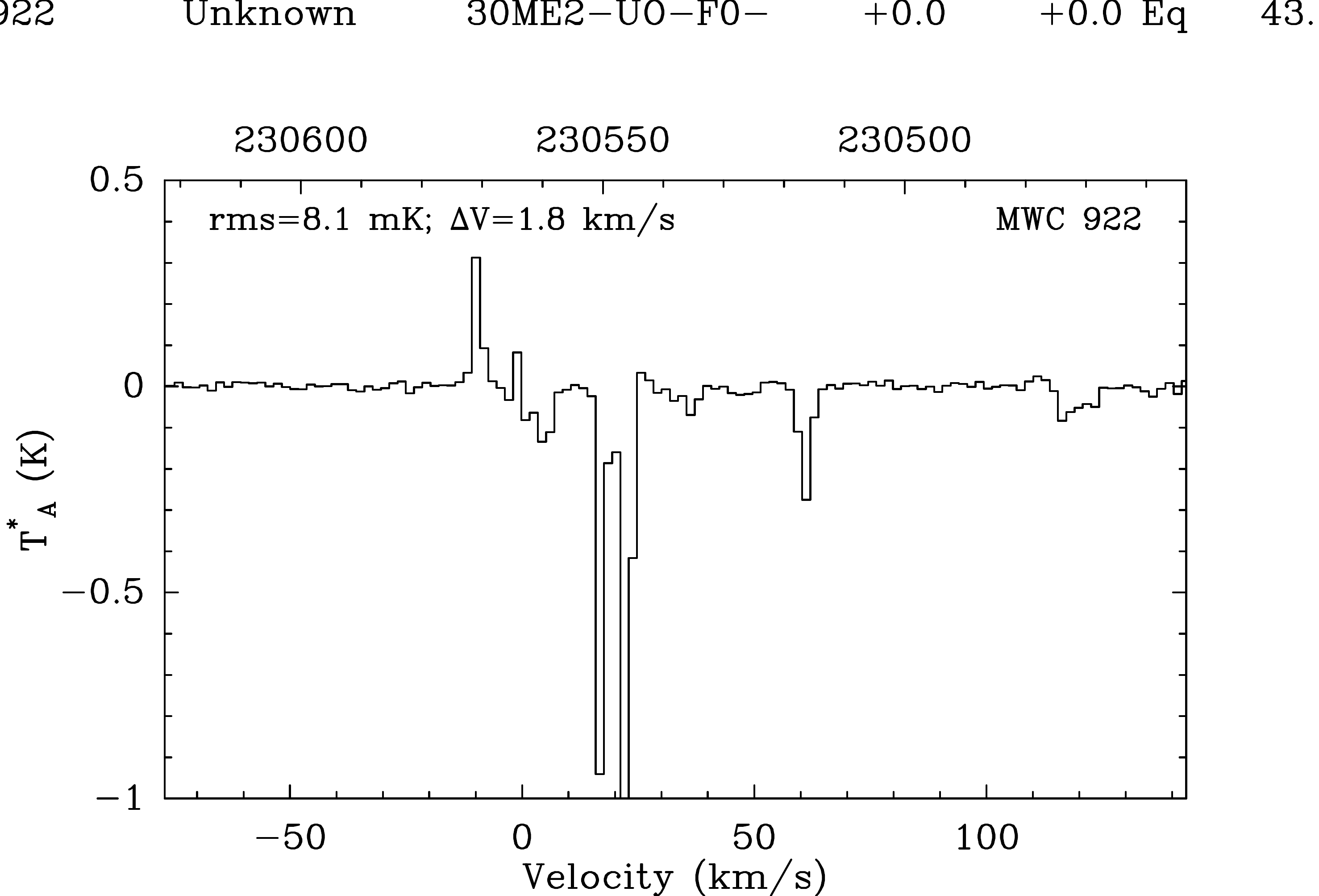} 
   \includegraphics*[bb=-10 0 820 470,width=0.45\textwidth]{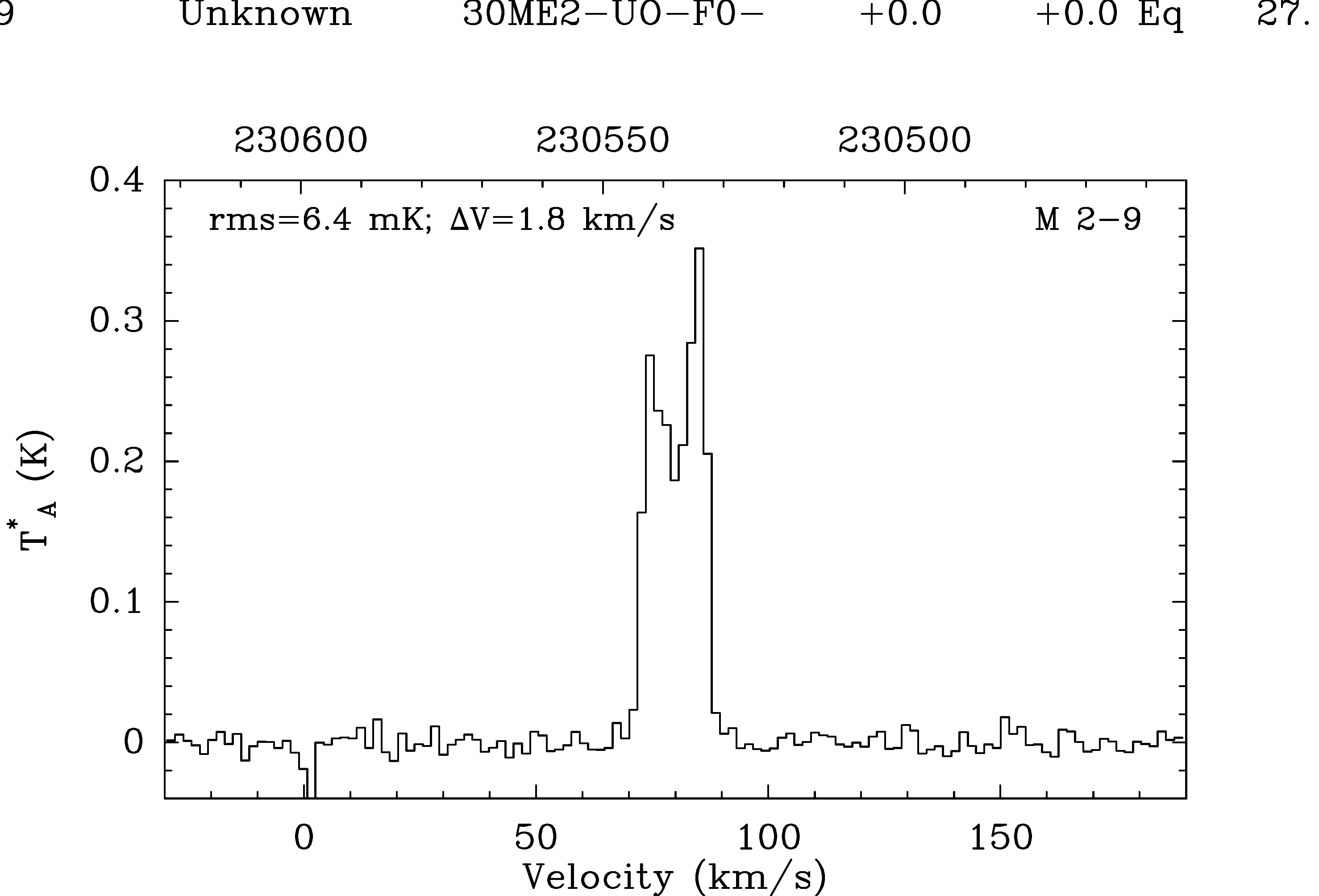}
   \includegraphics*[bb=-10 0 820 470,width=0.45\textwidth]{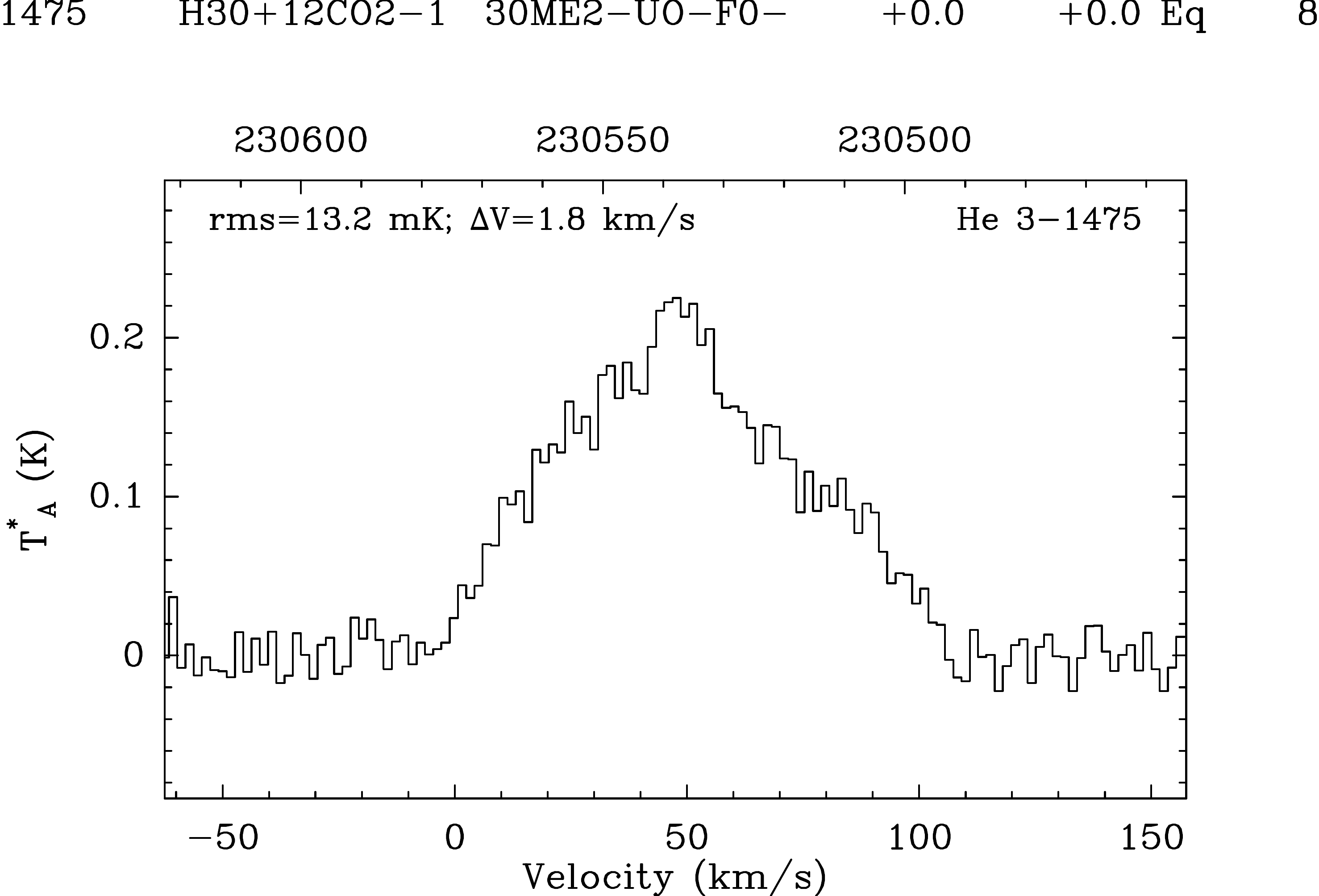} 
   \includegraphics*[bb=-10 0 820 470,width=0.45\textwidth]{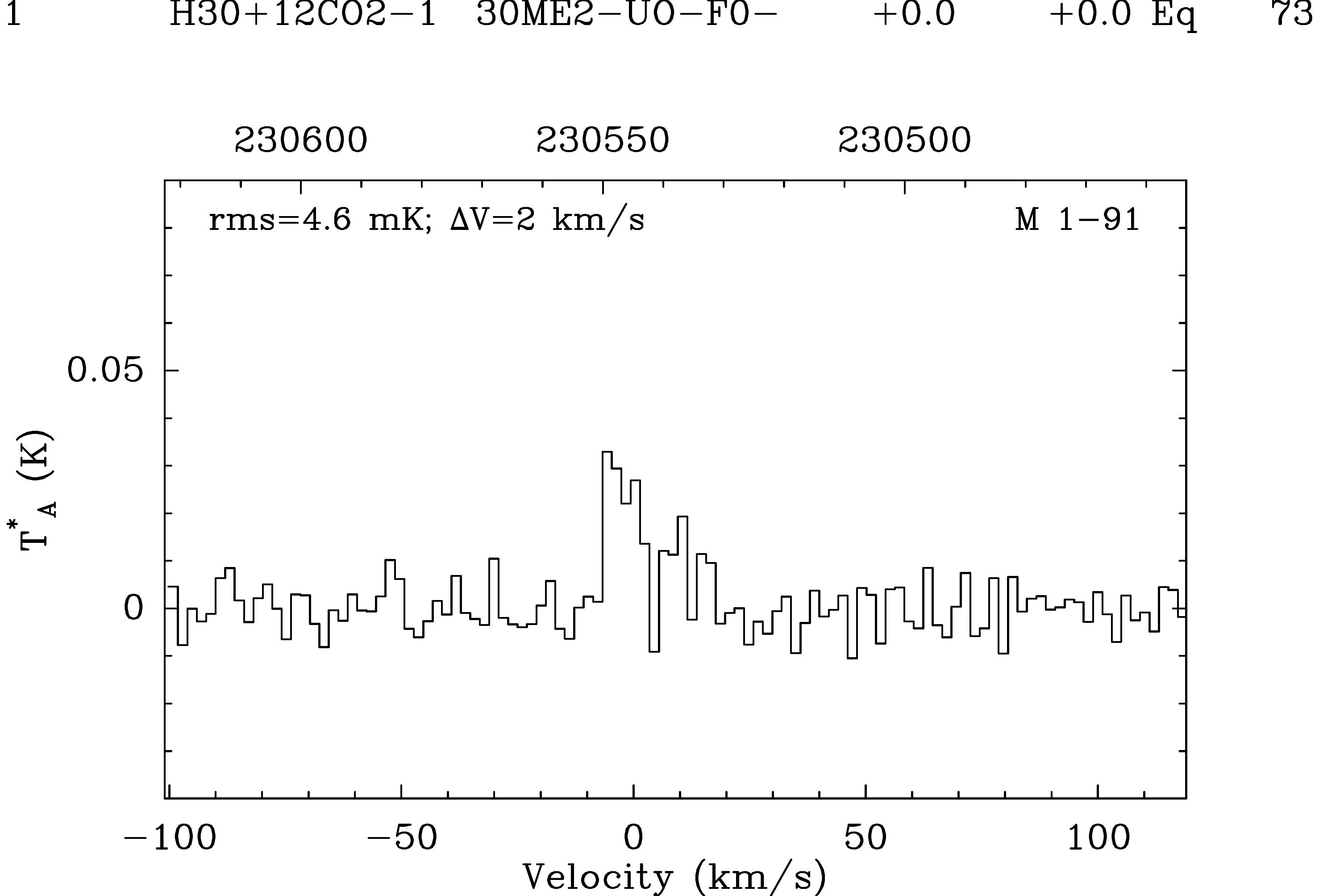}
   \includegraphics*[bb=-10 0 820 470,width=0.45\textwidth]{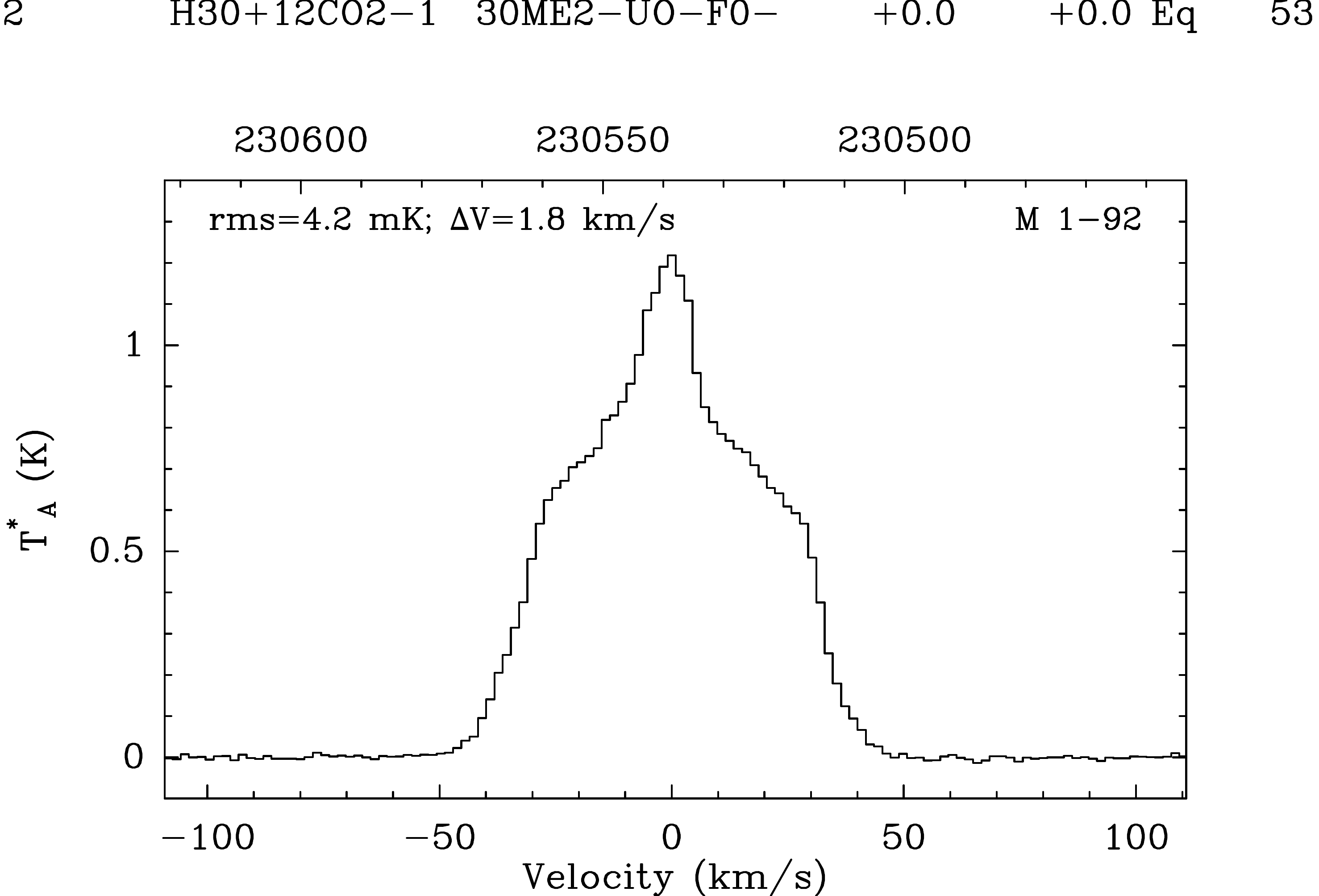} 
   \includegraphics*[bb=-10 0 820 470,width=0.45\textwidth]{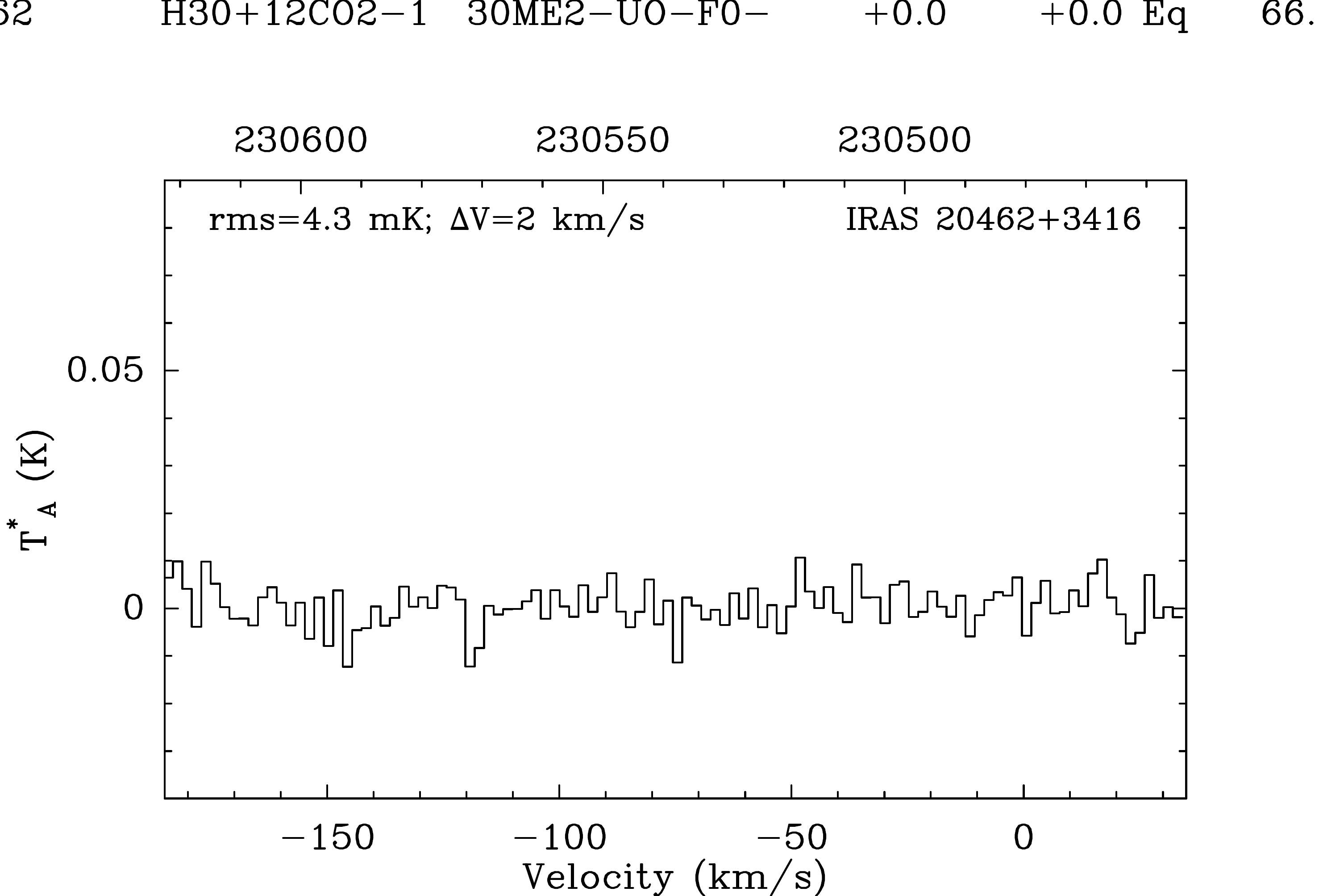}
   \includegraphics*[bb=-10 0 820 470,width=0.45\textwidth]{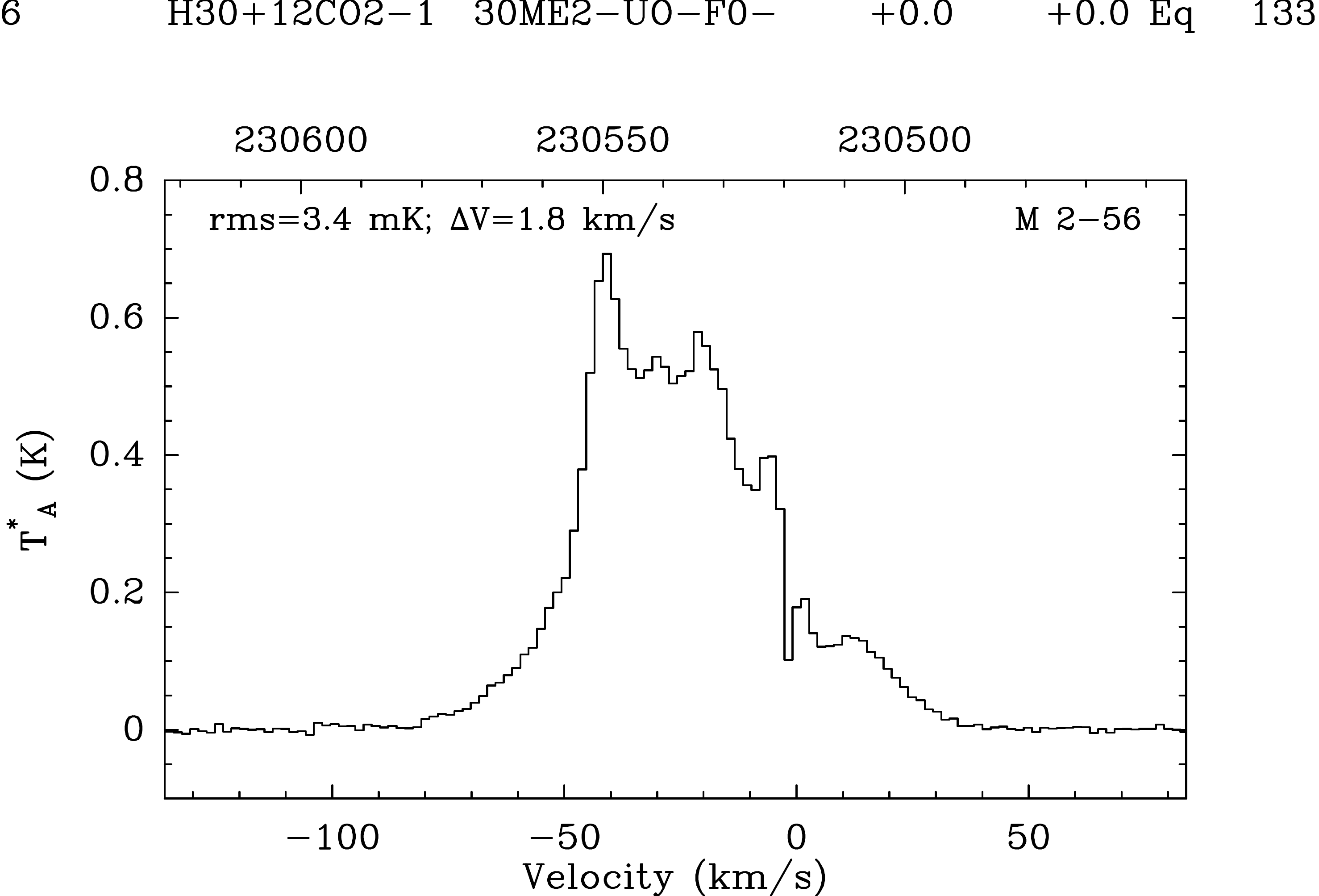} 

   \caption{Spectra of the \docetto\ transition observed toward our sample targets (Table\,\ref{t-buj}).}
   \label{f-co1mm}
   \end{figure*}

\begin{figure*}
   \centering 
   \includegraphics*[bb=-10 0 820 470,width=0.45\textwidth]{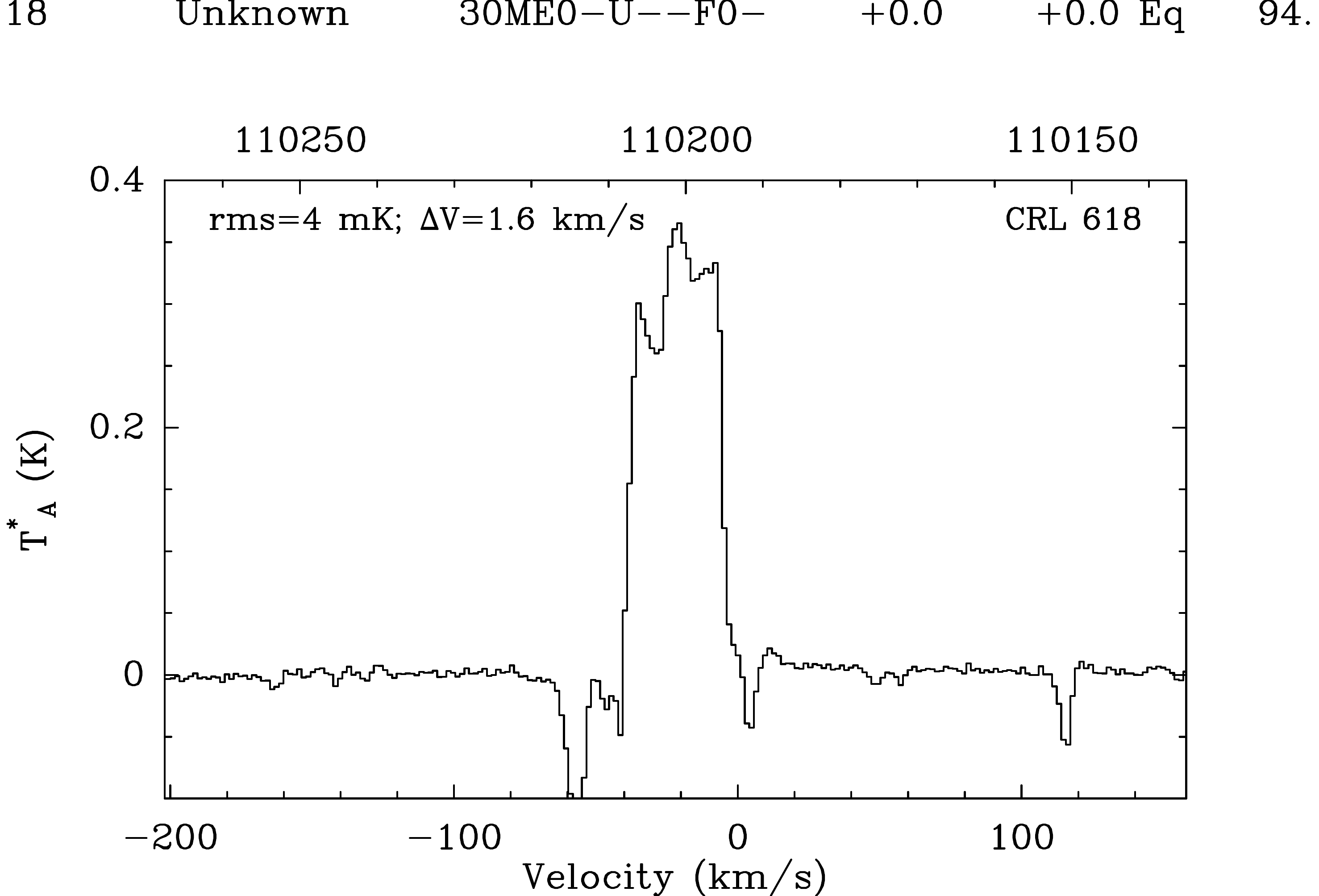}
   \includegraphics*[bb=-10 0 820 470,width=0.45\textwidth]{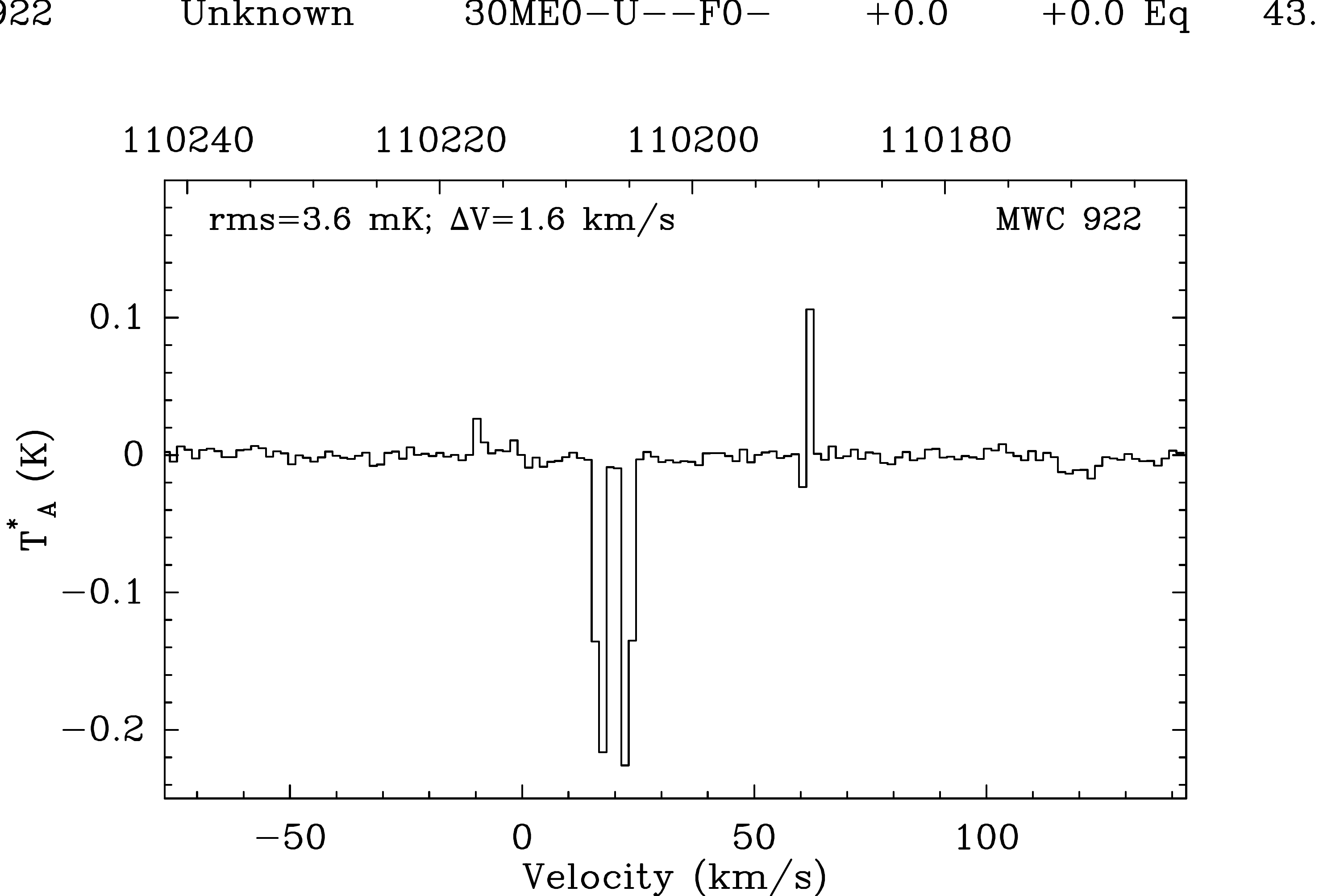} 
   \includegraphics*[bb=-10 0 820 470,width=0.45\textwidth]{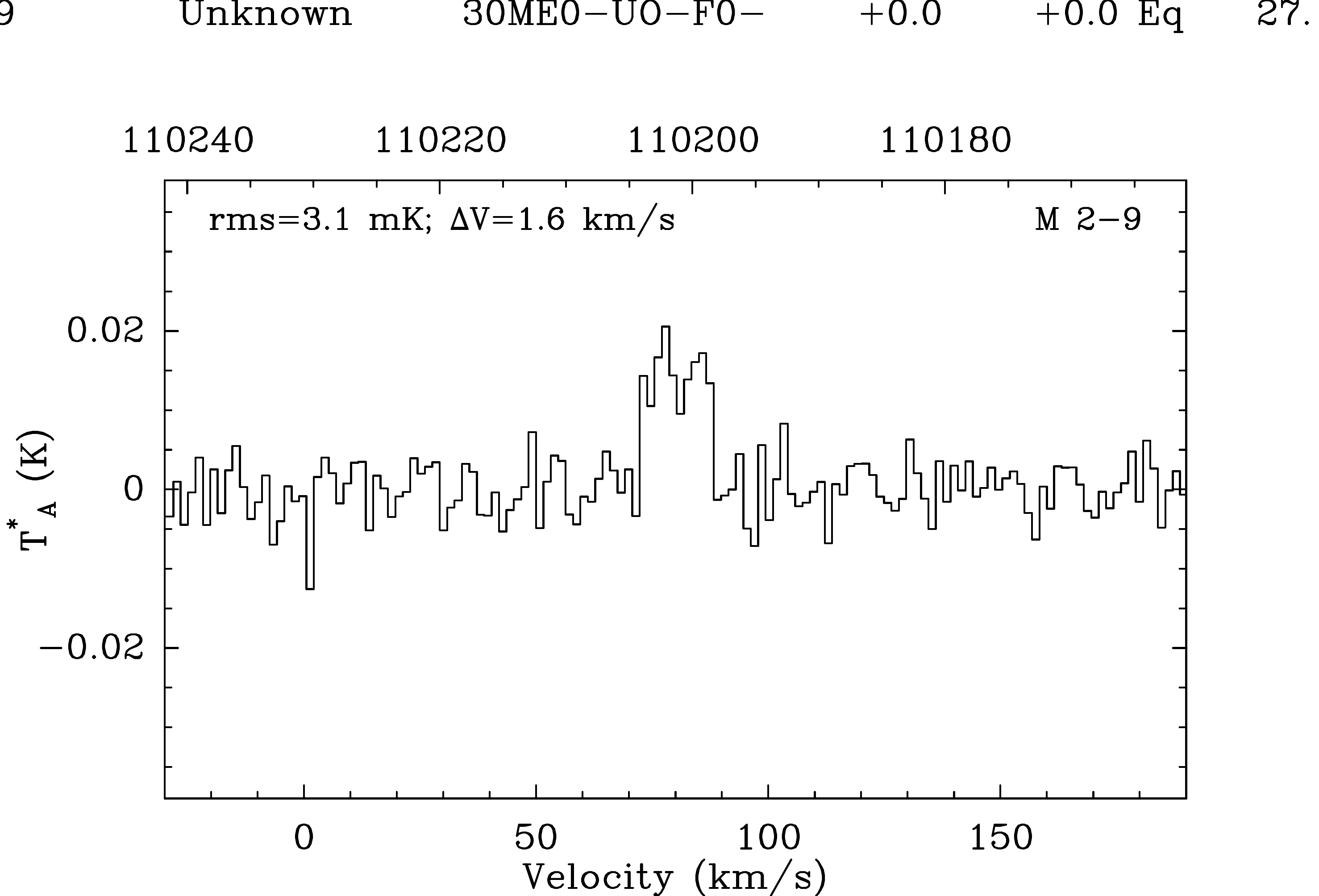}
   \includegraphics*[bb=-10 0 820 470,width=0.45\textwidth]{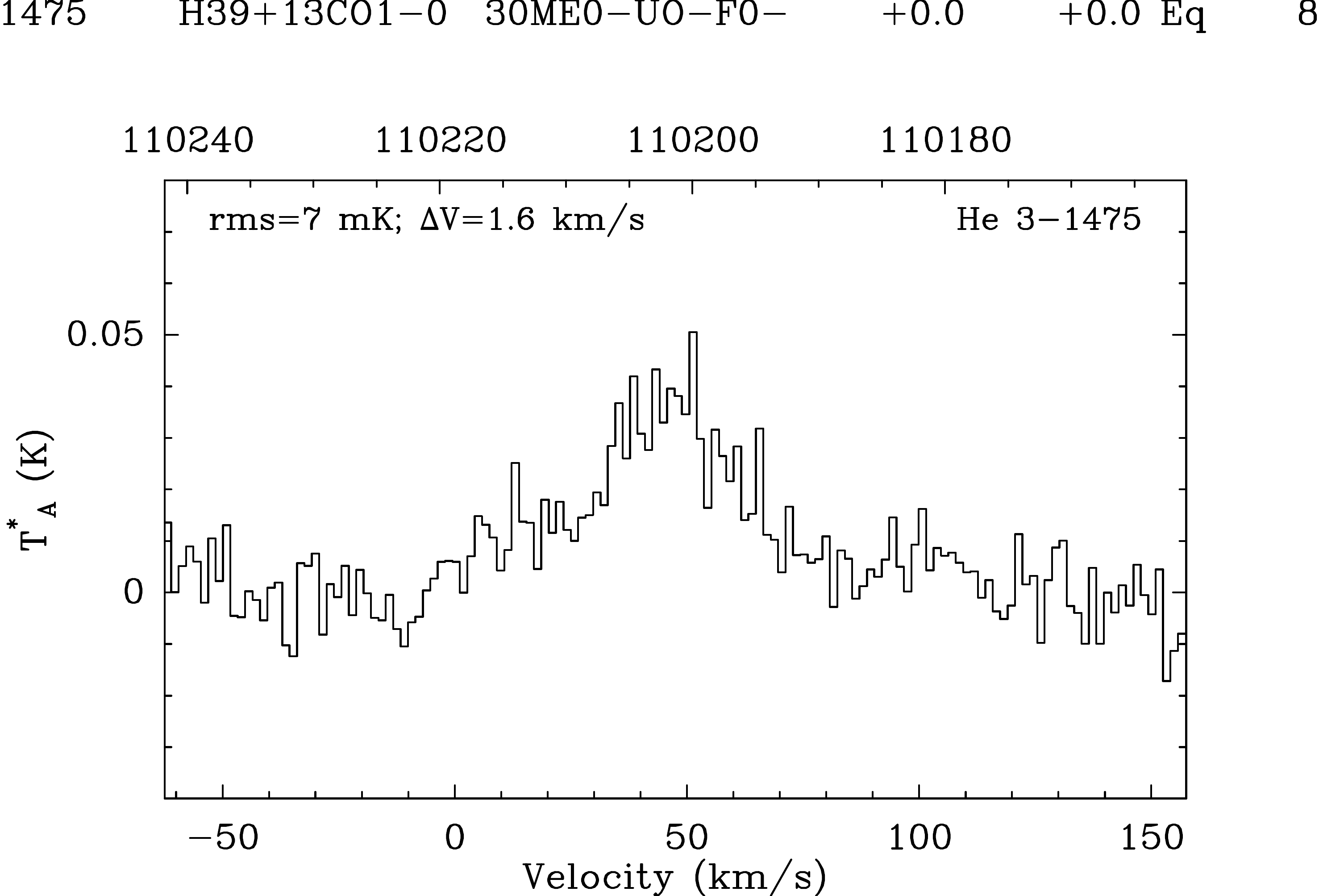} 
   \includegraphics*[bb=-10 0 820 470,width=0.45\textwidth]{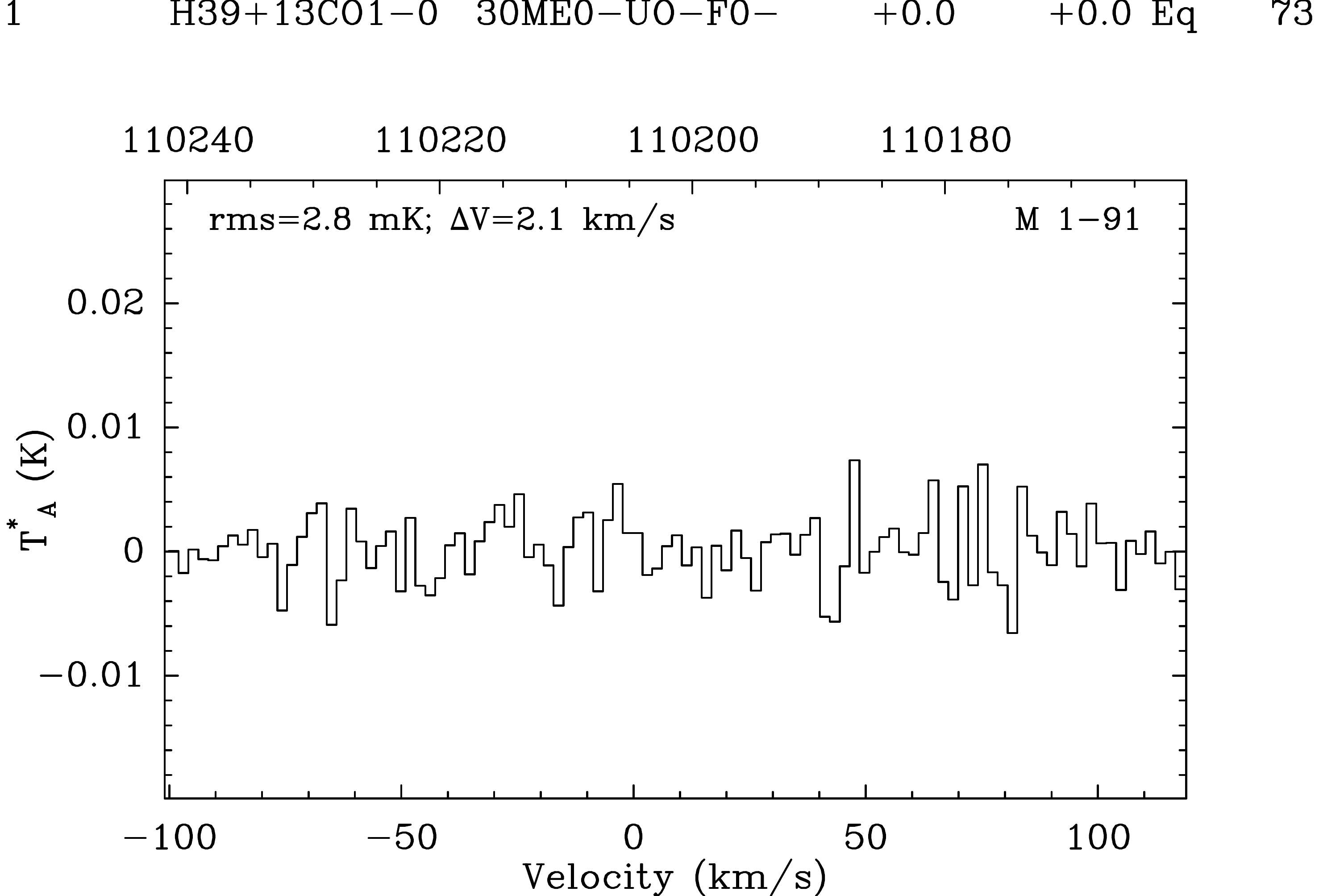}
   \includegraphics*[bb=-10 0 820 470,width=0.45\textwidth]{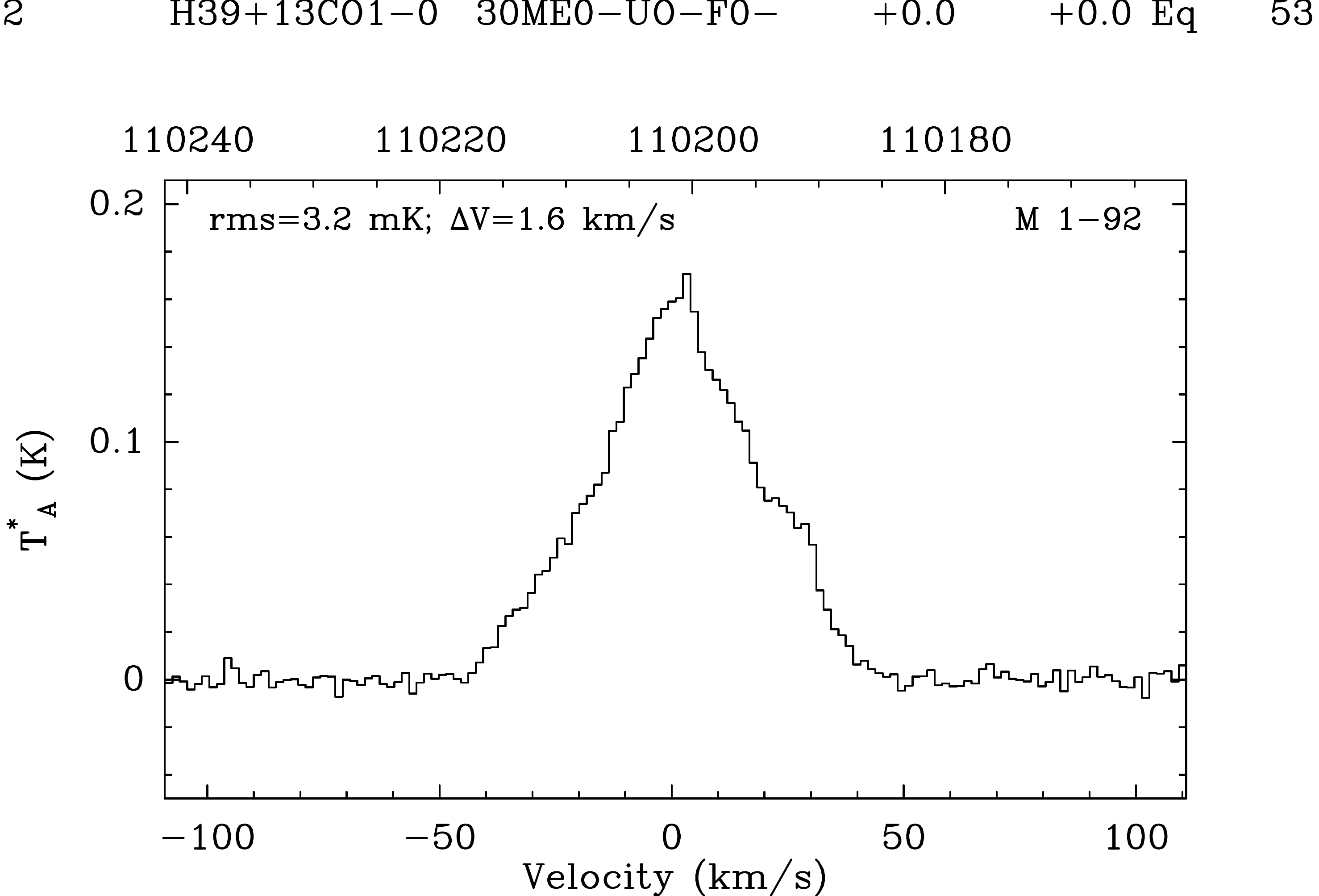} 
   \includegraphics*[bb=-10 0 820 470,width=0.45\textwidth]{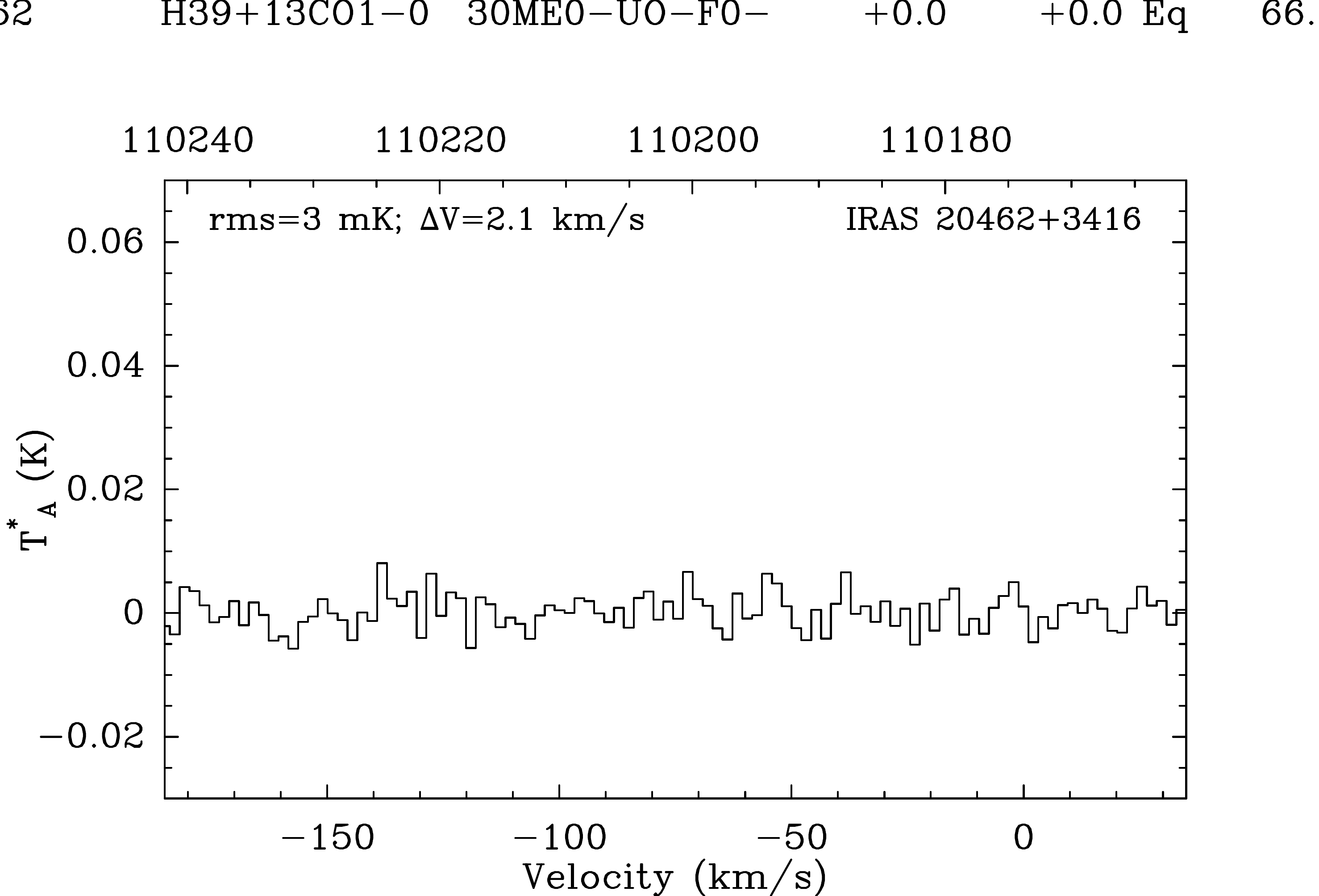}
   \includegraphics*[bb=-10 0 820 470,width=0.45\textwidth]{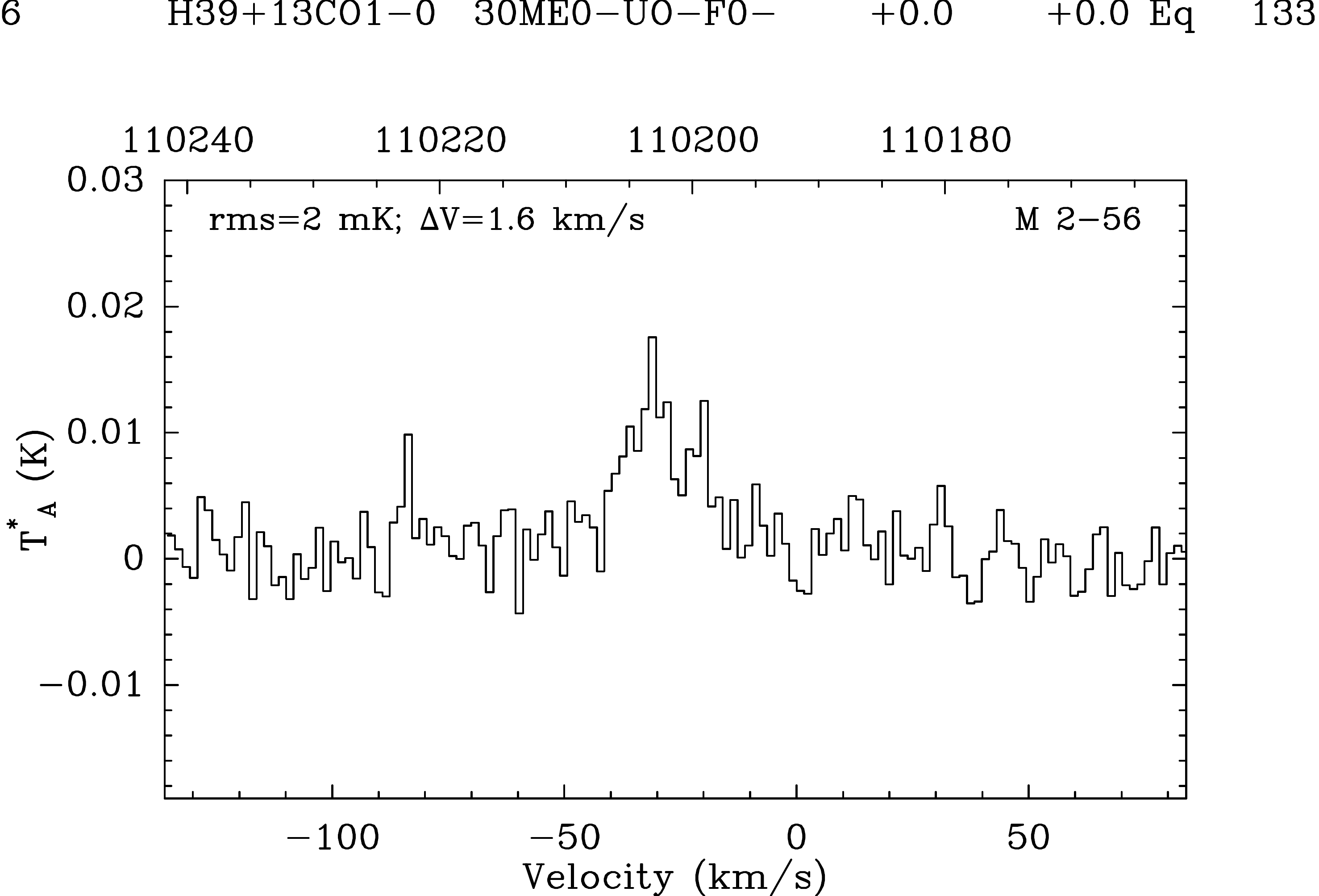} 

   \caption{Spectra of the \trece\ transition observed toward our sample targets (Table\,\ref{t-buj}).}
   \label{f-13co3mm}
   \end{figure*}

In this Appendix we report the \docetto\ and \trece\ spectral
     profiles observed toward our targets in this work (see
     Section\,\ref{obs}). Spectra are channel-smoothed to a
     final velocity resolution of $\sim$2\,\kms\ and are shown in
     units of \ta\ (K) and \vlsr\ (\kms) in Figs.\,\ref{f-co1mm} and
     \ref{f-13co3mm}.

     All our sample targets are CO emission detections except for
     IRAS\,20462+3416, which is also non-detected in previous surveys
     \citep[e.g.,][]{san12,he14}, and MWC\,922. This is the first
     detection of CO emission in the young PN M\,1-91 after some
     unsuccessful previous searches \citep[e.g.,][]{jos00,buj01}.  To our knowledge, these are
     also the first published CO spectra toward MWC\,922. The narrow
     absorption\ and fewer emission features observed toward this
     object are most likely produced by intervening ISM clouds along
     the line of sight. For CO detections, we find line profiles and
     line fluxes in good agreement with previous observations, whenever available within typical $\sim$20-30\%\ calibration uncertainties 
     \citep[e.g.,][]{buj01,san12,he14}.

%

%
%
\end{appendix}

   \listofobjects
\end{document}